\documentclass[superscriptaddress,preprintnumbers,amssymb]{revtex4}
\usepackage{graphicx}
\usepackage[utf8]{inputenc}
\usepackage{epsfig}
\usepackage{amsmath}
\usepackage{amsfonts}
\usepackage{amssymb}

\usepackage{braket}
\usepackage{cancel}
\usepackage{pstricks}
\usepackage{color}

\usepackage{epstopdf}

\begin{document}
{\tiny }\global\long\def\bra#1{\Bra{#1}}
{\tiny }\global\long\def\ket#1{\Ket{#1}}
{\tiny }\global\long\def\set#1{\Set{#1}}
{\tiny }\global\long\def\braket#1{\Braket{#1}}
{\tiny }\global\long\def\norm#1{\left\Vert #1\right\Vert }
{\tiny }\global\long\def\rmto#1#2{\cancelto{#2}{#1}}
{\tiny }\global\long\def\rmpart#1{\cancel{#1}}
{\tiny \par}

\title {Contact Interactions in Higgs-Vector Boson Associated Production at the ILC}
\author{Jonathan Cohen}
\email{jcohen@tx.technion.ac.il}
\affiliation{Physics Department, Technion-Institute of Technology, Haifa 32000, Israel}
\author{Shaouly Bar-Shalom}
\email{shaouly@physics.technion.ac.il}
\affiliation{Physics Department, Technion-Institute of Technology, Haifa 32000, Israel}
\author{Gad Eilam}
\email{eilam@physics.technion.ac.il}
\affiliation{Physics Department, Technion-Institute of Technology, Haifa 32000, Israel}

\date{\today}

\begin{abstract}
We explore new physics (NP) effects in Higgs-Vector boson associated
production at a future International Linear Collider (ILC)
via $e^{+}e^{-}\rightarrow Zh\,,Zhh$, using effective field theory (EFT)
techniques. In particular, we focus on a certain class of dimension 6 operators,
which are generated by tree-level exchanges of a new heavy vector field
in the underlying theory.
These operators induce new contact terms of the form $\psi\psi\phi D\phi$,
involving the Standard Model (SM) fermions ($\psi$),
gauge-bosons ($D$ is the covariant derivative) and the SM Higgs field ($\phi$).
We investigate the high-energy behaviour of these new effective interactions in
$e^{+}e^{-}\rightarrow Zh\,,Zhh$, imposing bounds from electroweak
precision measurements, and show that the
ILC is an excellent testing ground for probing this type of NP via $e^{+}e^{-}\rightarrow Zh\,,Zhh$.
We also address the validity of the EFT expansion and we
study the correlation between the $hZ$ and $hhZ$ signals, which can be utilized
in future searches for NP in these channels.
\end{abstract}


\maketitle

\section{Introduction}

The SM is by now a well established theory and has
been tested with an astounding accuracy. Nonetheless, since the SM
does not address some of the fundamental theoretical issues in particle physics,
such as the hierarchy problem, dark matter,
neutrino masses, flavor and CP violation, it is widely believed that the
NP which underlies the SM is around the corner, i.e., at the few TeV
scale. This has driven
physicists throughout the years to search for new theories beyond the SM, which,
in many cases, predict the existence of new particles.

In this paper we investigate NP effects in Higgs - Vector boson associated production at
a future $e^+ e^-$ collider, via $e^{+}e^{-}\rightarrow hZ$ \cite{Bjorken:1977wg,Ellis:1975ap} and $e^{+}e^{-}\rightarrow hhZ$ \cite{Gounaris:1979px}. These
processes are sensitive to a variety of underlying NP scenarios.
Of the many examples in the literature, let us briefly mention
studies of Higgs - Vector boson associated production processes in
Little Higgs models with T parity \cite{Tparityhz}, in supersymmetry where
the $e^{+}e^{-}\rightarrow hZ$ cross-section receives one-loop corrections
which are sensitive to the stop mass \cite{susyhz} and in models of extra
compact dimensions, in which strong gravitational interactions at the
TeV scale lead to virtual exchange of KK gravitons that
affect $hhZ$ production at the ILC \cite{kkhhz}.
The $e^{+}e^{-}\rightarrow hhZ$ cross-section can also be modified
in Two Higgs doublet models (2HDM) and in models with scalar leptoquarks
due to enhanced one-loop corrections to the triple Higgs coupling \cite{2hdmhhz}.

\begin{figure}
\begin{center}
\includegraphics[scale=0.25]{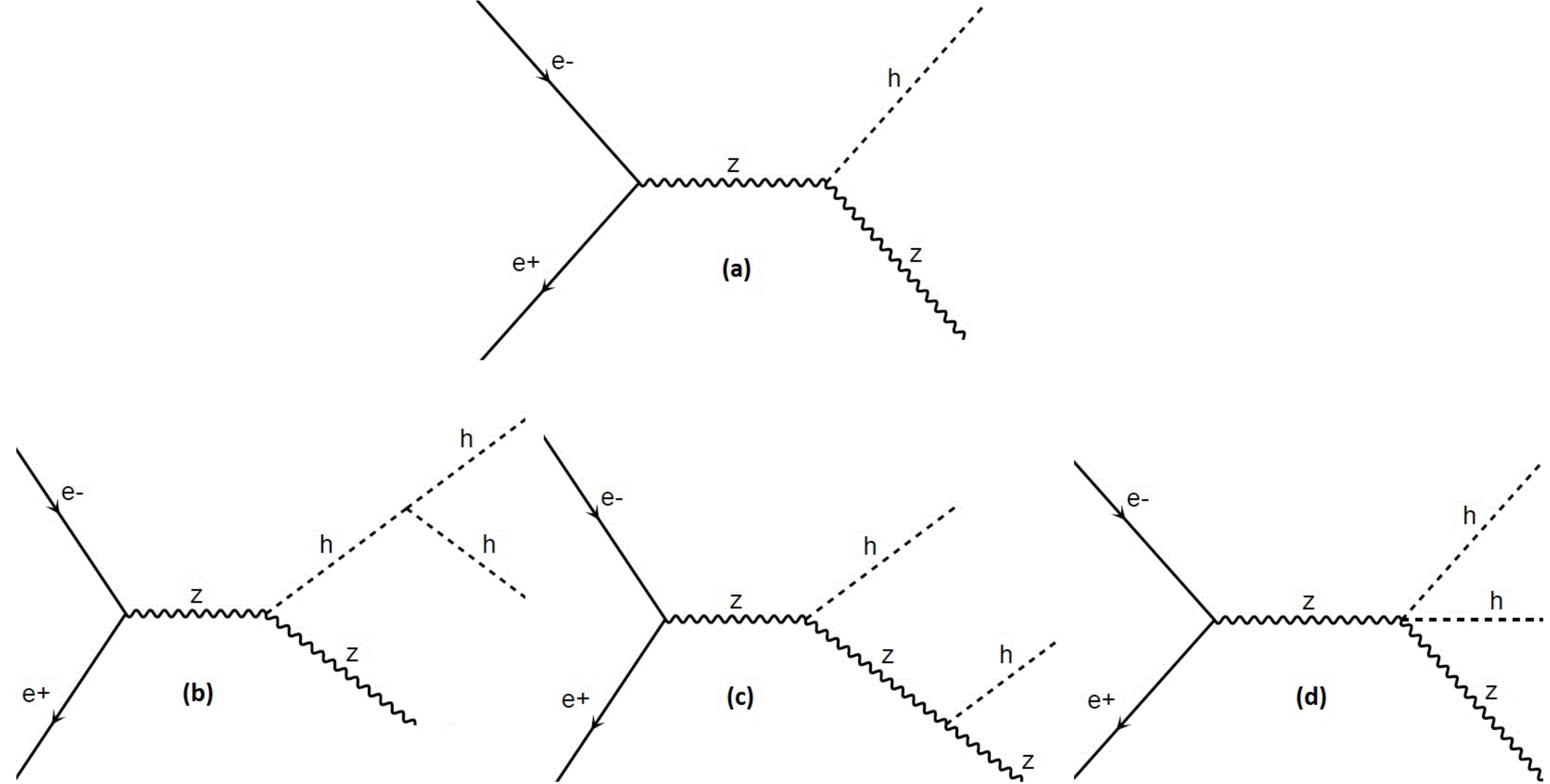}
\end{center}
\caption{Tree-level SM diagrams for $e^{+}e^{-}\rightarrow Z\rightarrow hZ$ (a) and for
$e^{+}e^{-}\rightarrow Z\rightarrow hhZ$ (b,c,d).
\label{SMdiag}}
\end{figure}

Over the years, the grueling task of the search for NP beyond the SM, also involved model-independent
studies, which utilize EFT techniques to explore new interactions among the SM
particles.
In this work we adopt the EFT approach and study the effects of new Higgs - Vector Boson
- fermion interactions in Higgs-Vector boson associated production
at the ILC via the processes $e^{+}e^{-}\rightarrow Zh\,,Zhh$, see Fig.~\ref{SMdiag}. We
parameterize the new effective interactions through higher-dimensional operators
assuming that:
\begin{itemize}
\item The new interactions obey the gauge symmetries of the SM:
$SU\left(3\right)_{C}\times SU\left(2\right)_{L}\times U\left(1\right)_{Y}$.

\item The underlying NP is weakly coupled, renormalizable and decoupled from the SM at low energies.

\item The light fields, i.e., the observable degrees of freedom below
the cutoff $\Lambda$ (see below), are the SM fields.
\end{itemize}

Within this EFT setup the SM is treated as a low-energy effective
theory and the new interactions are characterized by a new scale $\Lambda\gg v$,
which represents the scale (threshold) of the NP. The effective theory
is then described by:
\begin{eqnarray}
\mathcal{L} & = & \mathcal{L}_{SM}+\sum_{n}\frac{f_{i}^{\left(n\right)}}{\Lambda^{n-4}}\mathcal{O}_{i}^{\left(n\right)} ~,
\label{EFT}
\end{eqnarray}
where $i$ denotes the operator type, $n$ is its dimension and $f_{i}^{\left(n\right)}$
are the corresponding Wilson coefficients.

In principle, the $\mathcal{O}_{i}$'s are generated by integrating out the heavy
fields in the underlying theory; the different types of operators then
depend on the quantum numbers of
the exchanged heavy fields. Thus, a
generic EFT is, by construction, valid up to the scale $\Lambda$
(of the NP), so that by performing a measurement in a future collider
one can extract information on the ratio $f_{i}^{(n)}/\Lambda^{n-4}$ and,
therefore, hope to find clues regarding the underlying
theory \cite{Buchmu,patterns,GIMR,trott1,EFT1,EFT2,EFT3,trott2}
(for a comprehensive analysis of the renormalization of the dimension 6 operators
and its importance for precision studies of the SM EFT framework in (\ref{EFT}),
see \cite{ren1,ren2,ren3,ren4,ren5}). 
In that respect, we note that, under the assumption that the underlying NP is weakly coupled,
the dimensionless
coefficients $f_{i}^{(n)}$ in the underlying theory are expected to be
of $\mathcal{O}\left(1\right)$.

In this work we limit ourselves
to dimension 6 operators, $\mathcal{O}_{i}^{\left(6\right)}\,,$ which
contain the SM fields and derivatives, assuming that they represent
the leading NP effects.%
$^{[1]}$\footnotetext[1]{We will henceforth drop the subscript $n=6$ for the dimension 6 operators
$\mathcal{O}_{i}^{\left(6\right)}$.}
In particular, we consider a class
of operators, which, following the notation in \cite{GIMR},
will be denoted symbolically as the $\psi^{2}\varphi^{2}D$ class.
These operators contain a pair of fermions ($\psi$),
two Higgs fields ($\Phi$) and a SM covariant derivative ($D$)
and are generated by new heavy vector-boson exchanges in Higgs-fermion systems.

Consider for example the case where
a new heavy vector singlet field $V_{\mu}^{\prime}$, with a mass $M \gg v$, is added to the SM lagrangian
(the heavy vector can be thought of as some $U\left(1\right)^{\prime}$
remnant of a higher broken symmetry).
The lagrangian piece for $V_{\mu}^{\prime}$
then reads:
\begin{eqnarray}
\mathcal{L}  =  -\frac{1}{4}V_{\mu\nu}^{\prime}V^{\prime\mu\nu}+\frac{1}{2}M^{2}V_{\mu}^{\prime}V^{\prime\mu}
+ V_{\mu}^{\prime}\left(gi\Phi^{\dagger}\overleftrightarrow{D}^{\mu}\Phi+\tilde{g}\overline{\psi}\gamma^{\mu}\psi\right)\,,
\label{eq:5-1}
\end{eqnarray}
where, the {}``Hermitian derivative'' in (\ref{eq:5-1}) is defined as
$\Phi^{\dagger}\overleftrightarrow{D}_{\mu}\Phi  \equiv  \Phi^{\dagger}D_{\mu}\Phi-D_{\mu}\Phi^{\dagger}\Phi$.

Integrating out the heavy field $V_{\mu}^{\prime}$, by
using its Equation of Motion (EOM), we can
express $V_{\mu}^{\prime}$ in terms of the SM light fields:
\begin{eqnarray}
V_{\mu}^{\prime}  =-  \frac{1}{\left(\square-M^{2}\right)}\left(g\Phi^{\dagger}\overleftrightarrow{D}^{\mu}\Phi+\tilde{g}\overline{\psi}\gamma^{\mu}\psi\right) ~,
\label{eq:-13}
\end{eqnarray}
so that, performing the propagator expansion:
\begin{eqnarray}
\frac{1}{\left(\square-M^{2}\right)} & \underbrace{\approx}_{\square\ll M^{2}} & -\frac{1}{M^{2}}\sum_{k=0}^{\infty}\left(\frac{\square}{M^{2}}\right)^{k}\,,\label{prop}
\end{eqnarray}
and keeping only the first term, i.e, $k=0$, we obtain:
\begin{eqnarray}
V_{\mu}^{\prime} & \underbrace{\approx}_{\square\ll M^{2}} & \frac{1}{M^{2}}\left(g\Phi^{\dagger}\overleftrightarrow{D}^{\mu}\Phi+\tilde{g}\overline{\psi}\gamma^{\mu}\psi\right)\,.
\label{Vp}
\end{eqnarray}

Plugging now $V_{\mu}^{\prime}$ in (\ref{Vp}) back into the original lagrangian of (\ref{eq:5-1}),
we obtain the NP Lagrangian piece which emerges from the heavy vector-boson exchange:$^{[2]}$\footnotetext[2]{Note that integrating out the heavy vector field in (\ref{eq:5-1}) will
also induce new effective four-fermion contact operators.
The effects of such 4-Fermion
operators are not relevant for the Higgs-Vector boson production processes $e^{+}e^{-}\rightarrow Zh\,,Zhh$, and will, therefore,
not be considered here.}
\begin{eqnarray}
\Delta{\cal{L}}_{V^{\prime}} = \frac{f_{V^{\prime}}}{\Lambda^{2}} \mathcal{O}_{V^{\prime}}  ~,\label{OVprimelag}
\end{eqnarray}
where $f_{V^{\prime}} = g \tilde g$, $\Lambda = M$ and $\mathcal{O}_{V^{\prime}}$ is the dimension 6 heavy vector
singlet operator:
\begin{eqnarray}
\mathcal{O}_{V^{\prime}} = i \overline{\psi}\gamma^{\mu}\psi \Phi^{\dagger} \overleftrightarrow{D}^{\mu}\Phi ~.\label{OVprime}
\end{eqnarray}

In the case of a heavy vector triplet, one similarly obtains the operator:
\begin{eqnarray}
\mathcal{O}_{\tilde{V^{\prime}}} = i \overline{\psi}\sigma^{k}\gamma^{\mu}\psi \Phi^{\dagger}\sigma^{k}\overleftrightarrow{D}^{\mu}\Phi ~.\label{OVprimetilde}
\end{eqnarray}

Such $\psi^{2}\varphi^{2}D$ operators give rise to new contact interactions
of the form $l\overline{l}hZ$ and $l\overline{l}hhZ$ at scales lower than the
typical new heavy particle mass (see Appendix A) and, thus, contribute to the Higgs-Vector
boson associated production process $e^{+}e^{-}\rightarrow hZ, ~hhZ$ of interest in this work.
Examples of
Beyond the SM (BSM) constructions which involve new heavy vector fields that
can underly the $\psi^{2}\varphi^{2}D$ class operators include
TeV-scale $Z^{\prime}$ models (see e.g., \cite{zpmodels,zpmodels1}),
whose origin can be related to the breaking of grand unified theories
based on $SO\left(10\right)$ or $E_{6}$ symmetries, which may leave
one or several $U\left(1\right)$ remnants unbroken down to TeV energies,
before the symmetry is further broken to the SM symmetry.
Left-Right
twin Higgs models \cite{LRmodels} also introduce new heavy gauge bosons,
extra Higgs bosons and a top partner which can also affect the production
of $hZ$ and $hhZ$.

Similar contact interactions are also obtained for the $\psi^{2}\varphi^{2}D$ operators
involving quarks ($\psi=q$), i.e, $q\overline{q}hZ$, $q\overline{q}hhZ$, as well as contact interactions
involving the $W$: $\overline{u}dhW$, $\overline{u}dhhW$. These
may affect Higgs - Vector boson associated production
at the LHC and we leave that to a future work.
Nonetheless, it is
worth mentioning that Higgs-Vector boson associated production is
the third most dominant Higgs production channel at the LHC after
gluon-gluon fusion and vector boson fusion \cite{VBF1}. It has been found that operators containing derivative interactions can modify the kinematic distributions in Higgs Vector-boson associated production at the LHC \cite{Ellis}.

Note that other observables/processes can be utilized at the LHC to probe
the leptonic contact
interactions $l\overline{l}hZ$ and $l\overline{l}hhZ$, which are
generated by the
$\psi^{2}\varphi^{2}D$ operators. For example,
differential distributions
in the Higgs 3-body decay $h\rightarrow Zl^{+}l^{-}$
\cite{hz1,Isidori,EFT8} and the total Z-width \cite{EFT9} may
be useful for this purpose.
Nonetheless, as we will show here, a much higher sensitivity
to these leptonic contact terms can be obtained at a future ILC, via
$e^{+}e^{-}\rightarrow hZ, hhZ$. In particular,
an ILC will be able to probe the scale of the
$\psi^{2}\varphi^{2}D$ class operators, ranging from a few TeV to ${\cal O}(10)$ TeV,
depending upon its design (center of mass energy and luminosity).

Indeed, we wish to emphasize the underlying reasons and motivation for our choice of the 
$\psi^{2}\varphi^{2}D$ class of dim. 6 operators:
1. These operators are tree-level generated in the theory only by new heavy 
vector-boson exchanges and are, therefore, unique in that sense - probing a certain type of 
new physics which can be characterized by a single heavy scale $\Lambda$.
2. These operators give rise to new contact interactions of the form $eehZ$ and $eehhZ$ which 
will, therefore, give an effect proportional to $(E/\Lambda)^{2}$ in $\sigma(e^{+}e^{-}\rightarrow hhZ,hhZ)$, 
where $E$ is the c.m. energy of the process. They are, therefore, expected to give the dominant 
higher dimensional EFT effect in $e^{+}e^{-}\rightarrow hhZ,hhZ$, under the assumption of a weakly interacting 
underlying physics (i.e., with respect to other possible dim. 6 operators that can contribute to 
these processes). 

The paper is organized as follows: in sections \ref{sec2} we discuss the Higgs
Effective Lagrangian (HEL) framework and list the current bounds
from LEP/EW precision data on the $\psi^{2}\varphi^{2}D$ class operators.
In section \ref{sec3} we give a short overview of
Higgs - Vector boson associated production at the ILC. In sections
\ref{sec4} and \ref{sec5} we present analytical and
some benchmark numerical results
of the cross-sections
for the processes $e^{+}e^{-}\rightarrow hZ$ and $e^{+}e^{-}\rightarrow hhZ$
in the presence of the $\psi^{2}\varphi^{2}D$ class operators. In section \ref{sec5}
we also discuss the correlation between the $hZ$ and the $hhZ$ cross-sections as well as
the validity of the EFT expansion.
In section \ref{sec6} we present a more realistic
analysis of the sensitivity to the $\psi^{2}\varphi^{2}D$ class operators,
based on a more realistic background (BG) estimation.
In section \ref{sec7} we summarize.
Appendix A gives the Feynman rules associated with the $\psi^{2}\varphi^{2}D$
class operators and Appendix B contains intermediate steps of the
analytic calculation of $\sigma(e^{+}e^{-}\rightarrow hhZ)$.
In Appendices C and D we depict the tree-level Feynman diagrams
corresponding to $hZ$ and $hhZ$ signals, respectively, after
the $Z$-decays to a pair of fermions, which were calculated
for the realistic BG estimation.

\section{Higgs Effective Lagrangian - General Setup and constraints \label{sec2}}

In dealing with higher dimensional operators one often encounters
non trivial Lorentz structures. An efficient way to systematically
extract all the Feynman rules goes through a Mathematica package
software called FeynRules (FR) \cite{FRpaper}.
We have therefore used the Higgs Effective Lagrangian (HEL)
implementation in FR of \cite{HELpaper}.
Moreover,
the output from the HEL implementation in FR is
readable by the event generator software MadGraph 5 (MG5) \cite{MG5},
which further facilitates our analysis, allowing us to perform MG5
simulations in a straightforward manner.

The HEL setup of \cite{HELpaper} is defined by:
\begin{eqnarray}
\mathcal{L}  =  \mathcal{L}_{SM}+\sum\overline{c}_{i}\mathcal{O}_{i}
\equiv \mathcal{L}_{SM}+\mathcal{L}_{SILH}+
\mathcal{L}_{F_{1}}+\mathcal{L}_{F_{2}}+\mathcal{L}_{G}\,,\label{HEL}
\end{eqnarray}
where $\mathcal{L}_{SILH}$ (SILH=Strongly Interacting Light
Higgs) is inspired by scenarios where the Higgs field is part
of a strongly interacting sector, $\mathcal{L}_{F_{2}}$ contains interactions
among a pair of fermions with a single Higgs field and a gauge-boson
that originate from different NP scenarios (other than the heavy vector
exchanges), $\mathcal{L}_{G}$ contains new gauge-boson self-interactions
and $\mathcal{L}_{F_{1}}$ contains the $\psi^{2}\varphi^{2}D$ class operators of our interest
in (\ref{OVprime}) and (\ref{OVprimetilde}).
It is given by:
\begin{eqnarray}
\mathcal{L}_{F_{1}} & = & \frac{i\overline{c}_{HQ}}{v^{2}}\left[\overline{Q}_{L}\gamma^{\mu}Q_{L}\right]\left[\Phi^{\dagger}\overleftrightarrow{D}_{\mu}\Phi\right]+
\frac{4i\overline{c}_{HQ}^{\prime}}{v^{2}}\left[\overline{Q}_{L}\gamma^{\mu}T_{2k}Q_{L}\right]\left[\Phi^{\dagger}T_{2}^{k}\overleftrightarrow{D}_{\mu}\Phi\right]\nonumber \\
 &  & +\frac{i\overline{c}_{Hu}}{v^{2}}\left[\overline{u}_{R}\gamma^{\mu}u_{R}\right]\left[\Phi^{\dagger}\overleftrightarrow{D}_{\mu}\Phi\right]+\frac{i\overline{c}_{Hd}}{v^{2}}\left[\overline{d}_{R}\gamma^{\mu}d_{R}\right]\left[\Phi^{\dagger}\overleftrightarrow{D}_{\mu}\Phi\right] -\left[\frac{i\overline{c}_{Hud}}{v^{2}}\left[\overline{u}_{R}\gamma^{\mu}d_{R}\right]\left[\Phi\cdot\overleftrightarrow{D}_{\mu}\Phi\right]+h.c.\right]\label{LF1}\\
 &  & +\frac{i\overline{c}_{HL}}{v^{2}}\left[\overline{L}_{L}\gamma^{\mu}L_{L}\right]\left[\Phi^{\dagger}\overleftrightarrow{D}_{\mu}\Phi\right]+\frac{4i\overline{c}_{HL}^{\prime}}{v^{2}}\left[\overline{L}_{L}\gamma^{\mu}T_{2k}L_{L}\right]\left[\Phi^{\dagger}T_{2}^{k}\overleftrightarrow{D}_{\mu}\Phi\right] +\frac{i\overline{c}_{He}}{v^{2}}\left[\overline{e}_{R}\gamma^{\mu}e_{R}\right]\left[\Phi^{\dagger}\overleftrightarrow{D}_{\mu}\Phi\right],\nonumber 
\end{eqnarray}
where $Q_{L}=\left(\begin{array}{c}
u_{L}\\
d_{L}\end{array}\right),$ $u_{R}$ and $d_{R}$ are the three generations of left-handed and right-handed
quark fields, respectively and the corresponding lepton fields are $L_{L}=\left(\begin{array}{c}
\nu_{L}\\
l_{L}\end{array}\right)$ and $e_{R}$. Also, $T_{2k}$ are the $SU\left(2\right)$ generators in the fundamental
representation, $T_{2k}=\frac{\sigma_{k}}{2}$, where $\sigma_{k}$
are the Pauli matrices.

Furthermore, the coefficients $\overline{c}_{i}$ are normalized such that
they are related to $f_{i}$ in (\ref{EFT}) by
\begin{eqnarray}
\overline{c}_{i}  =  \frac{v^{2}}{\Lambda^{2}}f_{i}\,,\label{ci-to-fi}
\end{eqnarray}

In general, if one writes down all possible dimension 6 operators
consistent with the SM symmetries (which exhibit baryon and lepton
number conservation), one arrives to a finite number of operators.
However, the SM EOM along with the use of integration by parts and field redefinitions
may result in a redundancy of this description. That is, some of the
operators may be equivalent, up to total derivatives, to linear combinations
of other operators \cite{EFT2}. Indeed, as has been advocated in both \cite{GIMR} and \cite{HELpaper},
the $\psi^{2}\varphi^{2}D$ class operators, as they appear in $\mathcal{L}_{F_{1}}$, are equivalent
(using the EOM's) to linear combinations of some purely bosonic operators
in $\mathcal{L}_{SILH}$, and can, therefore, be eliminated by
trading them with these bosonic operators; the choice of basis may vary
depending on the analysis one wishes to carry out, see for example \cite{EFTbasis1,EFTbasis2}.
This might be part of the reason why the effects of these
$\psi^{2}\varphi^{2}D$ operators have not been thoroughly studied.
Moreover, the bosonic operators are loop generated in the underlying theory
(see \cite{EFT2}), so that their contribution is expected in general to be further
suppressed, typically by a factor of $1/16\pi^{2}$ (recall that
the $\psi^{2}\varphi^{2}D$ class operators
are generated by tree-level exchanges of heavy vector-bosons in the underlying theory).

Here we will follow
the GIMR basis of \cite{GIMR}, which explicitly includes the $\psi^{2}\varphi^{2}D$ class operators in
(\ref{OVprime}) and (\ref{OVprimetilde}), which we seek to explore in this work.
In particular, we will be interested in the leptonic operators corresponding to $\overline{c}_{HL},~\overline{c}_{HL}^{\prime}$ and
$\overline{c}_{He}$, which can affect the $hZ$ and $hhZ$ signals, $e^+ e^- \to hZ,hhZ$, at the ILC.
These leptonic operators and, in general all the $\psi^{2}\varphi^{2}D$ class operators, are tightly constrained
by Z-pole measurements, as they modify its vector and axial-vector couplings to a pair of fermions.
Indeed, a fit to ElectroWeak (EW) precision data \cite{EWPD1} has been performed
in \cite{EWPD2} to yield the following bounds on the Wilson coefficients of the leptonic
operators:$^{[3]}$\footnotetext[3]{The bounds
on the $\psi^{2}\varphi^{2}D$ operators involving the light(heavy)-quarks are comparable(weaker)
to those on the leptonic operators in (\ref{bounds}).}
\begin{eqnarray}
-0.0003< &\overline{c}_{HL}+\overline{c}_{HL}^{\prime} & <0.002
 \nonumber \\
-0.002< & \overline{c}_{HL}-\overline{c}_{HL}^{\prime} &<0.004 \nonumber \\
-0.0009< &\overline{c}_{He} & <0.003 ~. \label{bounds}
\end{eqnarray}

The bounds in (\ref{bounds}) can be translated into bounds on the ratio's
$f_{i}/\Lambda^{2}$ through (\ref{ci-to-fi}). In practice, the upper (lower)
bounds in (\ref{bounds}) correspond to $f_{i}>0$ ($f_{i}<0$). In
what follows we always set for simplicity  $f_{i}=1$ or $f_{i}=-1$ and
use $\Lambda$ as the only unknown with respect to the NP. For example
using the one coupling scheme for $\overline{c}_{HL}$:
\begin{equation}
-0.0003<\overline{c}_{HL}<0.002\,,\label{eq:20}
\end{equation}
the above upper and lower bounds correspond to $\Lambda\gtrsim5.5\,$
TeV and $\Lambda\gtrsim14\,$ TeV, for $f_{HL}=1$ and $f_{HL}=-1$, respectively.
Similarly, in the one parameter scheme for $\overline{c}_{He}$, we obtain $\Lambda\gtrsim4.5\,$
TeV and $\Lambda\gtrsim8\,$ TeV, for $f_{He}=1$ and $f_{He}=-1$, respectively.

\section{Higgs - Vector boson associated production at the ILC - a short overview \label{sec3}}

The future planned ILC, although having a lower energy reach, provides
a cleaner environment, as it doesn't suffer from the large QCD background
which is typical to hadron colliders. Moreover, the center of mass energy
at an ILC ($\sqrt{s}$) is precisely known and it also provides the possibility of
polarizing the colliding electron beams. The cross-sections of the main Higgs
production modes at the ILC are given for example in \cite{ILCCSX}.

\begin{figure}
\begin{center}
\includegraphics[scale=0.25]{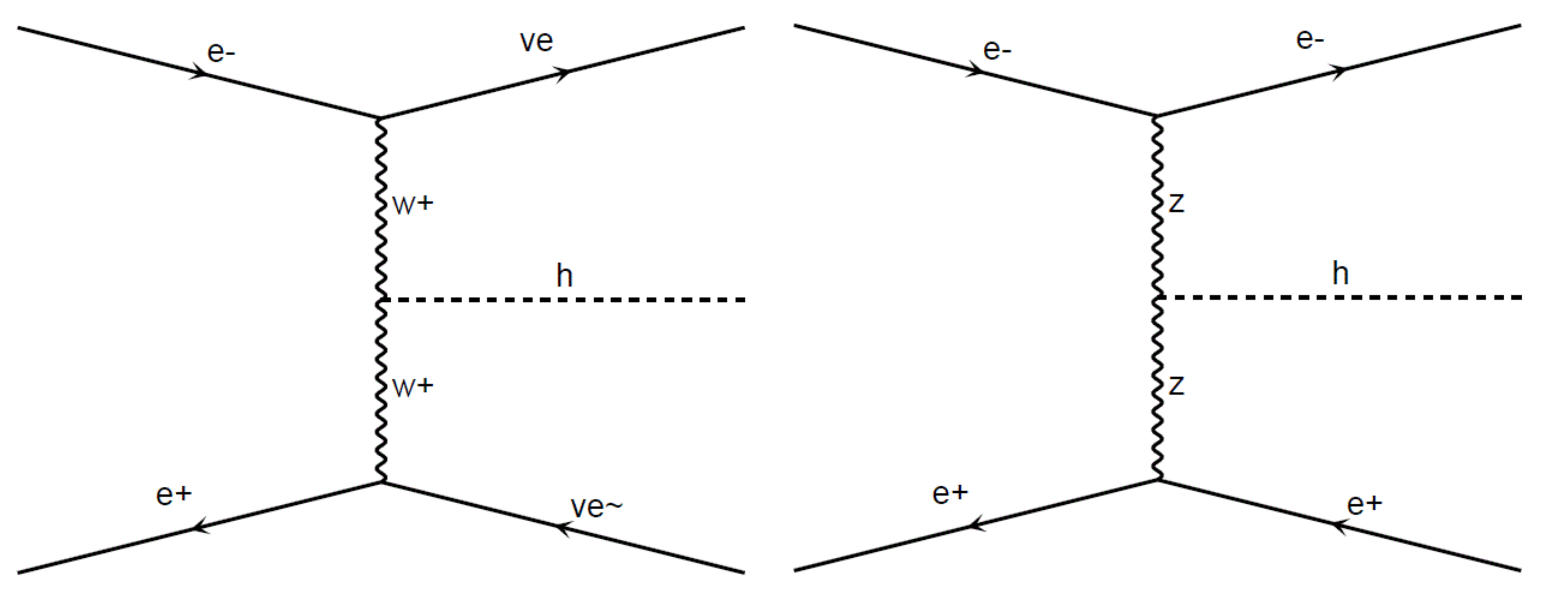}
\end{center}
\caption{Tree-level SM diagrams for the WW-fusion process $e^{+}e^{-}\rightarrow h\nu\overline{\nu}$
(left) and ZZ-fusion process $e^{+}e^{-}\rightarrow he^{+}e^{-}$
(right).
\label{figVVfusion}}
\end{figure}

Clearly, the BG for the process $e^{+}e^{-}\rightarrow hZ$ depends on the
subsequent decays of $Z$ and $h$. In the case of $Z\rightarrow\nu\overline{\nu}$
the leading BG is the WW-fusion process shown in Fig.~\ref{figVVfusion}, while
for the case where $Z\rightarrow e^{+}e^{-}$ ,i.e., $e^{+}e^{-}\rightarrow hZ\rightarrow he^{+}e^{-}$,
the leading BG is the ZZ-fusion process also shown in Fig.~\ref{figVVfusion} (see e.g., \cite{ZZfuse}).
Both the WW and ZZ-fusion processes grow logarithmically with $\sqrt{s}$, the
ZZ fusion being about 10 times smaller than the WW-fusion one due
to the different strengths between the W and the Z couplings to electrons.
In that respect, let us mention \cite{EFT4}, where a study of the dimension 6
operators $\mathcal{O}_{H}$ and $\mathcal{O}_{6}$
in $\mathcal{L}_{SILH}$ (see \cite{HELpaper}), which affect the $WWh$,
$ZZh$ and $hhh$ couplings in $e^{+}e^{-}\rightarrow hZ\rightarrow(anything)(l^{+}l^{-})$
was performed. They reconstructed the Higgs mass
using the recoil mass + acceptance cuts technique (see \cite{hz2}) and imposed additional
cuts, such as the invariant mass cut $\left|m_{Z}-\mathcal{M}\left(ll\right)\right|<10\,$~GeV,
to further reduce the BG for the $hZ$ signal. This allows, for example,
to reduce the effective number
of BG events for $e^{+}e^{-}\rightarrow hZ\rightarrow b\overline{b}l^{+}l^{-}$,
coming from $u$ and $t$ channel electron exchange in $e^{+}e^{-}\rightarrow ZZ\rightarrow b\overline{b}l^{+}l^{-}$,
to about 10\% of the signal (see also \cite{hz3}).

Indeed, as we will show below, the $hZ\rightarrow hf\overline{f}$ final state can
be distinguished from the gauge-boson fusion BG processes through an appropriate
set of cuts and choice of final states. For example,
the contribution of $\mathcal{O}_{He},\mathcal{O}_{HL},\mathcal{O}_{HL^{\prime}}$
from the interference with the SM can be substantial in the $hf\overline{f}$
channel with $f=\mu,\tau$ or $q$, i.e., giving rise to a correction
of more than 5\% (20\%) at a center of mass energy of $\sqrt{s}=500$ GeV
(1 TeV). On the other hand, the relative impact of $\mathcal{O}_{He},\mathcal{O}_{HL},\mathcal{O}_{HL^{\prime}}$
on $he^{+}e^{-}$ and $h+\rmpart E$ is  smaller, since
these signals suffer from the large WW-fusion and ZZ-fusion BG.

As mentioned earlier, an important feature of the future planned ILC
is the possibility of polarizing the
incoming electron-positron, which can be straightforwardly utilized
for various purposes. For example, for $hZ$ production followed by
Z decay to neutrinos, one can {}``switch off'' the WW fusion contribution
by choosing right-handed electrons (and left-handed positrons) \cite{hz4}.
The SM cross-sections for $e^{+}e^{-}\rightarrow Z\rightarrow hZ$
including initial state polarization effects can be found in \cite{VBF1}.

The process $e^+ e^- \to hZ$ at the ILC, in the presence of the contact
terms which are generated by the $\psi^{2}\varphi^{2}D$ class operators, was
initially considered in \cite{EFT5} and was found to show significant
deviations from the SM by choosing a combination of operators which
saturates the bounds on the corresponding Wilson coefficients, i.e, maximizing the statistical
significance. Later on, \cite{EFT6} expanded the analysis of \cite{EFT5}
by looking at angular distributions in $e^{+}e^{-}\rightarrow hZ\rightarrow h\overline{f}f$.
Recently, Craig \textit{et al.} in \cite{susyhz} have also considered the effects
of the $\psi^{2}\varphi^{2}D$ class operators in $e^{+}e^{-}\rightarrow hZ$
at an ILC with a center of mass
 energy of $\sqrt{s}=$250 GeV, taking into account only the interferences
with the SM (i.e. only the corrections of order $1/\Lambda^{2}$, see next section). They
found that the exclusion/discovery potential (of such a machine) to the scale these
operators, which heavily relies on the accuracy of the measurement, is
$\Lambda \sim {\rm few}$ TeV \cite{susyhz}.

As for $e^{+}e^{-}\rightarrow hhZ$ (see Fig.~\ref{SMdiag}), its cross-section
peaks at about 0.18 [fb] close to $\sqrt{s}=$500 GeV and then drops
as $1/s$ at high energies. On of the main motivations for measuring $e^+ e^- \to hhZ$ is the
feasibility of detecting the trilinear Higgs coupling $\lambda$ (see
diagram (b) in Fig.~\ref{SMdiag}), though there are other important diagrams
that do not contain $\lambda$ but still contribute to $hhZ$ (see
diagrams (c) and (d) in Fig.~\ref{SMdiag}). This results in a dilution of $\frac{\Delta\lambda}{\lambda}\simeq1.75\frac{\Delta\sigma_{hhZ}}{\sigma_{hhZ}}$
at $\sqrt{s}=$500 GeV, where $\Delta\lambda$ ($\Delta\sigma$) are
the measured accuracies \cite{hhz1}, \cite{hhz12}. The sensitivity to the Higgs
self coupling in $e^{+}e^{-}\rightarrow hhZ$ has, therefore, been
a subject of intense study throughout the years \cite{hhz2,hhz21,hhz22,hhz23}.

A primary example of an EFT analysis of $hhZ$ production at the ILC
was given in \cite{hhz3}, where
dimension 6 operators that give rise to anomalous Higgs self-couplings were considered.
They found that, at $\sqrt{s}=$800 GeV, the normalized $p_{T}\left(Z\right)$ distribution
and the $hh$ invariant mass distribution (in $e^{+}e^{-}\rightarrow hhZ$)
exhibit dramatic differences
from the SM ones in the presence of the new effective anomalous Higgs self-couplings. A similar
analysis, also involving CP violating effective operators, has been performed for $e^{+}e^{-}\rightarrow hW^{+}W^{-}$
in \cite{hww}.

\section{$e^{+}e^{-}\rightarrow hZ$ \& $e^{+}e^{-}\rightarrow hhZ$ - analytical analysis \label{sec4}}

In this section we present an analytical derivation of the
cross-sections for the processes $e^{+}e^{-}\rightarrow hZ$ and $e^{+}e^{-}\rightarrow hhZ$
in the presence of the $\psi^{2}\varphi^{2}D$ class operators$^{[4]}$\footnotetext[4]{We note that loop effects from the EFT are expected to be suppressed by at least a factor of 
$1/16\pi^{2}$ and, in some cases, by an additional factor of $(v/\sqrt{s})^2$, compared to the dominant 
tree-level contributions from the new contact terms. Their effect 
are, therefore, not important for the purpose of our investigation and they have a negligible effect on 
the results shown below (i.e., on the sensitivity plots to the scale of the new physics).
}.

\subsection{$e^{+}e^{-}\rightarrow hZ$ }

\begin{figure}
\begin{center}
\includegraphics[scale=0.25]{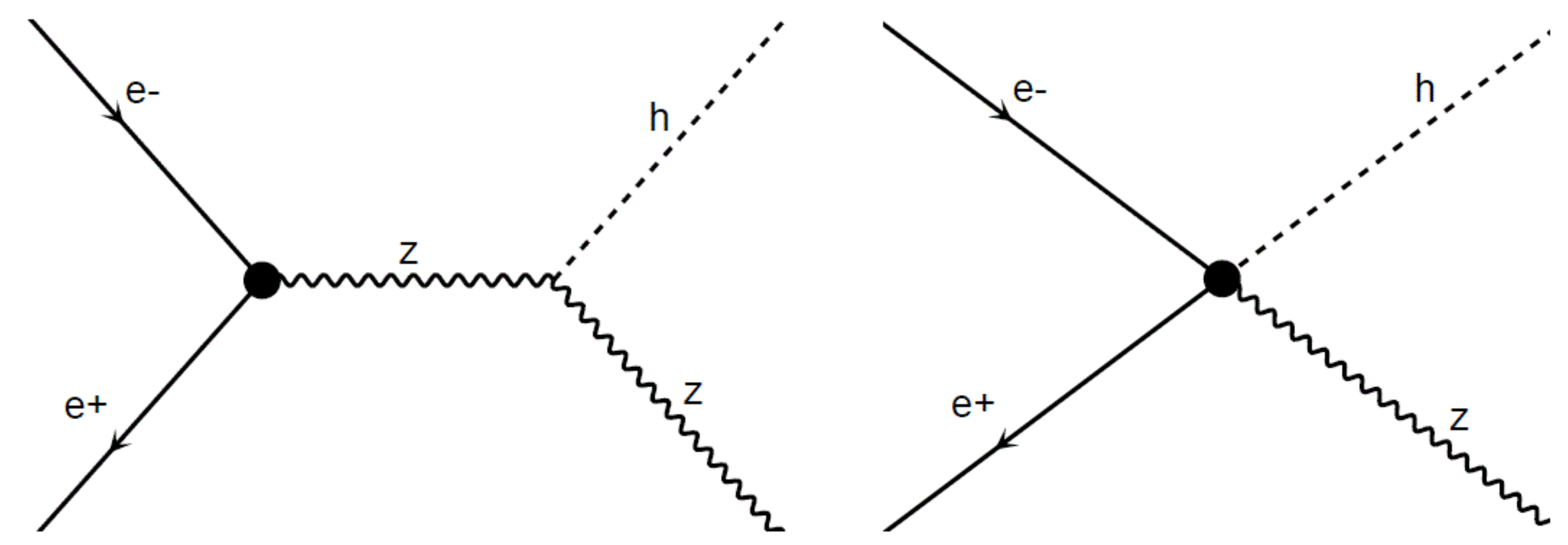}
\end{center}
\caption{Tree-level diagrams for $e^{+}e^{-}\rightarrow h\, Z$ in the presence of the
$\psi^{2}\varphi^{2}D$ operators, due to a shift in $Z$ coupling
to leptons (left) and due to the $eehZ$ contact term (right). The
relevant Feynman rules are given in Appendix A. \label{NPhZ}}
\end{figure}

The SM diagram leading to $e^{+}e^{-}\rightarrow hZ$ appears in Fig.~\ref{SMdiag}(a), while
the NP contributions to $e^{+}e^{-}\rightarrow hZ$ induced by the
$\psi^{2}\varphi^{2}D$ class operators $\mathcal{O}_{HL},\mathcal{O}_{HL}^{\prime}$
and $\mathcal{O}_{He}$ (in $\mathcal{L}_{F1}$) are depicted
in Fig.~\ref{NPhZ}.
Summing up the contributions from all orders (i.e. in the $\left(1/\Lambda^{2}\right)$
expansion) we obtain
\begin{eqnarray}
\sigma(hZ) & = & \sigma_{SM}(hZ)\left(1+\frac{\delta_{1}\left(s,f_{i}\right)}{\Lambda^{2}} +
\frac{ \delta_{2}\left(s,f_{i}\right)}{\Lambda^{4}} \right) ~,
\label{sigmahz}
\end{eqnarray}
where
\begin{eqnarray}
\sigma_{SM}(hZ) & = & \frac{\alpha m_{Z}^{2}}{12v^{2}}\frac{\left(a_{e}^{2}+v_{e}^{2}\right)}{\left(s-m_{Z}^{2}\right)^{2}}w\left(w^{2}+
\frac{12m_{Z}^{2}}{s}\right)\,, \label{eq:-29}
\end{eqnarray}
\begin{eqnarray}
w=\sqrt{\left(1-\frac{m_{h}^{2}}{s}-\frac{m_{Z}^{2}}{s}\right)^{2}-\frac{4m_{h}^{2}m_{Z}^{2}}{s^{2}}} ~,
\end{eqnarray}
and the ${\cal O}(1/\Lambda^2)$ and ${\cal O}(1/\Lambda^4)$ corrections are:
\begin{widetext}
\begin{eqnarray}
\delta_{1}\left(s,f_{i}\right) & = & \frac{s}{2 c_{W}s_{W}} \frac{a_{e} \left[ \left(f_{HL}+f_{HL}^{\prime}\right)-\frac{f_{He}}{2} \right]+v_{e} \left[ \left(f_{HL}+f_{HL}^{\prime}\right)+\frac{f_{He}}{2}\right]}{a_{e}^{2}+v_{e}^{2}} \frac{v^2}{M_Z^2} ~,
\label{delta1}\\
\delta_{2}\left(s,f_{i}\right) & = & \left(\frac{s}{4 c_{W}s_{W}} \right)^2 \frac{ \left[ \left( f_{HL}+f_{HL}^{\prime} \right)- \frac{f_{He}}{2} \right]^{2}+
\left[\left(f_{HL}+f_{HL}^{\prime}\right)+\frac{f_{He}}{2} \right]^{2}}{a_{e}^{2}+v_{e}^{2}} \left(\frac{v^2}{M_Z^2} \right)^2~.
\label{delta2}
\end{eqnarray}
\end{widetext}

Note that, as expected, the interference of the contact term with the
SM diagram grows with energy, i.e. the term $\propto\frac{\delta_{1}}{\Lambda^{2}}$.
For consistency matters, we included
the term $\propto\frac{\delta_{2}}{\Lambda^{4}}$, which corresponds to the squared amplitude
of the contact interaction NP diagram in Fig.~\ref{NPhZ} and
which is, therefore, expected to be small when $\sqrt{s} < \Lambda$ (i.e., compared to the leading
$\mathcal{O}\left(\frac{\delta_{1}}{\Lambda^{2}}\right)$
term). All other terms cancel out when
taking the sum of all contributions.

The term $\propto\frac{\delta_{2}}{\Lambda^{4}}$ is, however,
a vital ingredient of the full squared
amplitude, as it formally ensures a positive cross-section. The case
where the $\mathcal{O}\left(\frac{\delta_{2}}{\Lambda^{4}}\right)$ ``correction"
becomes comparable to the SM signifies the point where the validity
of our current EFT framework breaks and the necessity of considering
higher dimensional operators (i.e., in our case, dimension 8 ones). We will return to this
point later.

\begin{figure}
\begin{center}
\includegraphics[scale=0.35]{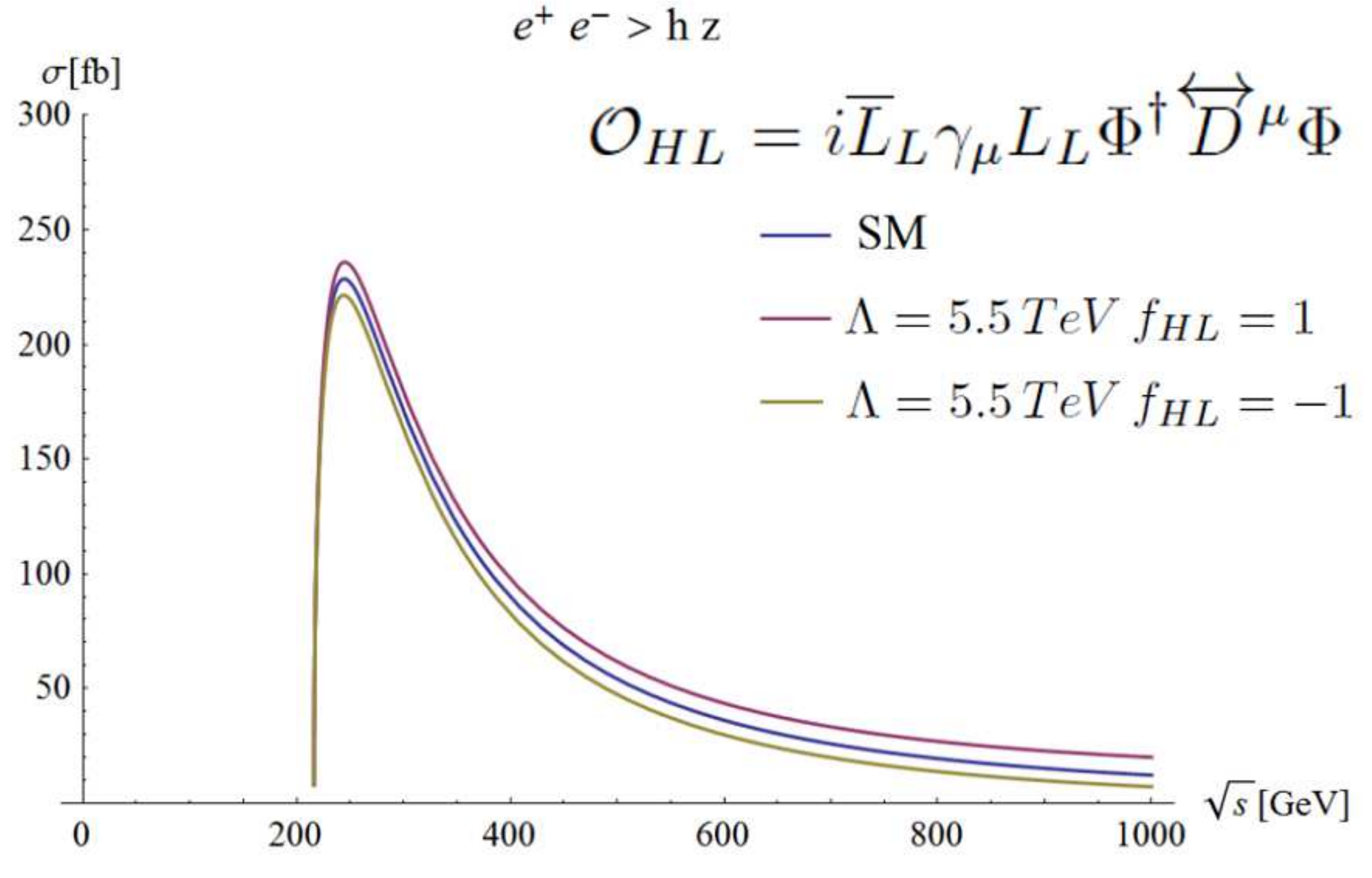}
\includegraphics[scale=0.47]{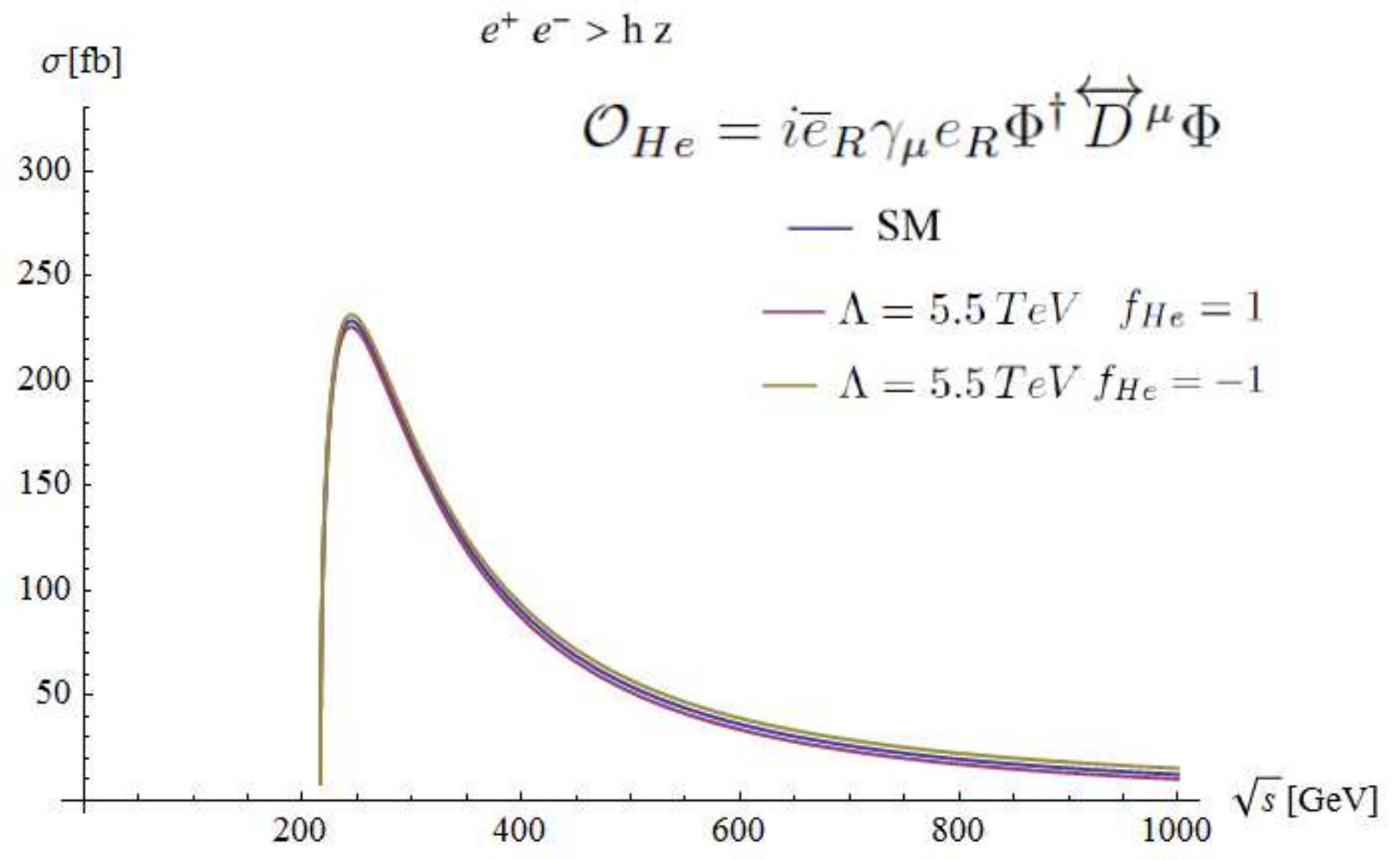}
\end{center}
\caption{$\sigma(e^{+}e^{-}\rightarrow hZ)$ as a function of the
center of mass energy in the presence of $\mathcal{O}_{HL}$ (left
figure) and $\mathcal{O}_{He}$ (right figure). The operator $\mathcal{O}_{HL}^{\prime}$
has the same effect as $\mathcal{O}_{HL}$ since the amplitude is
$\propto\left(f_{HL}+f_{HL}^{\prime}\right)$. \label{sigmahz1}}
\end{figure}

In Fig.~\ref{sigmahz1} we plot $\sigma\left(e^{+}e^{-}\rightarrow hZ\right)$
as a function of $\sqrt{s}$, where we switch on one operator at a
time (setting all others to zero).
In particular, we set $\Lambda=5.5$ TeV for both $f_{HL},f_{He}=\pm1$.
The operator $\mathcal{O}_{HL}^{\prime}$
has the same effect as $\mathcal{O}_{HL}$ since the amplitude is
$\propto\left(f_{HL}+f_{HL}^{\prime}\right)$.
We see
that the cross-section is less sensitive to $\mathcal{O}_{He}$,
for $\Lambda=5.5$ TeV, partly since $f_{He}$ appears to have an extra suppression
factor of $2$ w.r.t $f_{HL}$ in the
cross-section, see (\ref{delta1}) and (\ref{delta2}).

\subsection{$e^{+}e^{-}\rightarrow hhZ$ }

\begin{figure}[h]
\begin{center}
\vspace{0.5cm}
\includegraphics[scale=0.25]{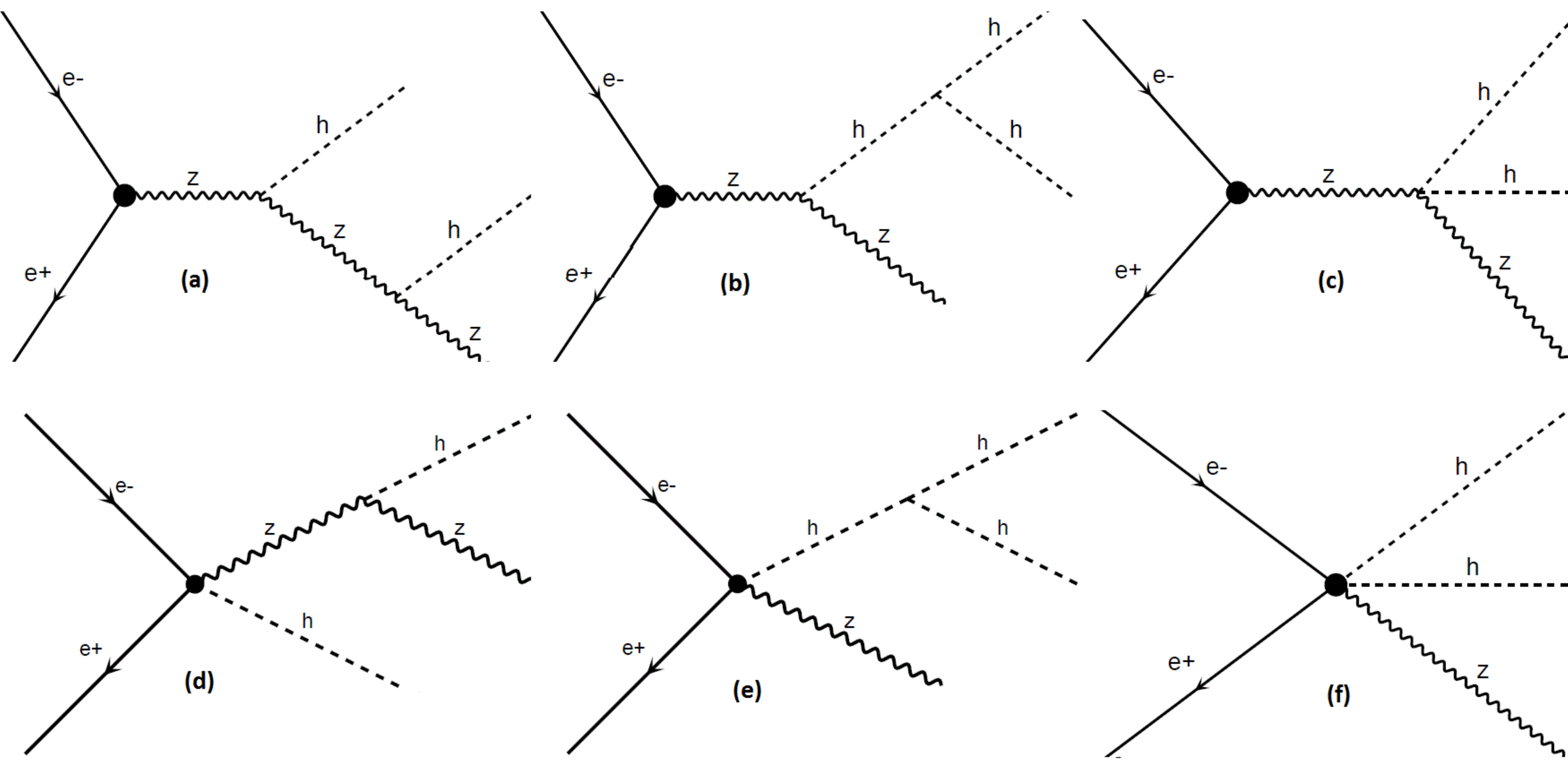}
\end{center}
\caption{Diagrams for $e^{+}e^{-}\rightarrow hhZ$ in the presence of $\psi^{2}\varphi^{2}D$ class operators.
\label{hhZNP}}
\end{figure}

As for the process $e^{+}e^{-}\rightarrow hhZ$,
the SM diagrams leading to $e^{+}e^{-}\rightarrow hhZ$ are shown
in Fig.~\ref{SMdiag}(b),(c),(d),
while the NP diagrams corresponding to the $\psi^{2}\varphi^{2}D$ class operators
are plotted in Fig.~\ref{hhZNP}. In particular,
there are two types of new contributions:
a shift in the Z coupling to leptons and the
new contact terms $eehZ$ and $eehhZ$.
The NP diagrams exhibit similar patterns as the three SM ones,
so that we can express them in terms of the SM one.
Moreover, the expansions in $1/\Lambda^2$ exactly coincides with
the one given above for the $e^+e^- \to hZ$ case, allowing us to
conveniently write the total squared amplitude as:
\begin{eqnarray}
\left|\mathcal{M}\right|^{2} & = & \left|\mathcal{M}_{SM}\right|^{2} \left(1+\frac{\delta_{1}\left(s,f_{i}\right)}{\Lambda^{2}} +
\frac{ \delta_{2}\left(s,f_{i}\right)}{\Lambda^{4}} \right) ~, \label{amp2hhZ}
\end{eqnarray}
where $\delta_1$ and $\delta_2$ are given in (\ref{delta1}) and (\ref{delta2}), respectively,
and $\left|\mathcal{M}_{SM}\right|^{2}$ is given in appendix B. The differential cross-section is
\begin{widetext}
\begin{eqnarray}
d\sigma & = & \frac{1}{2}\frac{1}{2s}\frac{1}{\left(2\pi\right)^{5}}\frac{1}{4}\sum\left|\mathcal{M}\right|^{2}\delta^{4}
\left(\left(-l_{1}\right)+l_{2}-p_{3}-p_{4}-p_{5}\right)\frac{d^{3}p_{3}}{2E_{3}}\frac{d^{3}p_{4}}{2E_{4}}
\frac{d^{3}p_{5}}{2E_{5}}\,,\label{eq:-54}
\end{eqnarray}
\end{widetext}
where $\left(-l_{1},l_{2}\right)$ denote the $\left(e^{+},e^{-}\right)$ momenta
and $\left(p_{3},p_{4},p_{5}\right)$ denote the $\left(h,h,z\right)$ momenta
(the extra $1/2$ factor which accounts for the two identical particles $hh$
in the final state is explicitly factored out).

As will be shown below, the resemblance between the
$hZ$ and $hhZ$ expansions at the level of the matrix element squared in the presence of
our $\psi^2 \phi^2 D$ class operators, may be come handy for the study of NP in
$e^+ e^- \to hZ, hhZ$.

\section{$e^{+}e^{-}\rightarrow hZ$ \& $e^{+}e^{-}\rightarrow hhZ$ - numerical analysis \label{sec5}}

\subsection{Naive sensitivities and benchmark values}

Let us define
\begin{eqnarray}
N_{SD}\left(\sqrt{s},f_{i},\Lambda\right) & \equiv & \frac{N^{T}-N^{SM}}{\sqrt{N^{T}}} ~,\label{NSD}
\end{eqnarray}
where $N_{SD}$ is the statistical significance of the signal,
$\sigma^{T}$ ($\sigma^{SM}$)
is the cross-section in the presence of the new effective operators (in
the SM), $N^{T,SM} = \sigma^{T,SM}L$
is the corresponding total number of events and
$L=\intop\mathcal{L}dt$ is the integrated luminosity,
which may vary, depending on the given center of mass energy and design
of the ILC. In particular, we have performed MADGRAPH simulations (using MG5 \cite{MG5})
for both $e^{+}e^{-}\rightarrow hZ$ and
$e^{+}e^{-}\rightarrow hhZ$, in the presence of $\mathcal{O}_{HL}$ -
using the HEL model implementation of \cite{HELpaper} in MG5 -
at an ILC with the benchmark designs \cite{ILCCSX}:
$\left\{ \sqrt{s}~[{\rm TeV}],L~[{\rm fb}^{-1}]\right\}= \left\{0.5,500\right\}, \left\{1,1000\right\}, \left\{2,2000\right\}, \left\{3,2500\right\}$.

\begin{figure}
\begin{center}
\includegraphics[scale=0.43]{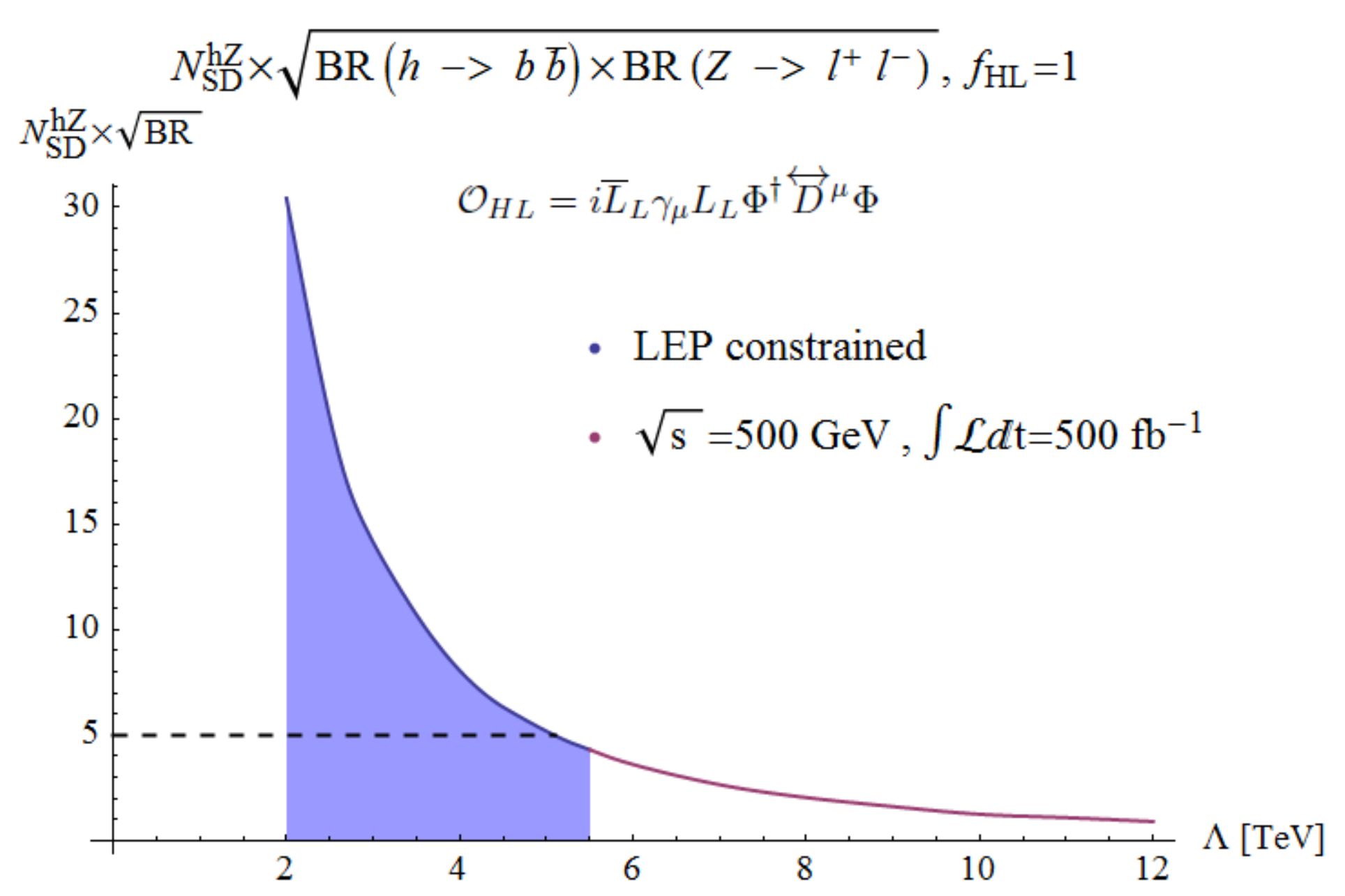}
\includegraphics[scale=0.43]{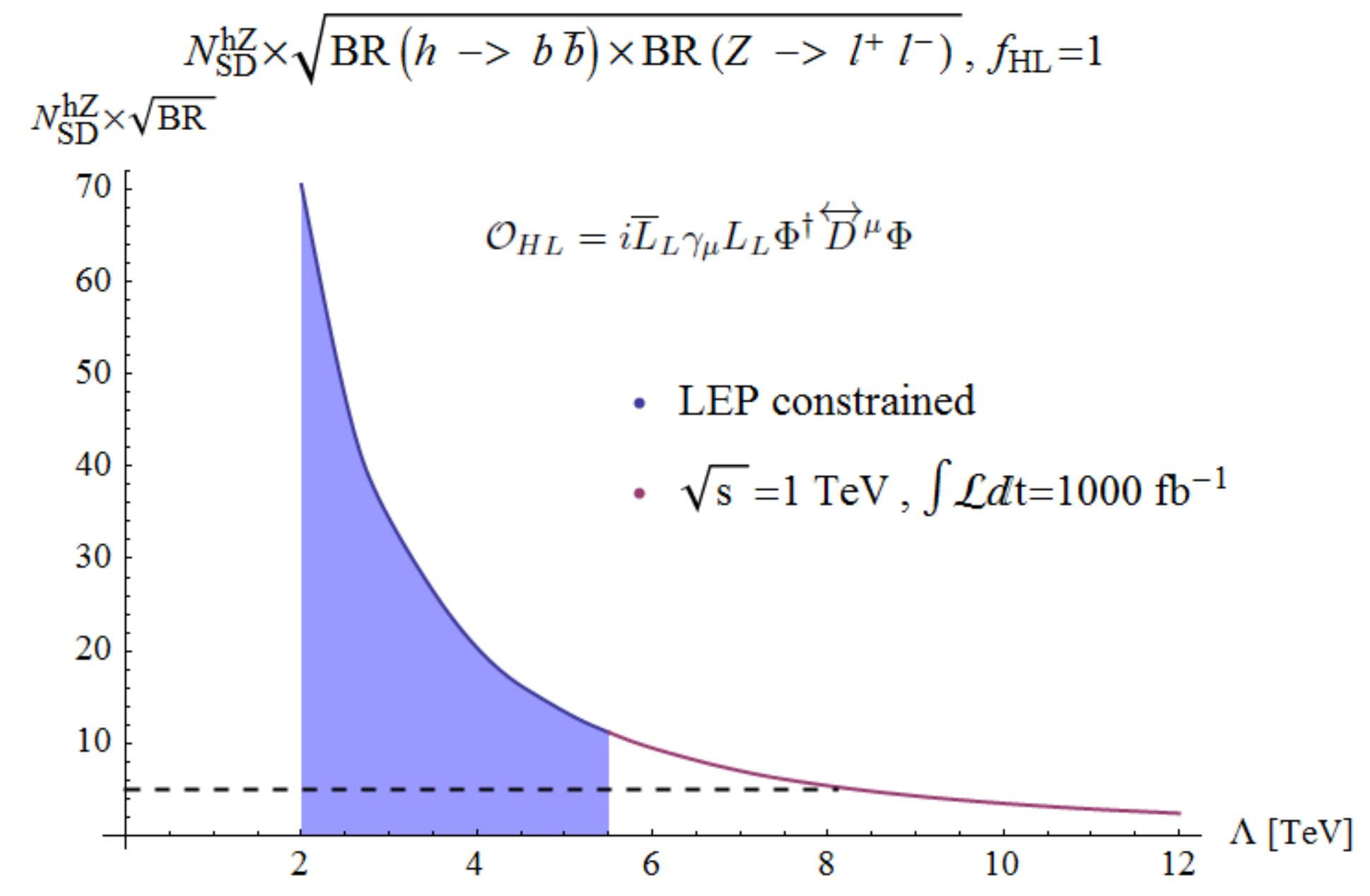}
\includegraphics[scale=0.43]{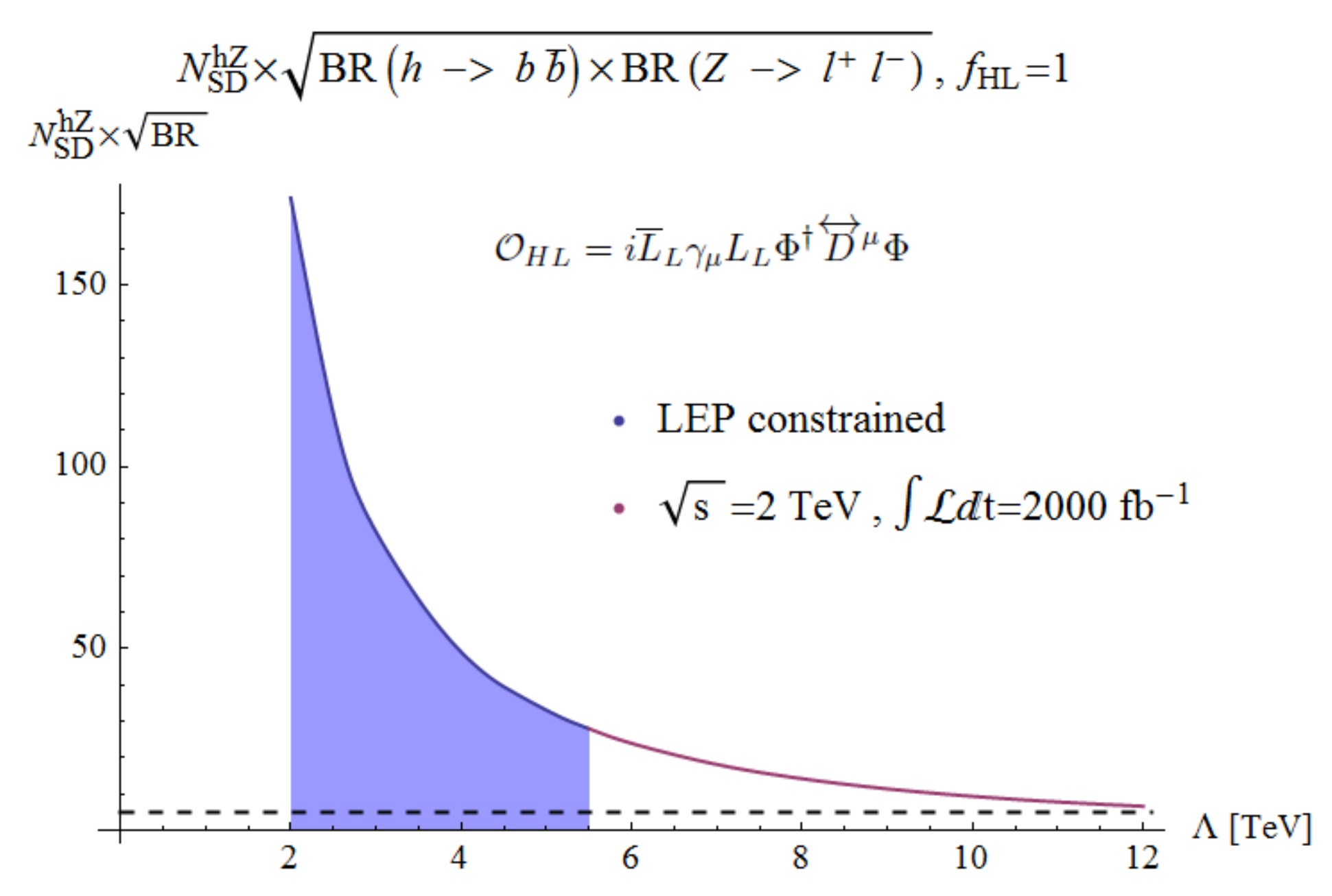}
\includegraphics[scale=0.43]{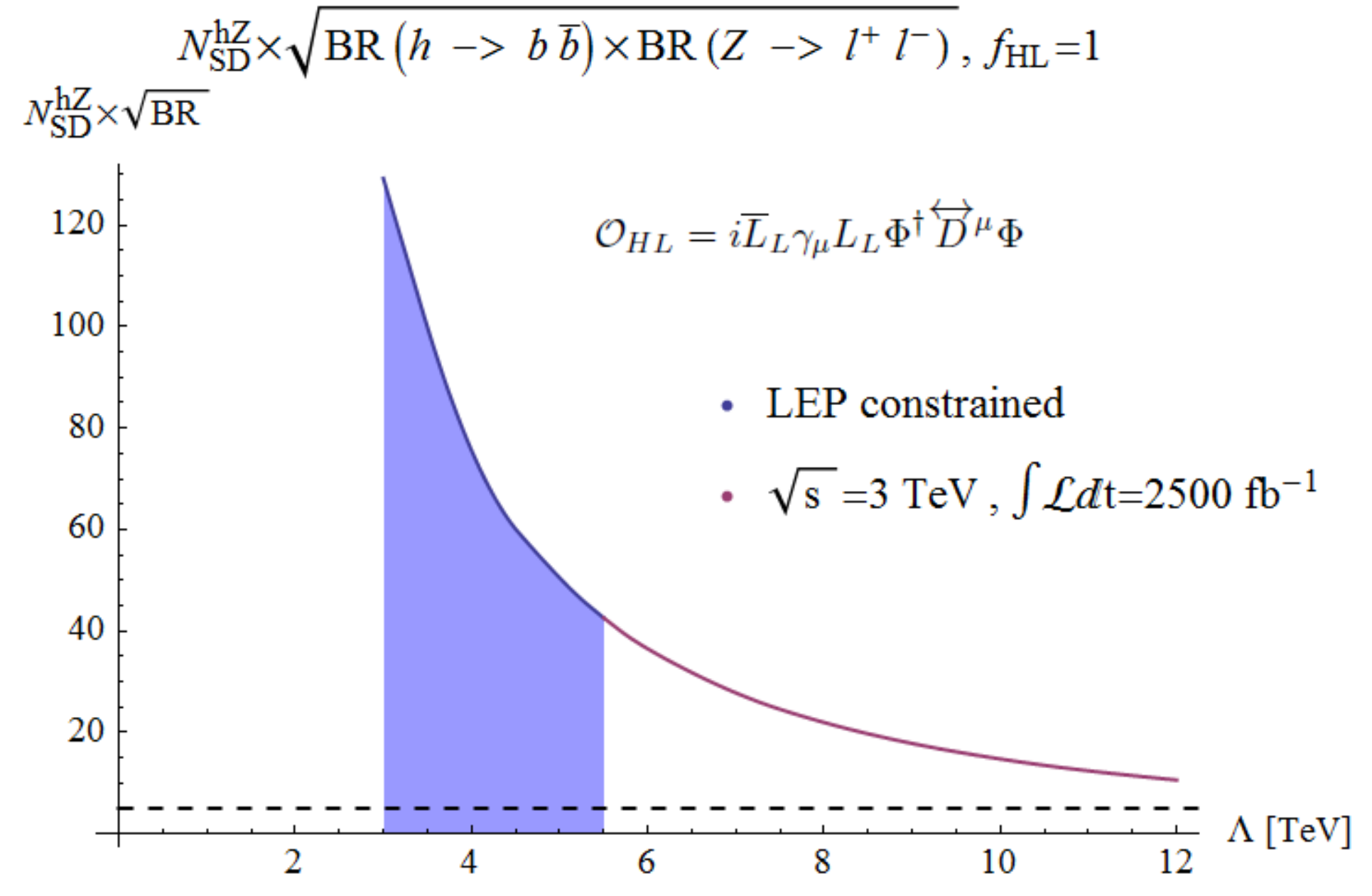}
\end{center}
\caption{Sensitivity to the NP scale $\left(\Lambda\right)$ for $e^{+}e^{-}\rightarrow hZ\rightarrow b\overline{b}l^{+}l^{-}$
at $\sqrt{s}=500$ GeV with $L=500\, fb^{-1}$, $\sqrt{s}=1$
TeV with $L=1000\, fb^{-1}$, $\sqrt{s}=2$ TeV with
$L=2000\, fb^{-1}$ and $\sqrt{s}=3$ TeV with $L=2500\, fb^{-1}$.
The blue region is constrained by LEP. The dashed horizontal
line is the naive $5\sigma$ sensitivity. \label{NSD1}}
\end{figure}

In Fig.~\ref{NSD1} we show the expected sensitivity $N_{SD}$
as a function of $\Lambda$ for the $hZ$ signal, where
we examine the specific decay mode of the $hZ$ final state $h\rightarrow b\overline{b}$ and $Z\rightarrow l^{+}l^{-}$,
with $l^{\pm}=e^{\pm},\mu^{\pm}$ (we sum over the electrons and muons final states, i.e., $l^{+}l^{-}=e^{+}e^{-}+\mu^{+}\mu^{-}$),
by multiplying the number of events $N^{T,SM}$ with the corresponding SM Branching
Ratios (BRs): $BR\left(h\rightarrow b\overline{b}\right) \sim 60\%$ and
$BR\left(Z\rightarrow l^{+}l^{-}\right)=BR\left(Z\rightarrow e^{+}e^{-}\right)+BR\left(Z\rightarrow \mu^{+}\mu^{-}\right) \sim 6.8\%$.
We see, for instance, that a 1 TeV ILC will be sensitive to the NP scale (associated
with the $\psi^2 \phi^2 D$ operator $\mathcal{O}_{HL}$)
$\Lambda\simeq 6$ TeV at the $\sim10\sigma$ level, in the channel
$hZ\rightarrow b\overline{b}l^{+}l^{-}$ (efficiencies such as b-tagging, etc. are not included).
As can be seen, at larger center of mass energies the effect
of $\mathcal{O}_{HL}$ is more pronounced. %
\begin{figure}
\begin{center}
\includegraphics[scale=0.58]{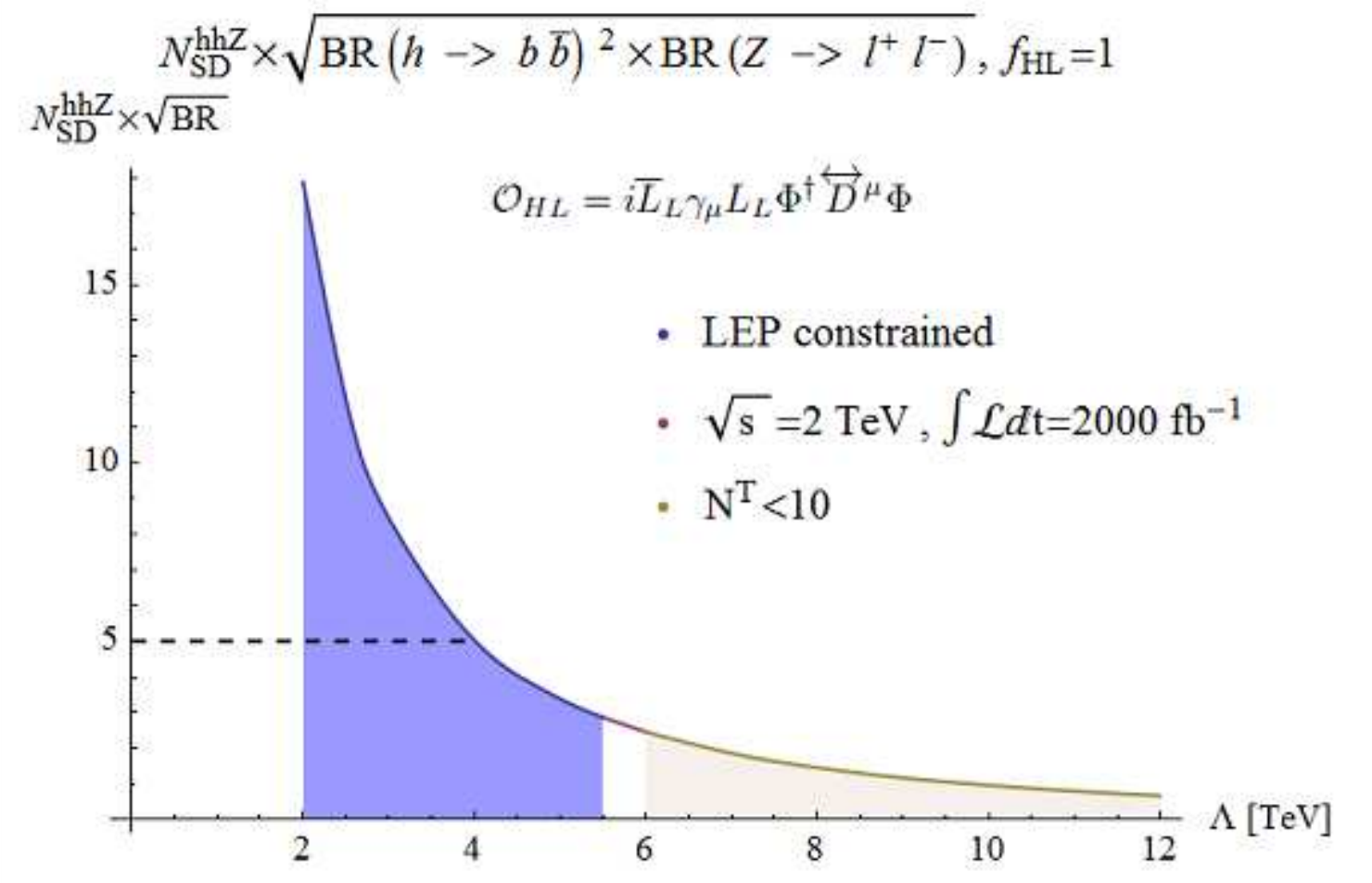}
\includegraphics[scale=0.58]{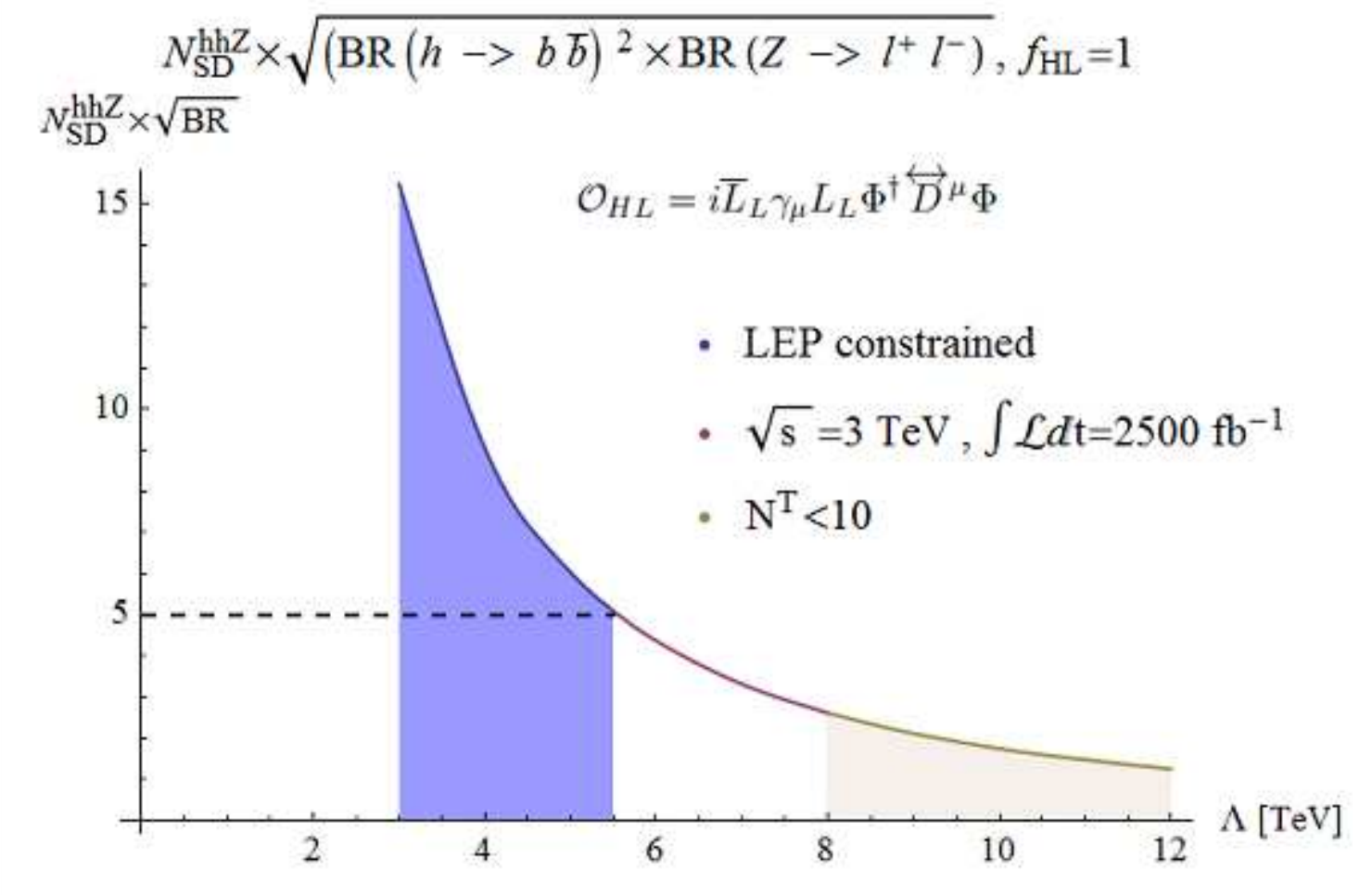}
\end{center}
\caption{Sensitivity to NP $\left(\Lambda\right)$ for $e^{+}e^{-}\rightarrow hhZ\rightarrow b\overline{b}b\overline{b}l^{+}l^{-}$
at $\sqrt{s}=2$ TeV with $L=2000\, fb^{-1}$ (left plot) and $\sqrt{s}=3$
TeV with $L=2500\, fb^{-1}$ (right plot). The dark blue region
is constrained by LEP. The shaded brown region has $N^{T}<10$. In
between lies a {}``window of opportunity'' for $\Lambda$. The dashed
horizontal line is the naive $5\sigma$ sensitivity. \label{NSD2}}
\end{figure}
As for $hhZ$ production, we find no sensitivity to $\mathcal{O}_{HL}$ when $\Lambda > 5.5$
TeV (which is the LEP bound on this operator) at an ILC with $\sqrt{s}=500$
GeV and $\sqrt{s}=1$ TeV, where the total number of events
is $N^{T}<10$. We, therefore, show in Fig.~\ref{NSD2} the sensitivity plots
for the $hhZ$ signal only in the case where the center of mass energy is $\sqrt{s}=2$ TeV and 3 TeV.
We see that, for the 2 TeV ILC, a $5\sigma$ effect can be obtained
for the signal $hhZ\rightarrow b\overline{b}b\overline{b}l^{+}l^{-}$
if $\Lambda=4$ TeV. The
sensitivity reach is extended to $\Lambda\sim8$ TeV at a $\sqrt{s}=3$ TeV machine,
a $5\sigma$ effect is plausible in the case $hhZ\rightarrow b\overline{b}b\overline{b}l^{+}l^{-}$
for $\Lambda=5.5$ TeV.

In Table 3.1 we summarize the highlights
of the analysis presented in this section. %

\begin{table}[h]
\begin{tabular}{|c|c|c|c|c|}
\hline
\multicolumn{5}{|c|}{$e^{+}e^{-}\rightarrow hZ\rightarrow b\overline{b}+l^{+}l^{-}$}\tabularnewline
\hline
\hline
$\sqrt{s}$ & 500 GeV & 1 TeV & 2 TeV & 3 TeV\tabularnewline
\hline
$\Lambda>$ & 5 TeV & 8 TeV & 12 TeV & $\gtrsim14$ TeV\tabularnewline
\hline
\end{tabular} \\
\bigskip
\begin{tabular}{|c|c|c|}
\hline
\multicolumn{3}{|c|}{$e^{+}e^{-}\rightarrow hhZ\rightarrow b\overline{b}+b\overline{b}+l^{+}l^{-}$}\tabularnewline
\hline
\hline
$\sqrt{s}$ & 2 TeV & 3 TeV\tabularnewline
\hline
$\Lambda>$ & 4 TeV & 5.5 TeV\tabularnewline
\hline
\end{tabular}

\caption{$5\sigma$ reach on $\Lambda$ for each ILC design: naive estimates
for $hZ$ production (upper table) and $hhZ$ production (lower table), followed
by the Higgs decay to $b\overline{b}$ and the $Z$ decay to leptons.}
\end{table}

\subsection{Validity of the EFT expansion}

Let us further define the ``validity'' function, $\mathcal{R}$:
\begin{eqnarray}
\mathcal{R} & \equiv & \frac{\Delta\sigma_{2}}{\sigma_{1}}\,,\label{validR}
\end{eqnarray}
where
\begin{eqnarray*}
\Delta\sigma_{2} & \equiv & \frac{\delta_{2}\left(s,f_{i}\right)}{\Lambda^{4}}\,,\,\sigma_{1}\equiv1+
\frac{\delta_{1}\left(s,f_{i}\right)}{\Lambda^{2}}\,,
\end{eqnarray*}
in accordance with the expansion used for the $hZ$ and $hhZ$ cross-sections, see (\ref{sigmahz}). That is,
$\Delta\sigma_{2}$ corresponds to the square of the NP amplitude.

\begin{figure}
\begin{center}
\includegraphics[scale=0.45]{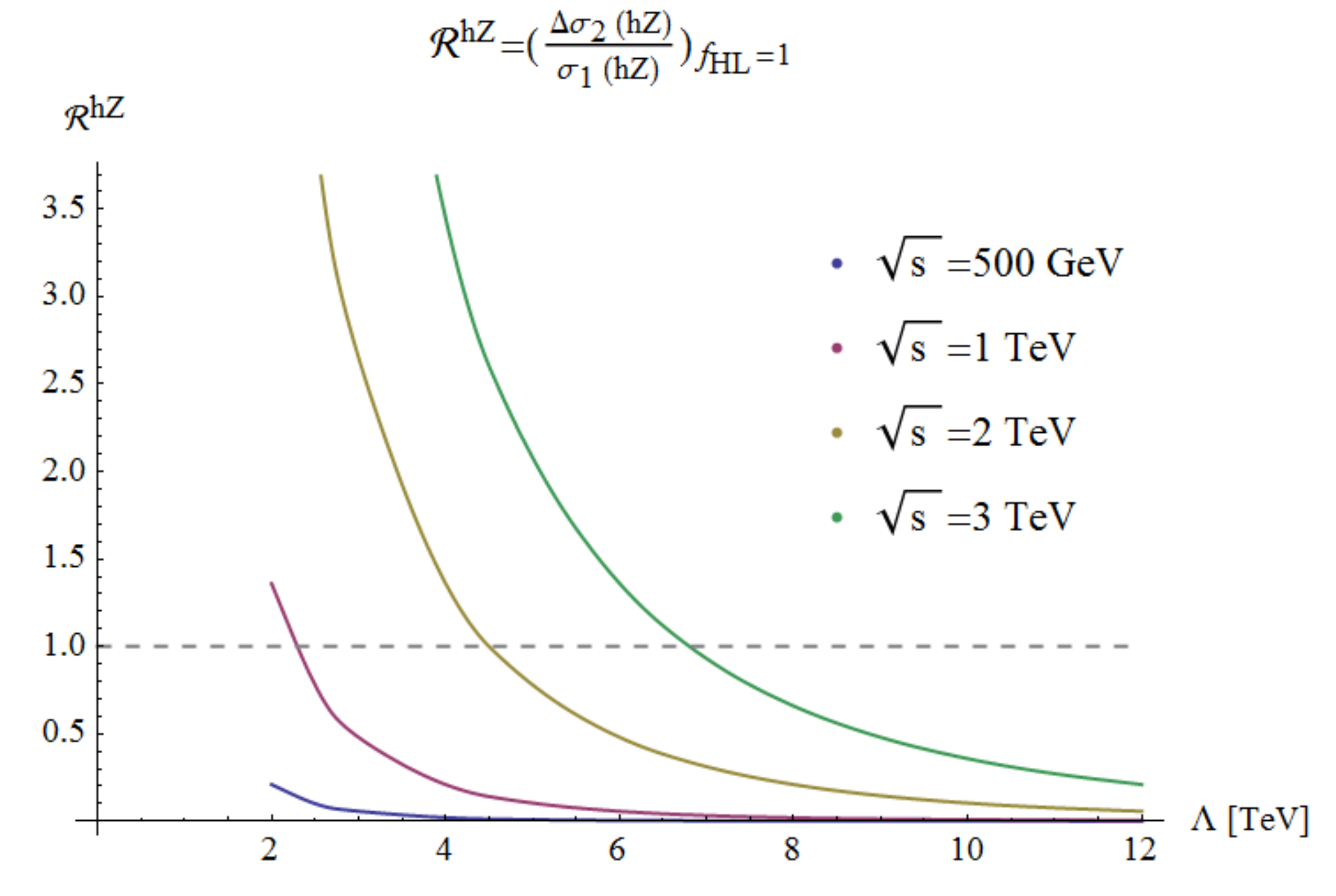}
\includegraphics[scale=0.45]{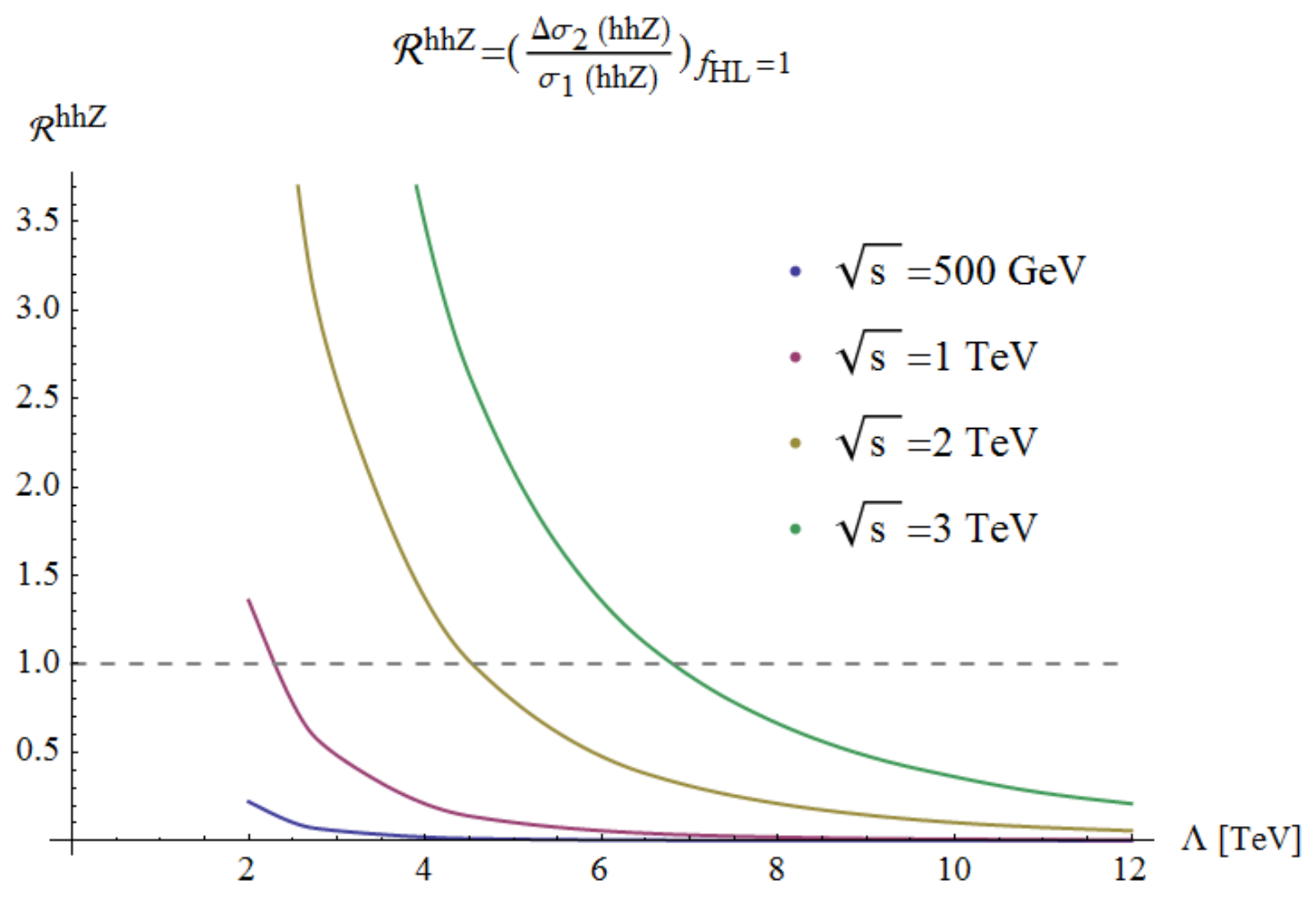}
\end{center}
\caption{Validity functions $\mathcal{R}^{hZ}\equiv\frac{\Delta\sigma_{2}\left(e^{+}e^{-}\rightarrow hZ\right)}{\sigma_{1}\left(e^{+}e^{-}\rightarrow hZ\right)}$
(left plot) and $\mathcal{R}^{hhZ}\equiv\frac{\Delta\sigma_{2}\left(e^{+}e^{-}\rightarrow hhZ\right)}{\sigma_{1}\left(e^{+}e^{-}\rightarrow hhZ\right)}$
(right plot), as a function of $\Lambda$ for all energies under consideration. \label{Rfunc}}
\end{figure}

In Fig.~\ref{Rfunc} we examine the validity functions $\mathcal{R}$
as a function of the scale of NP, $\Lambda$, for both
$hZ$ and $hhZ$ production in the presence of $\mathcal{O}_{HL}$
and for all center of mass energies under consideration.
The similarity between the validity
functions for the $hZ$ and $hhZ$ signals, shown in Fig.~\ref{Rfunc}, is a consequence
of the resemblance of the expansions in $1/\Lambda^{2}$ that was found in the previous section for
the corresponding cross-sections (see also below).
We see that the $1/\Lambda^{4}$ correction term
$\Delta\sigma_{2}$ exceeds the $1/\Lambda^{2}$ interference term,
i.e., giving $\mathcal{R}>1$,
as we go to lower values of $\Lambda$ (in which case $s/\Lambda^2$ increases).
We consider all values of
$\Lambda$ that lie above the dashed line $\mathcal{R}=1$ in Fig.~\ref{Rfunc}
to be inconsistent with our EFT expansion to dimension 6 operators.
This is prompted from our ignorance of the value of $\Delta\sigma_{2}$,
as it is subject to corrections from dimension 8 operators which we didn't consider
in our analysis. In particular, dimension
8 operators in the expansion
\begin{eqnarray}
\mathcal{L} & = & \mathcal{L}_{SM}+\sum_{i}\frac{f_{i}^{\left(6\right)}}{\Lambda^{2}}\mathcal{O}_{i}^{\left(6\right)}+
\sum_{j}\frac{f_{j}^{\left(8\right)}}{\Lambda^{4}}\mathcal{O}_{i}^{\left(8\right)}+
\cdots\label{dim8}\end{eqnarray}
may contribute to the  $eehZ$ and/or $eehhz$ contact interactions through their interference
with the SM diagrams, yielding a contribution
of the same order ($1/\Lambda^{4}$) as $\Delta\sigma_{2}$.

A natural dimension 8 operator, $\mathcal{O}_{i}^{\left(8\right)}$, that
comes to mind is $\Phi^{\dagger}\Phi\times\mathcal{O}_{HL}$, which
contributes to the same contact interactions but with a small suppression factor
of $v^{2}/\Lambda^{2}$ and is, thus, negligible. Another interesting dimension 8 operator
arises from the propagator expansion in (\ref{prop}) from
the $k=1$ term, yielding the same $\psi^{2}\varphi^{2}D$
class operators but with extra higher derivatives. That is, the operator
$i\overline{\psi}\gamma^{\mu}\psi{\color{black}{\color{blue}\square}}\Phi^{\dagger}\overleftrightarrow{D}_{\mu}\Phi$,
which, in the case of $\mathcal{R}>1$, may play an important role.

\subsection{An $hZ-hhZ$ Correlation \label{sec32}}

\begin{figure}
\begin{center}
\includegraphics[scale=0.45]{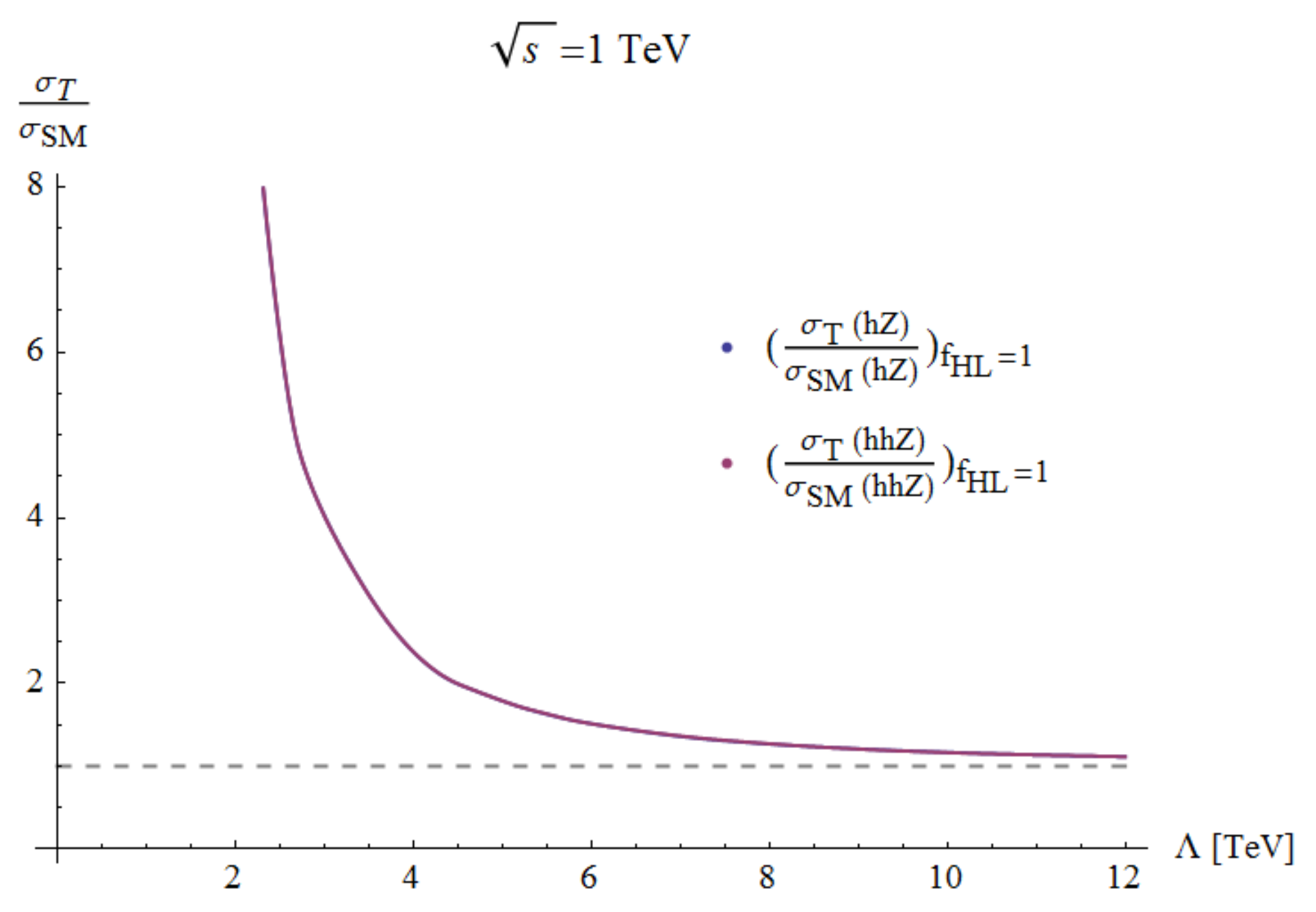}\quad
\end{center}
\caption{$\left(\frac{\sigma^{T}}{\sigma^{SM}}\right)_{f_{HL}=1}$ for $e^{+}e^{-}\rightarrow hZ$
and $e^{+}e^{-}\rightarrow hhZ$,
at $\sqrt{s}=1$ TeV. \label{fig_cor}}
\end{figure}

Using the results of section \ref{sec4}, we plot in Fig.~\ref{fig_cor}
the ratio $\frac{\sigma^{T}}{\sigma^{SM}}$ as a function of $\Lambda$,
in the presence of the operator $\mathcal{O}_{HL}$,
for both the $hZ$ and $hhZ$ signals at a 1 TeV ILC.
As expected, we find $\frac{\sigma^{T}\left(hZ\right)}{\sigma^{SM}\left(hZ\right)}=
\frac{\sigma^{T}\left(hhZ\right)}{\sigma^{SM}\left(hhZ\right)}$, which holds for any center of mass
energy. This validates numerically the similarity of the expansions
in $1/\Lambda^{2}$ for $\sigma^{T}\left(e^+ e^- \to hZ\right)$ and
$\sigma^{T}\left(e^+ e^- \to hhZ\right)$ that was found in the analytic derivation of section \ref{sec4}.

This property could play a key role
in distinguishing between different NP scenarios. For example anomalous
Higgs self-couplings are expected to exhibit a different behavior in $e^{+}e^{-}\rightarrow hZ$
versus $e^{+}e^{-}\rightarrow hhZ$.

\begin{figure}[h]
\begin{center}
\includegraphics[scale=0.45]{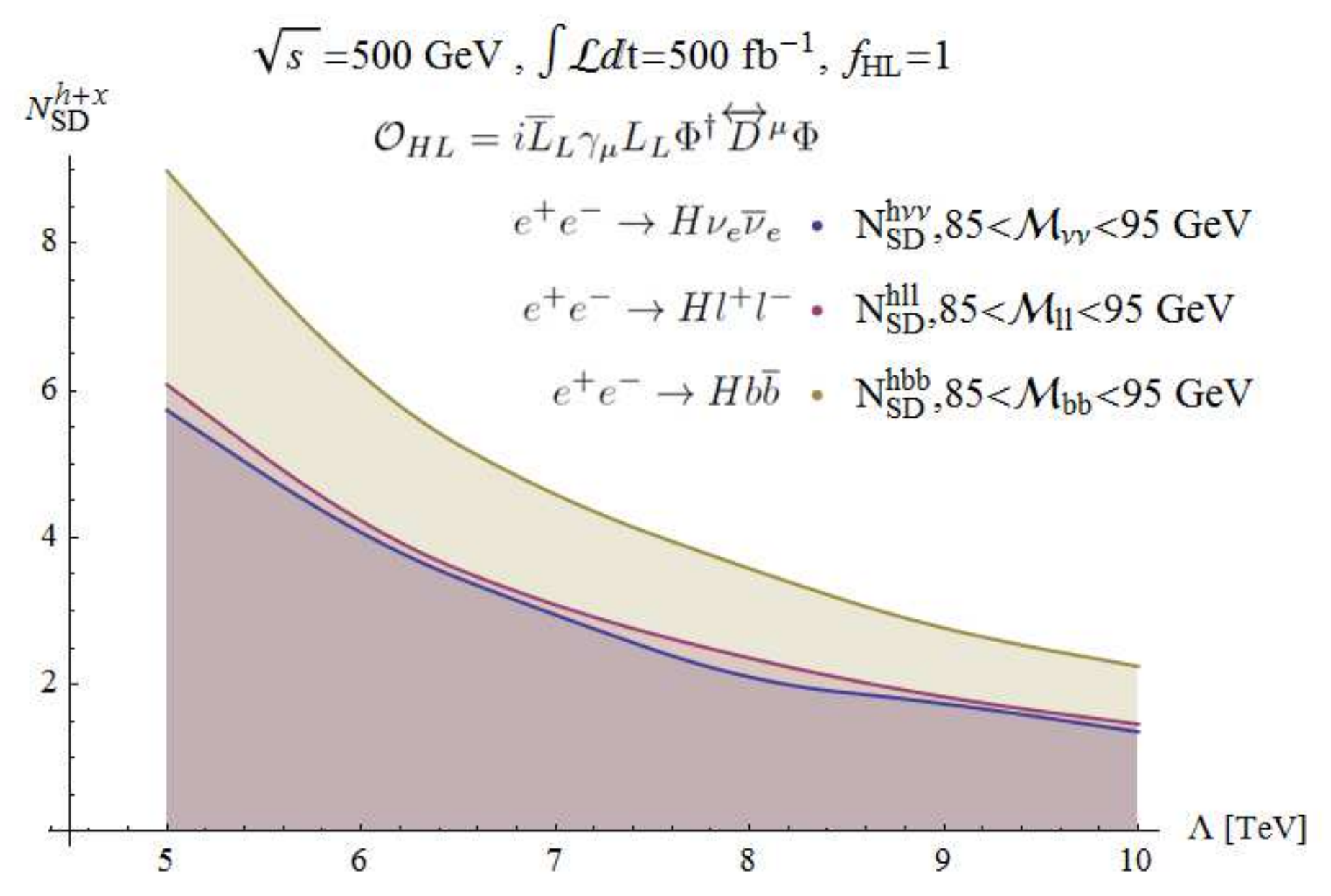}
\includegraphics[scale=0.45]{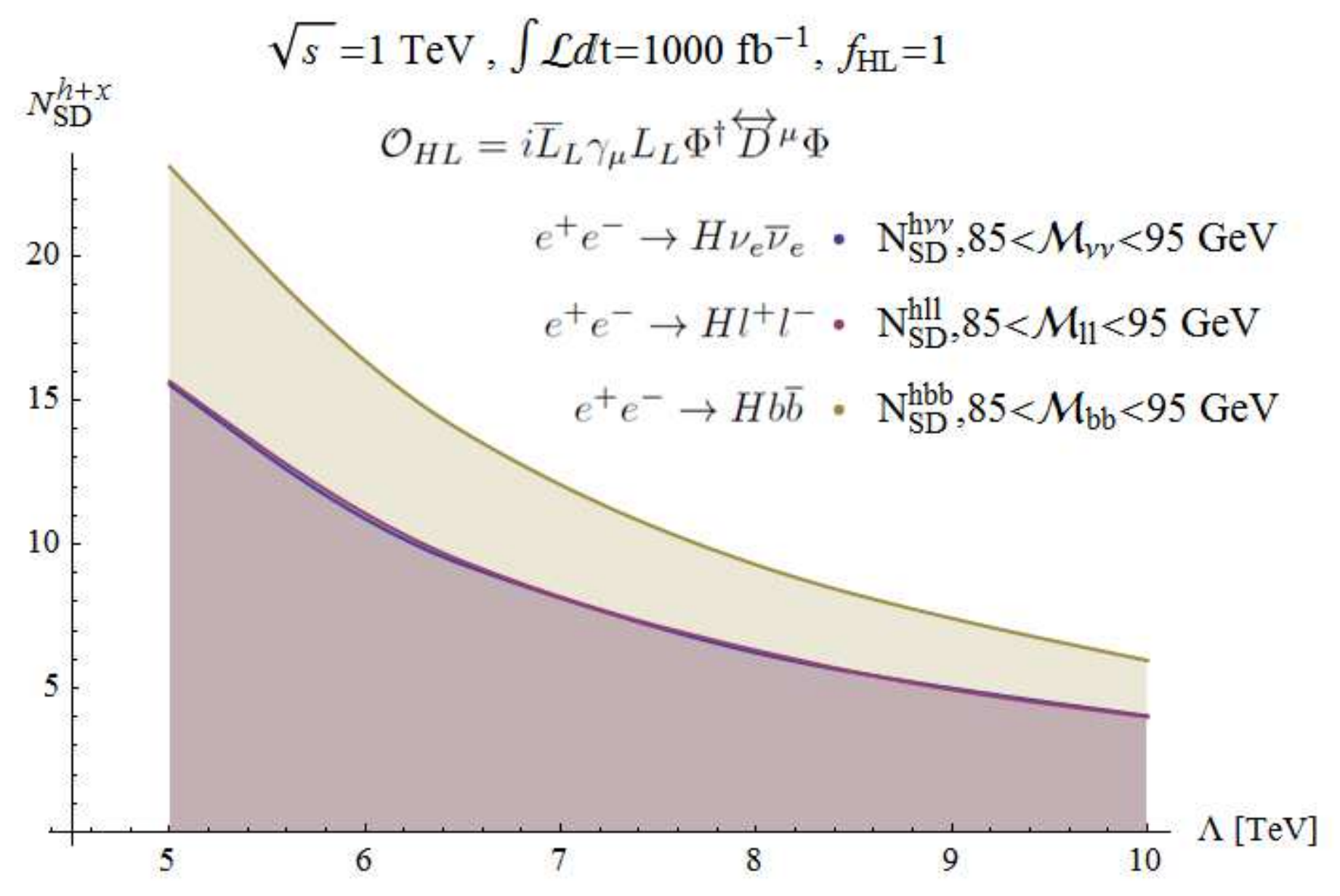}
\includegraphics[scale=0.45]{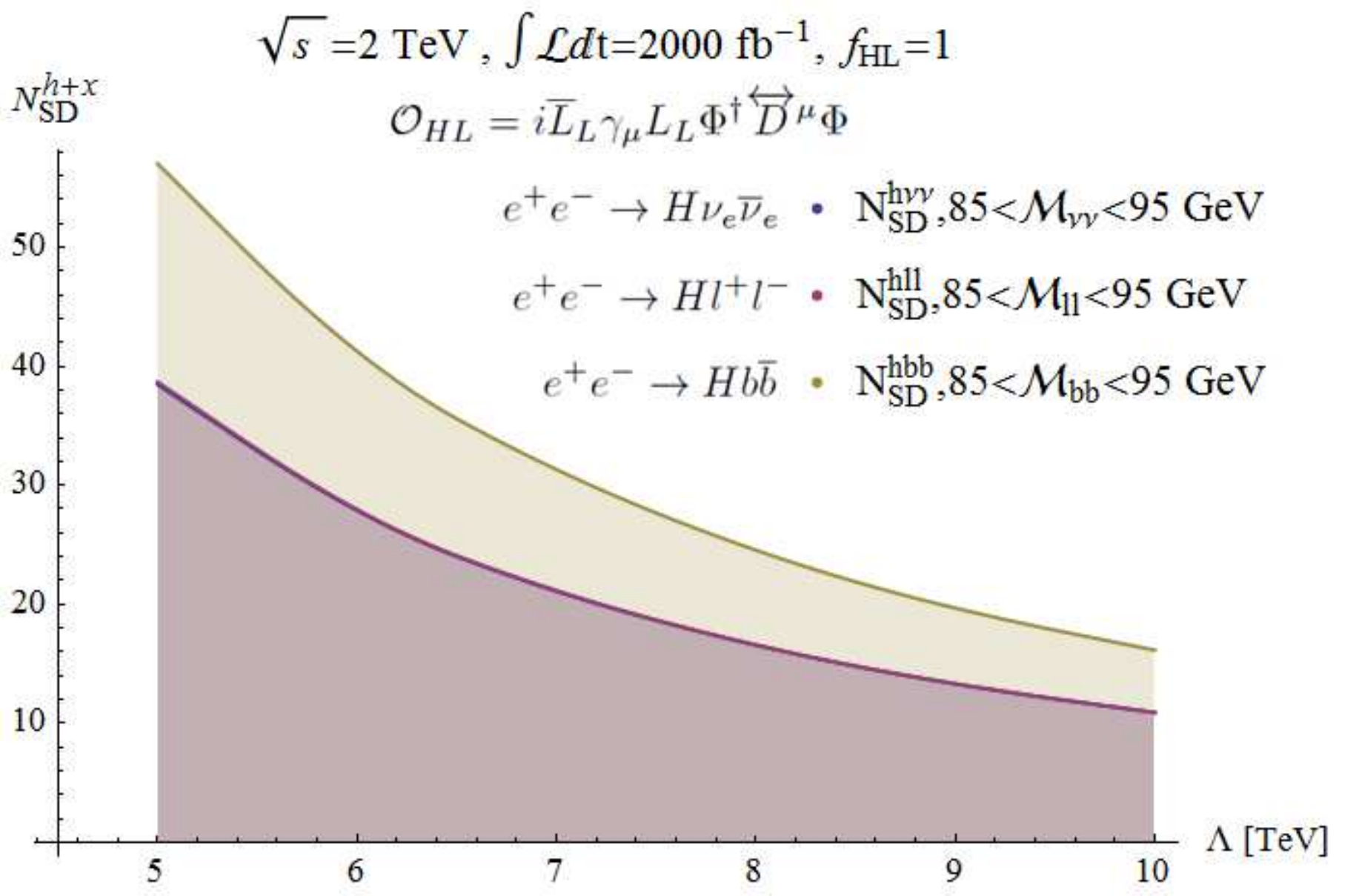}
\includegraphics[scale=0.45]{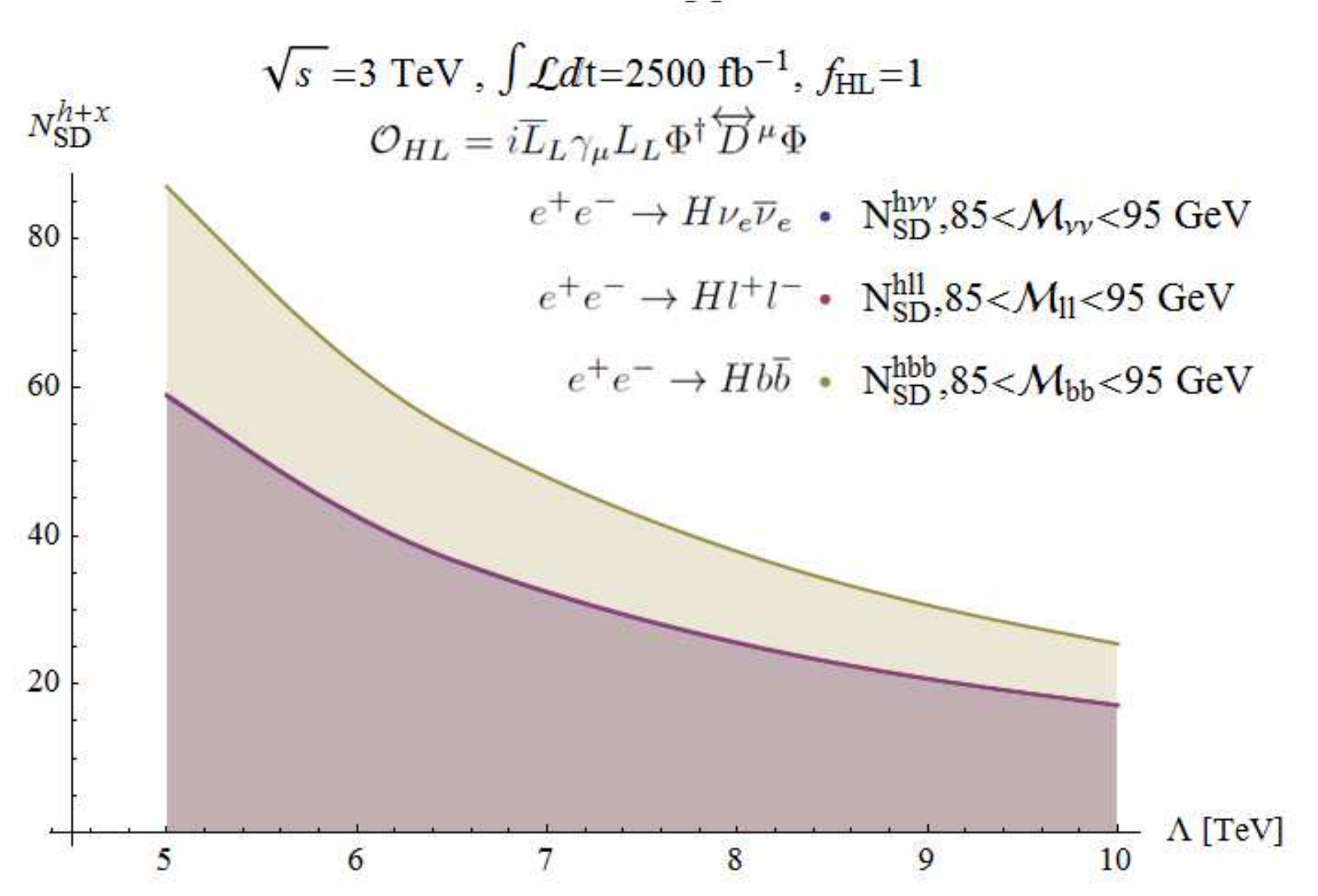}
\end{center}
\caption{Sensitivity ($N_{SD}$) to the NP scale $\Lambda$,
for $e^{+}e^{-}\rightarrow hZ\rightarrow h\nu_{e}\overline{\nu}_{e}, hl^{+}l^{-},hb\overline{b}$,
at $\sqrt{s} = 500$ GeV with $L=500\, fb^{-1}$, $\sqrt{s} = 1$ TeV with $L=1000\, fb^{-1}$,
$\sqrt{s} = 2$ TeV with $L=2000\, fb^{-1}$ and $\sqrt{s} = 3$ TeV
with $L=2500\, fb^{-1}$. \label{RS1}}
\end{figure}
\begin{table}[h]
\begin{tabular}{|c|c|c|c|c|c|c|}
\hline
\multicolumn{7}{|c|}{$e^{+}e^{-}\rightarrow h\nu_{e}\overline{\nu}_{e}$ comparison with
naive estimates of section 3.1 }\tabularnewline
\hline
\hline
 & \multicolumn{3}{c|}{$\sqrt{s}$=500~GeV} & \multicolumn{3}{c|}{$\sqrt{s}$=1~TeV}\tabularnewline
\hline
 & Before cuts & After cuts & $\sigma_{hZ}\times\mathcal{BR}_{Z}$ & Before cuts & After cuts & $\sigma_{hZ}\times\mathcal{BR}_{Z}$\tabularnewline
\hline
$\sigma^{SM}$ {[}fb{]} & 757 & 3.64 & 3.74 & 205.7 & 0.7874 & 0.840\tabularnewline
\hline
$\left(\sigma^{T}\right)_{\Lambda=6\, TeV}$ {[}fb{]} & 759 & 4.029 & 4.16 & 206.7 & 1.182 & 1.127\tabularnewline
\hline
\end{tabular} \\
\bigskip
\begin{tabular}{|c|c|c|c|c|c|c|}
\hline
 & \multicolumn{3}{c|}{$\sqrt{s}$=2~TeV} & \multicolumn{3}{c|}{$\sqrt{s}$=3~TeV}\tabularnewline
\hline
 & Before cuts & After cuts & $\sigma_{hZ}\times\mathcal{BR}_{Z}$ & Before cuts & After cuts & $\sigma_{hZ}\times\mathcal{BR}_{Z}$\tabularnewline
\hline
$\sigma^{SM}$ {[}fb{]} & 374.9 & 0.1889 & 0.203 & 483.6 & 0.0834 & 0.0898\tabularnewline
\hline
$\left(\sigma^{T}\right)_{\Lambda=6\, TeV}$ {[}fb{]} & 376 & 0.7574 & 0.8190 & 485.8 & 0.9561 & 1.03\tabularnewline
\hline
\end{tabular}
\caption{SM and SM+NP cross-sections in the $e^{+}e^{-}\rightarrow h\nu_{e}\overline{\nu}_{e}$
channel including all SM and NP diagrams, after imposing the cut on the missing
invariant mass of the two neutrinos $m_{Z}-4\Gamma_{Z}<\mathcal{M}_{\nu_{e}\overline{\nu}_{e}}<m_{Z}+4\Gamma_{Z}$
(in order to suppress the WW-fusion BG, see Fig.~\ref{figVVfusion} and Appendix C).
Also shown are the corresponding naive cross-sections of section \ref{sec5}:
$\sigma_{hZ}\times\mathcal{BR}_{Z}$,
where $\sigma_{hZ}\equiv\sigma\left(e^{+}e^{-}\rightarrow hZ\right)$
and $\mathcal{BR}_{Z}\equiv\mathcal{BR}\left(Z\rightarrow\nu_{e}\overline{\nu}_{e}\right)=6.6\%$.
Results are given for $\sqrt{s}$=500 GeV, 1 TeV (upper table) and
2 ,3 TeV (lower table). For the NP cross-section we take $\Lambda=$6
TeV. \label{tab41}}
\end{table}

\section{Realistic background estimation and sensitivities \label{sec6}}

In this section we present a realistic calculation of the signal to BG ratio for $e^{+}e^{-}\rightarrow hZ,hhZ$, 
including all possible diagrams that are generated by the SM + $\psi^{2}\varphi^{2}D$ type operators. We note again that possible effects from diagrams containing EFT
insertions of other dim. 6 operators
(e.g., effective triple Higgs couplings) are expected to be much smaller,
than the leading SM + $\psi^{2}\varphi^{2}D$ contribution (for a fixed value of
$\Lambda$)
and are, therefore, neglected in both signal and BG calculation.

\subsection{$e^{+}e^{-}\rightarrow hZ$ }

Let us consider the process
$e^{+}e^{-}\rightarrow hZ$ followed by $Z \rightarrow \nu_{e}\overline{\nu}_{e},l^{+}l^{-},b\overline{b}$
in the presence of $\mathcal{O}_{HL}$, taking into account all the
possible signal+BG diagrams which lead to
$e^{+}e^{-}\rightarrow h\nu_{e}\overline{\nu}_{e},hl^{+}l^{-},hb\overline{b}$.
As before, we sum over the electrons and muons final states ($l^{+}l^{-}=e^{+}e^{-}+\mu^{+}\mu^{-}$)
in both the signal and BG calculations.
In particular,
there are 16 diagrams in the $h\nu_{e}\overline{\nu}_{e}$ channel,
24 diagrams in the $hl^{+}l^{-}$ channel and 18 diagrams in $hb\overline{b}$ channel,
all depicted in Appendix C.
We impose kinematical cuts to suppress
the BG (in particular the WW and ZZ-fusion processes)
and perform a more realistic estimate of the sensitivity
to the $\psi^2 \varphi^2 D$ operators
in $e^{+}e^{-}\rightarrow hZ\rightarrow h\nu_{e}\overline{\nu}_{e}, hl^{+}l^{-},hb\overline{b}$.

For example, for the $he^{+}e^{-}$ final state, we reject most of the
BG from the ZZ-fusion process (see Fig.~\ref{figVVfusion}),
by imposing a cut on the invariant mass of the $e^{+}e^{-}$ system
to lie within $m_{Z}-2\Gamma_{Z}<\mathcal{M}_{ee}<m_{Z}+2\Gamma_{Z}$,
together with the acceptance cuts $p_{T}\left(e\right)>15$ GeV, $p_{T}\left(ee\right)>80$
GeV.

Our results for the case of $e^+ e^- \to h\nu_{e}\overline{\nu}_{e}$
are shown in Table \ref{tab41}, where we also
compare the naive cross-section estimates of section \ref{sec5} [i.e.,
$\sigma(e^+e^- \to h \nu_{e}\overline{\nu}_{e}) \approx \sigma(e^+ e^- \to hZ) \times \mathcal{BR}\left(Z\rightarrow\nu_{e}\overline{\nu}_{e}\right)$]
with the full cross-section calculation, including all diagrams and
imposing the appropriate invariant-mass cut to reduce the WW-fusion BG.
Evidently, our naive estimates hold even in the presence
of the irreducible BG,
indicating that the WW-fusion BG has been significantly reduced.

We note that the $h \nu_{e}\overline{\nu}_{e} = h+\rmpart{E_{T}}$ signal is of particular interest since it
resembles dark matter (DM) searches in both $e^+ e^-$ machines, see e.g., \cite{hzILC1,hzILC2}, and at the LHC, see e.g., \cite{hzLHC}.
For example, if the DM interacts via a Higgs
portal operator $\chi^{2}\Phi^{2}$
(where $\chi=$DM and $\Phi$ is the SM Higgs doublet),
then the production of
an off-shell Higgs via $gg\rightarrow h^{\ast} \rightarrow h\,\chi\overline{\chi}$
will gives rise
to the $h+\rmpart{E_{T}}$ signature at the LHC \cite{hzLHC}.

As in section \ref{sec5}, we plot in Fig.~\ref{RS1} the sensitivity, $N_{SD}$, as
a function of the scale of the NP scale $\Lambda$, for the three proposed scenarios/decay
modes: $hZ \to h\nu_{e}\overline{\nu}_{e}, hl^{+}l^{-},hb\overline{b}$.
The $hb\overline{b}$ channel is expected to have a higher sensitivity to $\Lambda$
since it doesn't suffer from WW/ZZ-fusion BG (see the corresponding diagrams in Appendix C), see also \cite{hztohbb}.
The neutrino and lepton channels (blue and red respectively) exhibit
the same behaviour, albeit, with a lower sensitivity than the $b\overline{b}$
channel (yellow). In particular, we see that, at a 1 TeV collider, the $h+\rmpart{E_{T}}$
channel (neutrino channel) is sensitive to $\Lambda=6$ TeV at a $\sim 10\sigma$ level,
whereas the $h b\overline{b}$ channel will reach this sensitivity ($10\sigma$)
for a higher NP threshold of $\sim \Lambda=8$ TeV.
\begin{table}
\begin{tabular}{|c|c|c|}
\hline
$\sqrt{s}$ & $N_{SD}^{hll}\,,N_{SD}^{h\nu\nu}$ & $\Lambda$\tabularnewline
\hline
\hline
500 GeV & $6\sigma$ & 5 TeV\tabularnewline
\hline
1 TeV & $10\sigma$ & 6 TeV\tabularnewline
\hline
2 TeV & $20\sigma$ & 7 TeV\tabularnewline
\hline
3 TeV & $25\sigma$ & 8 TeV\tabularnewline
\hline
\end{tabular}
\bigskip
\begin{tabular}{|c|c|c|}
\hline
$\sqrt{s}$ & $N_{SD}^{hbb}$ & $\Lambda$\tabularnewline
\hline
\hline
500 GeV & $6\sigma$ & 6 TeV\tabularnewline
\hline
1 TeV & $10\sigma$ & 8 TeV\tabularnewline
\hline
2 TeV & $20\sigma$ & 9 TeV\tabularnewline
\hline
3 TeV & $25\sigma$ & 10 TeV\tabularnewline
\hline
\end{tabular}
\caption{The expected sensitivity on the scale of NP, $\Lambda$, for selected values of $N_{SD}^{hll}\,,N_{SD}^{h\nu\nu}$
(left table) and $N_{SD}^{hbb}$ (right table), see also text. \label{tab42}}
\end{table}

In Table \ref{tab42} we list some selected realistic results for the expected sensitivity
of the ILC to the scale of NP, $\Lambda$, in $e^+ e^- \to hZ \to h\nu_{e}\overline{\nu}_{e}, hl^{+}l^{-},hb\overline{b}$.

\subsection{$e^+ e^- \to hhZ$}

We repeat the same analysis for the $hhZ$ signals $e^{+}e^{-}\rightarrow hhZ\rightarrow hh\nu_{e}\overline{\nu}_{e},
hhl^{+}l^{-}, hhb\overline{b}$
at $\sqrt{s}=3$ TeV, taking into account all SM+NP diagrams and imposing
similar kinematical cuts to reduce the BG. There are $\sim100$ diagrams
for the $hhl^{+}l^{-}$ and $hhb\overline{b}$ channels and $\sim70$
diagrams for the $hh\nu_{e}\overline{\nu}_{e}$ channel; a sample
of these diagrams is shown in Appendix D.
\begin{figure}[h]
\begin{center}
\includegraphics[scale=0.4]{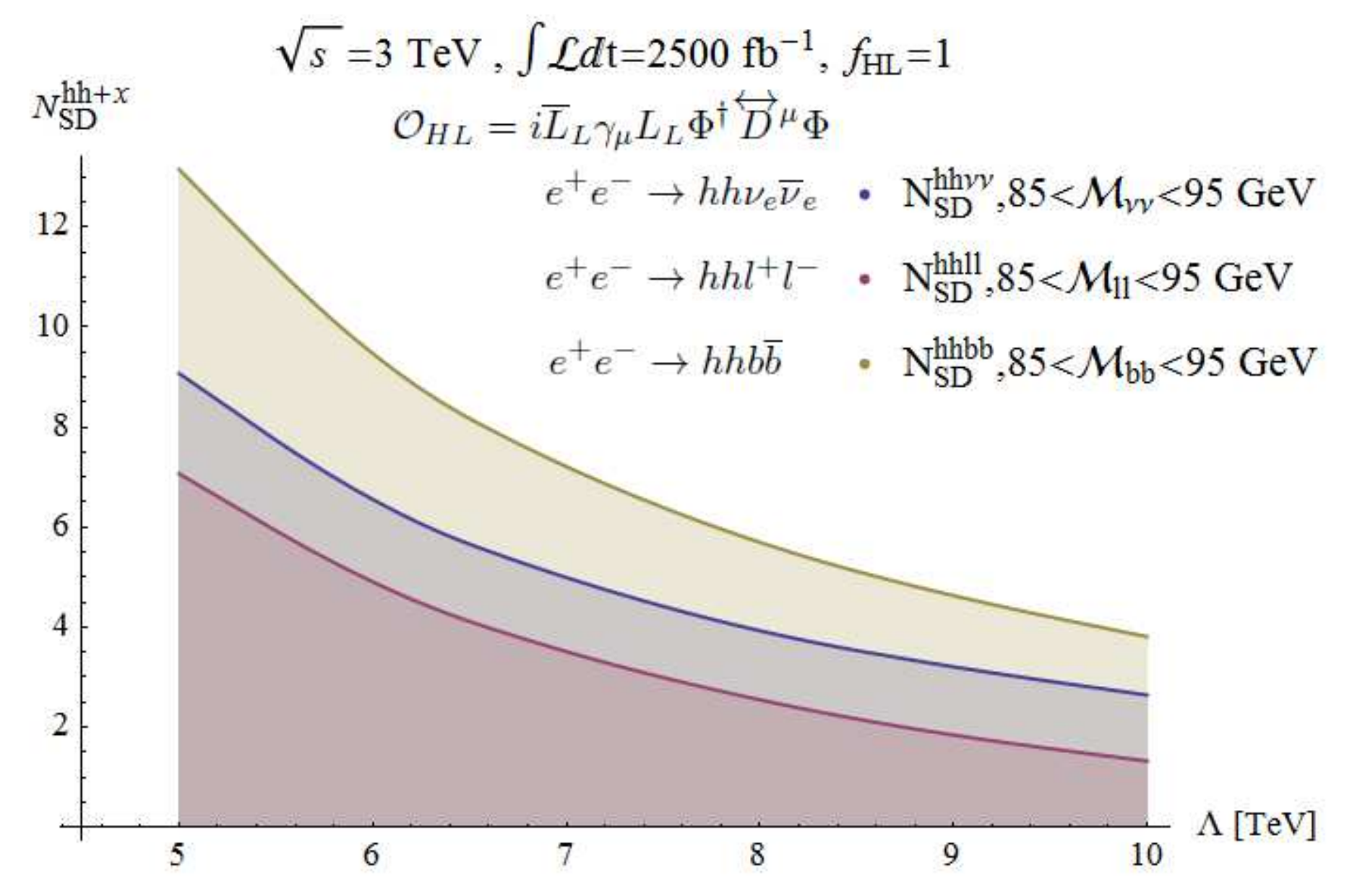}
\includegraphics[scale=0.4]{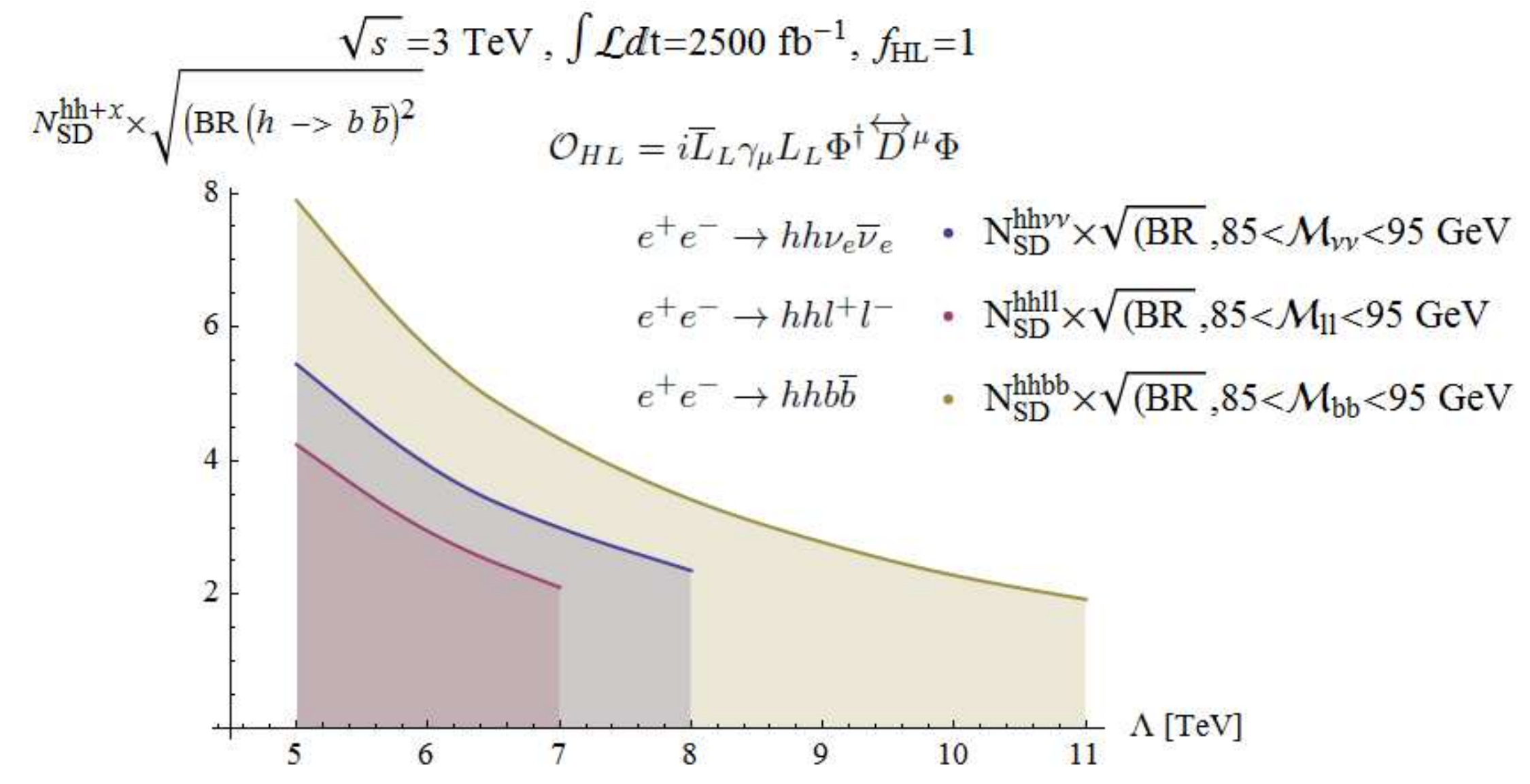}
\end{center}
\caption{Sensitivity ($N_{SD}$) to the NP scale $\Lambda$, at a $\sqrt{s}=3$ TeV ILC with $L=2500\, fb^{-1}$. 
Left figure: for $e^{+}e^{-}\rightarrow hhZ\rightarrow hh\nu_{e}\overline{\nu}_{e}, hhl^{+}l^{-}, hhb\overline{b}$, 
right figure: for $e^{+}e^{-}\rightarrow hhZ$ followed by
$hh \rightarrow b\overline{b}b\overline{b}$ and $Z \to \nu_{e}\overline{\nu}_{e}, l^{+}l^{-}, b\overline{b}$, 
where the curves are cut when $N^{T}<10$ (i.e., less than 10 events).
\label{RS2}}
\end{figure}
\begin{figure}[h]
\begin{center}
\includegraphics[scale=0.5]{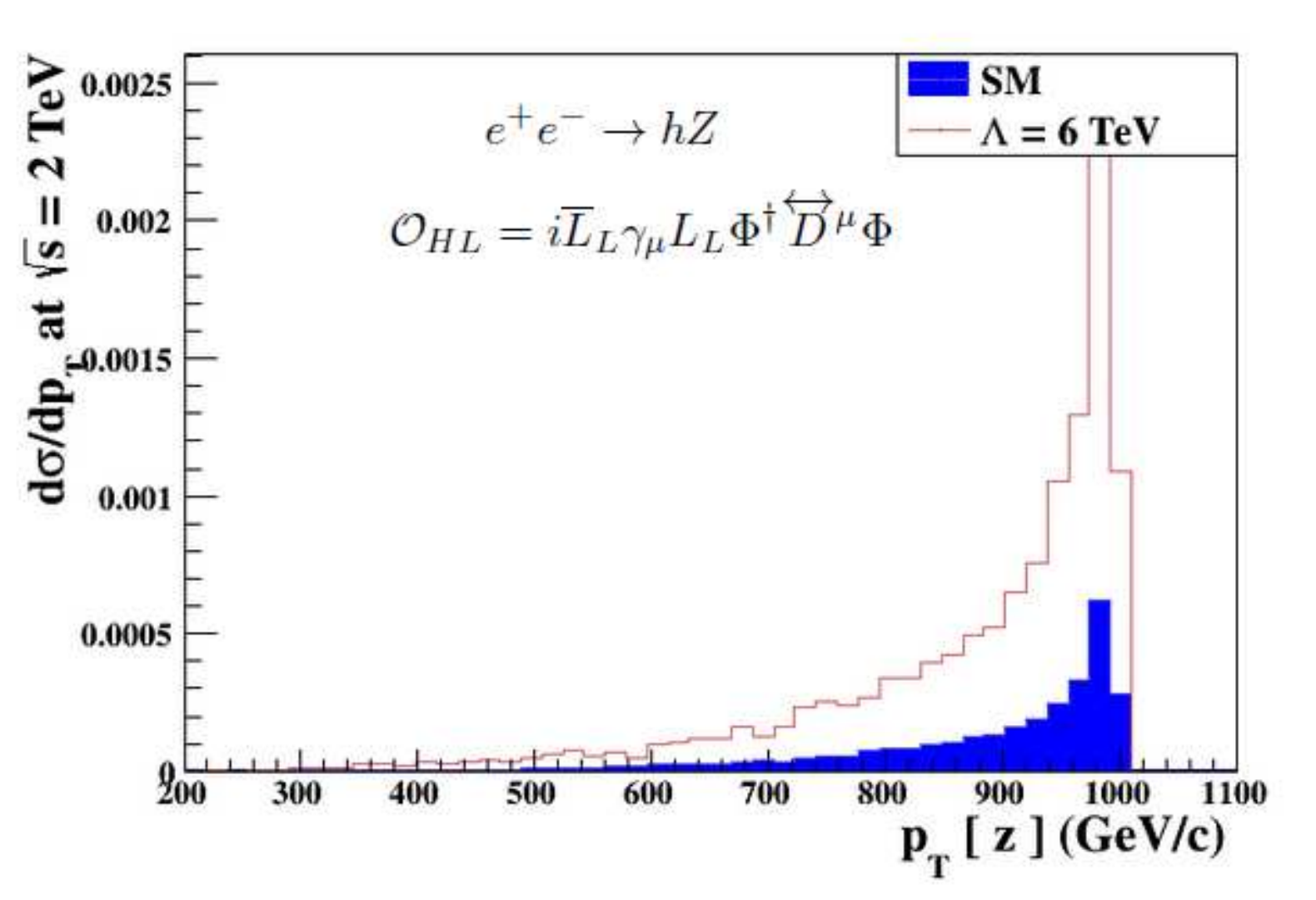}
\includegraphics[scale=0.5]{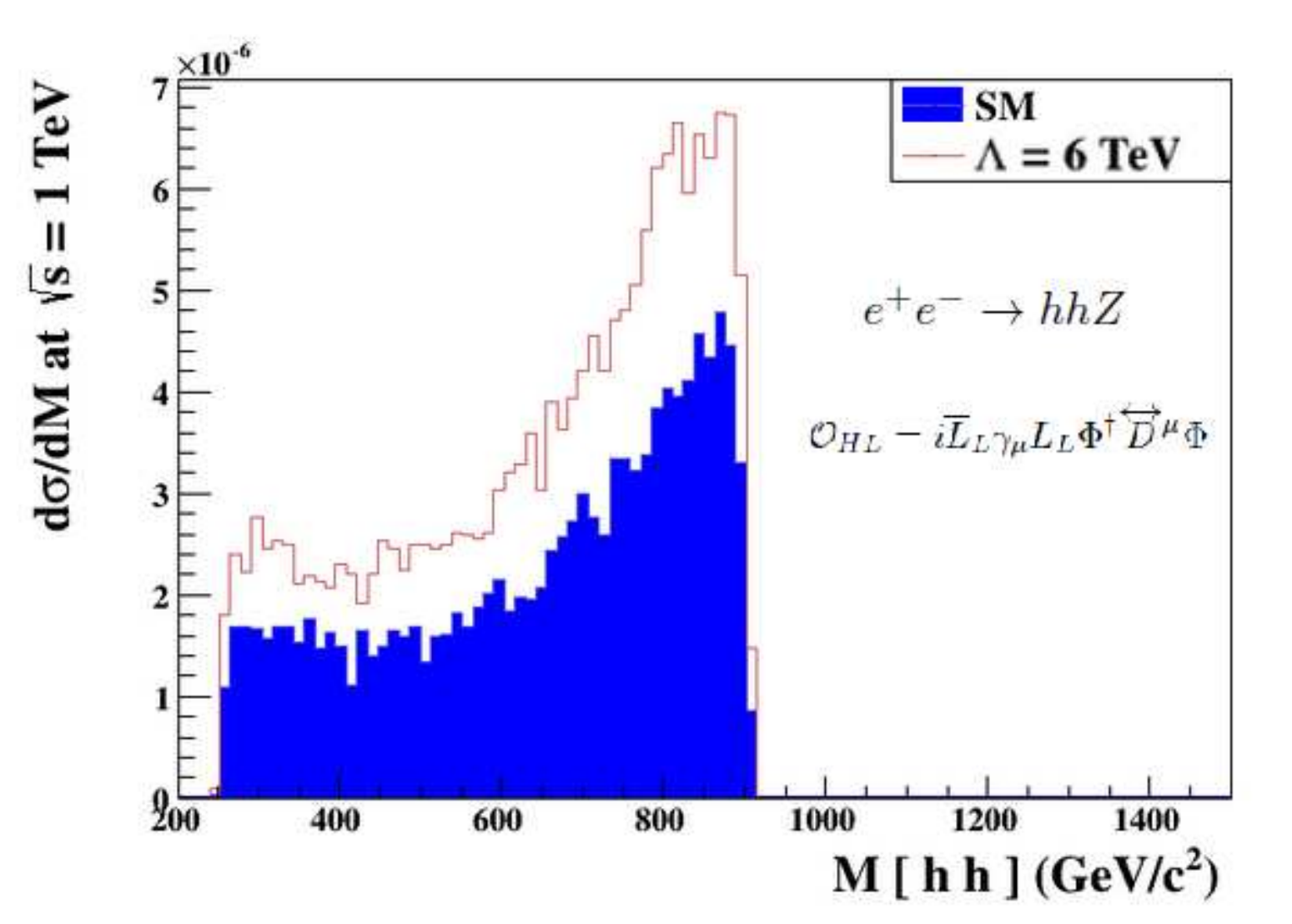}
\includegraphics[scale=0.5]{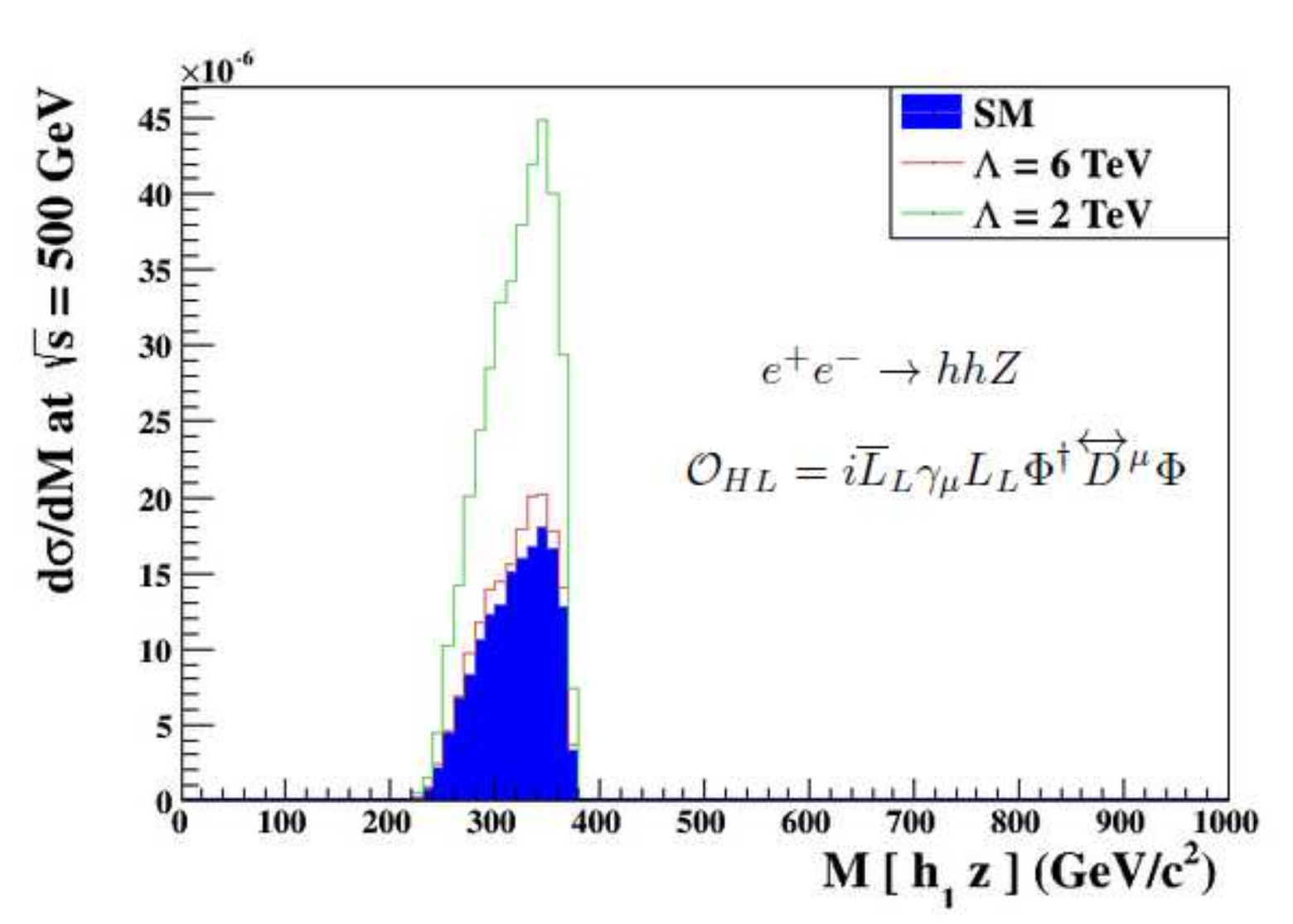}
\includegraphics[scale=0.5]{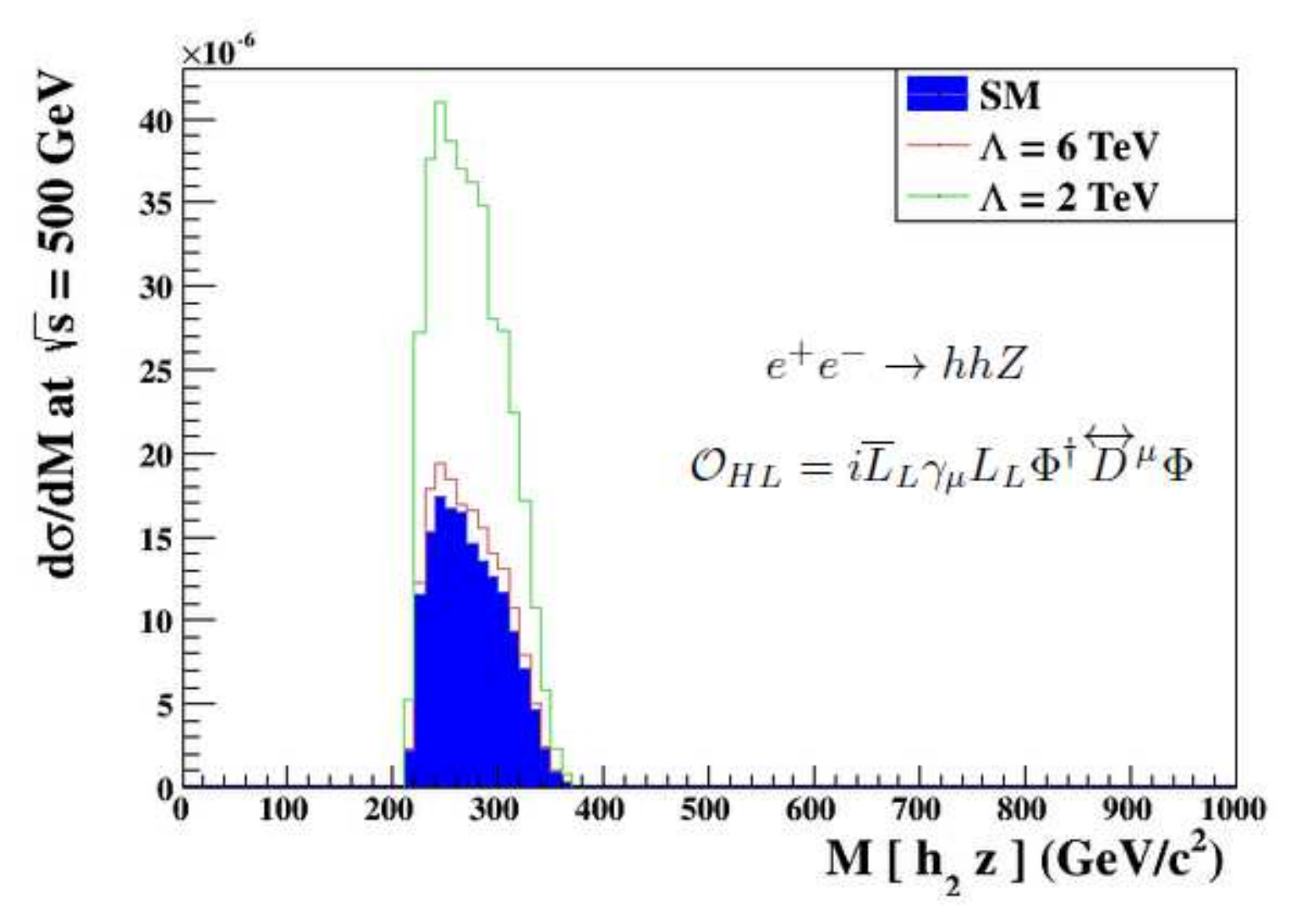}
\end{center}
\caption{The $p_{T}\left(Z\right)$ distribution in $e^{+}e^{-}\rightarrow hZ$
at $\sqrt{s}=$2 TeV (upper left), the invariant mass distribution of $hh$ in
$e^{+}e^{-}\rightarrow hhZ$ at $\sqrt{s}=$1 TeV (upper right), the
invariant mass distribution of $Z$ + the Higgs with the largest-$p_{T}$ in
$e^{+}e^{-}\rightarrow hhZ$ at $\sqrt{s}=$500 GeV (lower left)
and the invariant mass of $Z$ + the Higgs with the 2nd largest-$p_{T}$
in $e^{+}e^{-}\rightarrow hhZ$ at $\sqrt{s}=$500 GeV (lower right).
The blue-solid histogram depicts the SM predictions while the red
(green) solid lines correspond to the total cross-section including
the effect of $\mathcal{O}_{HL}$, where $f_{HL}=1$ and $\Lambda=6$
TeV ($\Lambda=2$ TeV). \label{RS4}}
\end{figure}

Our results for the $e^{+}e^{-}\rightarrow hhZ\rightarrow hh\nu_{e}\overline{\nu}_{e},
hhl^{+}l^{-}, hhb\overline{b}$ cases are shown in Fig.~\ref{RS2}.
We see that the difference between
the $hh\nu_{e}\overline{\nu}_{e}$ and $hhl^{+}l^{-}$ signals (blue
and red, respectively) is more pronounced in the case of the Higgs pair production channel $e^+ e^- \to hhZ$
(than the single Higgs production one $e^+ e^- \to hZ$).
Also, as in the case of $hZ$ production, the $hhb\overline{b}$
channel (yellow) is more sensitive than the leptonic channels to the scale of the NP $\Lambda$.
In fact, in the case of the lepton channels, $hh\nu_{e}\overline{\nu}_{e}$ and $hhl^{+}l^{-}$,
the results shown in Fig.~\ref{RS2}
are very similar to our naive estimates of section \ref{sec5}.

In Fig.~\ref{RS2}
we also show the sensitivity for the case in which each of the Higgs in
the final state further decays via $h\rightarrow b\overline{b}$, i.e.,
$N_{SD}^{hh+x}\times\sqrt{BR\left(h\rightarrow b\overline{b}\right)^{2}}$,
and the corresponding reach on $\Lambda$ for all $Z$ decay channels.
We omit the region excluded by LEP ($\Lambda\lesssim5$ TeV) and
the cases where there are less than 10 events, i.e., the region where $N^{T}<10$.
We see that, in the case of $hh\rightarrow b\overline{b}b\overline{b}$,
the ``window" of reach on $\Lambda$ depends on the $Z$ decay channel. In particular,
when $Z$ decays to leptons the sensitivity window reaches $\Lambda=7$ TeV
(smaller by $\sim 1$ TeV compared to the naive estimate of section \ref{sec5}),
whereas for the other channels the sensitivity reach on $\Lambda$ extends
to $8$ TeV (neutrino channel) and $11$ TeV ($b\overline{b}$ channel).

\subsection{Differential Distributions}

We have processed the numerical results of section \ref{sec5} using
MadAnalysis5 \cite{MGA}, to study some useful differential distributions in
both $hZ$ and $hhZ$ production channels in the presence of $\mathcal{O}_{HL}$.
This is shown in Fig.~\ref{RS4}, where we see that no particular
new behaviour is exhibited, in the sense that all distributions are magnified w.r.t
the SM, whereas the shape of the distributions remains intact.

\pagebreak

\section{Summary \label{sec7}}

We have investigated possible NP effects in Higgs - Vector boson associated
production at a future ILC in a model independent approach,
using a certain class of dimension 6 operators - the $\psi^{2}\varphi^{2}D$ class operators,
where $\psi$ is a fermion, $\varphi$ is the SM Higgs field and $D$ is the SM
covariant derivative.
These operators are generated by new heavy vector-boson exchanges
at high energy scales $\Lambda \gg v$ in the underlying
theory and they
give rise to new contact interactions of the form $e^+e^- hZ$ and
$e^+e^- hhZ$.

\begin{table}[h]
\begin{centering}
\begin{tabular}{|c|c|c|c|}
\hline
\multicolumn{4}{|c|}{$e^{+}e^{-}\rightarrow hZ\rightarrow h+x$ for $\Lambda=6$ TeV }\tabularnewline
\hline
\hline
 & $N_{SD}^{hll}$ & $N_{SD}^{h\nu\nu}$ & $N_{SD}^{hbb}$\tabularnewline
\hline
$\sqrt{s}=500$ GeV & $4\sigma$ & $4\sigma$ & $6\sigma$\tabularnewline
\hline
$\sqrt{s}=1$ TeV & $10\sigma$ & $10\sigma$ & $16\sigma$\tabularnewline
\hline
\end{tabular} \\
\bigskip
\begin{tabular}{|c|c|c|c|}
\hline
\multicolumn{4}{|c|}{$e^{+}e^{-}\rightarrow hhZ\rightarrow hh+x$ for $\Lambda=7$ TeV }\tabularnewline
\hline
\hline
 & $N_{SD}^{hhll}$ & $N_{SD}^{hh\nu\nu}$ & $N_{SD}^{hhbb}$\tabularnewline
\hline
$\sqrt{s}=3$ TeV & $3.5\sigma$ & $5\sigma$ & $7\sigma$\tabularnewline
\hline
\end{tabular}
\par\end{centering}

\caption{An example of realistic estimates for the expected
statistical significance $N_{SD}^{hx}$ for probing NP with $\Lambda=6$
TeV in $e^+e^- \to hZ$ followed by $Z \to x$, at $\sqrt{s}=$500 GeV, 1 TeV (upper table)
and, similarly, $N_{SD}^{hhx}$ for probing NP with $\Lambda=7$
TeV in $e^+e^- \to hhZ$ followed by $Z \to x$, at $\sqrt{s}=3$ TeV (lower table). \label{tabsummary}}

\end{table}

We have presented full analytical expressions of the
tree-level cross-sections for $e^+e^- \to hZ$ and $e^+e^- \to hhZ$ in the presence
of the $\psi^{2}\varphi^{2}D$ class operators, and
showed that they have an identical dependence in the $1/\Lambda^{2}$ expansion.
As a result, we found an interesting correlation between the $hZ$ and
the $hhZ$ signals which can be utilized in future NP searches in these
channels.

We performed MadGraph simulations for both $e^{+}e^{-}\rightarrow hZ$
and $e^{+}e^{-}\rightarrow hhZ$ at an ILC with center of mass energies
$\sqrt{s}=500$ GeV, 1 TeV, 2 TeV and 3 TeV and obtained realistic
estimates of the sensitivity to the NP scale $\Lambda$, based on the full
set of SM + NP diagrams which includes the irreducible background
in the cases where the Z-boson decays via
$Z\rightarrow\nu_{e}\overline{\nu}_{e}, l^{+}l^{-}$ and $Z \to b\overline{b}$.

We have also considered the constraints on the $\psi^{2}\varphi^{2}D$ class operators,
primarily from LEP, since these operators modify the Z-couplings to
fermions. Our results show that a TeV-scale ILC will be able to probe
NP in $e^+e^- \to hZ,hhZ$ in the form of the $\psi^{2}\varphi^{2}D$ class operators
at scales beyond the LEP bounds and the LHC 14 reach.
A sample of our results is given in Table \ref{tabsummary}.

\bigskip
\bigskip
\bigskip
\bigskip

{\bf Acknowledgments:}
We thank Amarjit Soni and Jose Wudka for useful discussions.

\pagebreak

\bibliographystyle{h-physrev3}
\bibliography{bibfile}

\newpage

\appendix

\section{Feynman rules}

The following Feynman rules corresponding to $\mathcal{L}_{F_{1}}$ in (\ref{LF1})
were obtained via the HEL model implementation of \cite{HELpaper} in FeynRules, in the
physical basis by, i.e., without the Goldstone bosons $G^{0},G^{\pm}$.
The left and right projection operators are denoted by $P_{\pm}=\frac{1\pm\gamma_{5}}{2}$. Also,
$e=g s_{w}=g^{\prime}c_{w}$, where $s_w(c_w)$ is the sin(cos) of the Weinberg angle.
The charged leptons are denoted by $l$ and the neutrinos by $\nu_{l}$ ,while $u_{q}$
and $d_{q}$ are the up-type and down-type quarks, respectively.
$V^{CKM}$ is the CKM mixing matrix.

The Feynmann rules involving leptons read:
\begin{eqnarray}
\left\{ \overline{\nu}_{l},l,W^{\mu+}\right\}  & : & \frac{i\overline{c}_{HL}^{\prime}e\gamma^{\mu}P_{-}}{\sqrt{2}s_{w}}\label{eq:-57}\\
\left\{ \overline{\nu}_{l},l,h,h,W^{\mu+}\right\}  & : & \frac{i\sqrt{2}\overline{c}_{HL}^{\prime}e\gamma^{\mu}P_{-}}{s_{w}v^{2}}\label{eq:-59}\\
\left\{ \overline{\nu}_{l},l,h,W^{\mu+}\right\}  & : & \frac{i\sqrt{2}\overline{c}_{HL}^{\prime}e\gamma^{\mu}P_{-}}{s_{w}v}\label{eq:-60}\\
\left\{ \overline{\nu}_{l},\nu_{l},Z^{\mu}\right\}  & : & -\frac{i\overline{c}_{HL}e\gamma^{\mu}P_{-}}{2s_{w}c_{w}}+\frac{i\overline{c}_{HL}^{\prime}e\gamma^{\mu}P_{-}}{2s_{w}c_{w}}\label{eq:-65}\\
\left\{ \overline{\nu}_{l},\nu_{l},h,h,Z^{\mu}\right\}  & : & -\frac{i\overline{c}_{HL}e\gamma^{\mu}P_{-}}{s_{w}c_{w}v^{2}}+\frac{i\overline{c}_{HL}^{\prime}e\gamma^{\mu}P_{-}}{s_{w}c_{w}v^{2}}\label{eq:-66}\\
\left\{ \overline{\nu}_{l},\nu_{l},h,Z^{\mu}\right\}  & : & -\frac{i\overline{c}_{HL}e\gamma^{\mu}P_{-}}{s_{w}c_{w}v}+\frac{i\overline{c}_{HL}^{\prime}e\gamma^{\mu}P_{-}}{s_{w}c_{w}v}\label{eq:-67}\\
\left\{ \overline{l},\nu_{l},W^{\mu-}\right\}  & : & \frac{i\overline{c}_{HL}^{\prime}e\gamma^{\mu}P_{-}}{\sqrt{2}s_{w}}\label{eq:-68}\\
\left\{ \overline{l},\nu_{l},h,h,W^{\mu-}\right\}  & : & \frac{i\sqrt{2}\overline{c}_{HL}^{\prime}e\gamma^{\mu}P_{-}}{s_{w}v^{2}}\label{eq:-69}\\
\left\{ \overline{l},\nu_{l},h,W^{\mu-}\right\}  & : & \frac{i\sqrt{2}\overline{c}_{HL}^{\prime}e\gamma^{\mu}P_{-}}{s_{w}v}\label{eq:-70}\\
\left\{ \overline{l},l,Z^{\mu}\right\}  & : & -\frac{i\overline{c}_{HL}e\gamma^{\mu}P_{-}}{2s_{w}c_{w}}-\frac{i\overline{c}_{HL}^{\prime}e\gamma^{\mu}P_{-}}{2s_{w}c_{w}}-\frac{i\overline{c}_{He}e\gamma^{\mu}P_{+}}{4s_{w}c_{w}}\label{eq:-71}\\
\left\{ \overline{l},l,h,h,Z^{\mu}\right\}  & : & -\frac{i\overline{c}_{HL}e\gamma^{\mu}P_{-}}{s_{w}c_{w}v^{2}}-\frac{i\overline{c}_{HL}^{\prime}e\gamma^{\mu}P_{-}}{s_{w}c_{w}v^{2}}-\frac{i\overline{c}_{He}e\gamma^{\mu}P_{+}}{2s_{w}c_{w}v^{2}}\label{eq:-72}\\
\left\{ \overline{l},l,h,Z^{\mu}\right\}  & : & -\frac{i\overline{c}_{HL}e\gamma^{\mu}P_{-}}{s_{w}c_{w}v}-\frac{i\overline{c}_{HL}^{\prime}e\gamma^{\mu}P_{-}}{s_{w}c_{w}v}-
\frac{i\overline{c}_{He}e\gamma^{\mu}P_{+}}{2s_{w}c_{w}v}\label{eq:-73}
\end{eqnarray}
The Feynman rules involving quarks read:
\begin{eqnarray}
\left\{ \overline{u}_{q},d_{q},W^{\mu+}\right\}  & : & \frac{i\overline{c}_{HQ}^{\prime}eV^{CKM}\gamma^{\mu}P_{-}}{\sqrt{2}s_{w}}+\frac{i\overline{c}_{Hud}e\gamma^{\mu}P_{+}}{\sqrt{2}s_{w}}\label{eq:-74}\\
\left\{ \overline{u}_{q},d_{q},h,h,W^{\mu+}\right\}  & : & \frac{i\sqrt{2}\overline{c}_{HQ}^{\prime}eV^{CKM}\gamma^{\mu}P_{-}}{s_{w}v^{2}}+\frac{i\sqrt{2}\overline{c}_{Hud}e\gamma^{\mu}P_{+}}{s_{w}v^{2}}\label{eq:-75}\\
\left\{ \overline{u}_{q},d_{q},h,W^{\mu+}\right\}  & : & \frac{i\sqrt{2}\overline{c}_{HQ}^{\prime}eV^{CKM}\gamma^{\mu}P_{-}}{s_{w}v}+\frac{i\sqrt{2}\overline{c}_{Hud}e\gamma^{\mu}P_{+}}{s_{w}v}\label{eq:-76}\\
\left\{ \overline{u}_{q},u_{q},Z^{\mu}\right\}  & : & -\frac{i\overline{c}_{HQ}e\gamma^{\mu}P_{-}}{2s_{w}c_{w}}+\frac{i\overline{c}_{HQ}^{\prime}e\gamma^{\mu}P_{-}}{2s_{w}c_{w}}-\frac{i\overline{c}_{Hu}e\gamma^{\mu}P_{+}}{2s_{w}c_{w}}\label{eq:-77}\\
\left\{ \overline{u}_{q},u_{q},h,h,Z^{\mu}\right\}  & : & -\frac{i\overline{c}_{HQ}e\gamma^{\mu}P_{-}}{s_{w}c_{w}v^{2}}+\frac{i\overline{c}_{HQ}^{\prime}e\gamma^{\mu}P_{-}}{s_{w}c_{w}v^{2}}-\frac{i\overline{c}_{Hu}e\gamma^{\mu}P_{+}}{s_{w}c_{w}v^{2}}\label{eq:-78}\\
\left\{ \overline{u}_{q},u_{q},h,Z^{\mu}\right\}  & : & -\frac{i\overline{c}_{HQ}e\gamma^{\mu}P_{-}}{s_{w}c_{w}v}+\frac{i\overline{c}_{HQ}^{\prime}e\gamma^{\mu}P_{-}}{s_{w}c_{w}v}-\frac{i\overline{c}_{Hu}e\gamma^{\mu}P_{+}}{s_{w}c_{w}v}\label{eq:-79}\\
\left\{ \overline{d}_{q},u_{q},W^{\mu-}\right\}  & : & \frac{i\overline{c}_{HQ}^{\prime}e\left(V^{CKM}\right)^{\ast}\gamma^{\mu}P_{-}}{\sqrt{2}s_{w}}+\frac{i\overline{c}_{Hud}e\gamma^{\mu}P_{+}}{\sqrt{2}s_{w}}\label{eq:-80}\\
\left\{ \overline{d}_{q},u_{q},h,h,W^{\mu-}\right\}  & : & \frac{i\sqrt{2}\overline{c}_{HQ}^{\prime}e\left(V^{CKM}\right)^{\ast}\gamma^{\mu}P_{-}}{s_{w}v^{2}}+\frac{i\sqrt{2}\overline{c}_{Hud}e\gamma^{\mu}P_{+}}{s_{w}v^{2}}\label{eq:-81}\\
\left\{ \overline{d}_{q},u_{q},h,W^{\mu-}\right\}  & : & \frac{i\sqrt{2}\overline{c}_{HQ}^{\prime}e\left(V^{CKM}\right)^{\ast}\gamma^{\mu}P_{-}}{s_{w}v}+\frac{i\sqrt{2}\overline{c}_{Hud}e\gamma^{\mu}P_{+}}{s_{w}v}\label{eq:-82}\\
\left\{ \overline{d}_{q},d_{q},Z^{\mu}\right\}  & : & -\frac{i\overline{c}_{HQ}e\gamma^{\mu}P_{-}}{2s_{w}c_{w}}-\frac{i\overline{c}_{HQ}^{\prime}e\gamma^{\mu}P_{-}}{2s_{w}c_{w}}-\frac{i\overline{c}_{Hd}e\gamma^{\mu}P_{+}}{2s_{w}c_{w}}\label{eq:-83}\\
\left\{ \overline{d}_{q},d_{q},h,h,Z^{\mu}\right\}  & : & -\frac{i\overline{c}_{HQ}e\gamma^{\mu}P_{-}}{s_{w}c_{w}v^{2}}-\frac{i\overline{c}_{HQ}^{\prime}e\gamma^{\mu}P_{-}}{s_{w}c_{w}v^{2}}-\frac{i\overline{c}_{Hd}e\gamma^{\mu}P_{+}}{s_{w}c_{w}v^{2}}\label{eq:-84}\\
\left\{ \overline{d}_{q},d_{q},h,Z^{\mu}\right\}  & : & -\frac{i\overline{c}_{HQ}e\gamma^{\mu}P_{-}}{s_{w}c_{w}v}-\frac{i\overline{c}_{HQ}^{\prime}e\gamma^{\mu}P_{-}}{s_{w}c_{w}v}-
\frac{i\overline{c}_{Hd}e\gamma^{\mu}P_{+}}{s_{w}c_{w}v}\label{eq:-85}
\end{eqnarray}

\pagebreak
\section{$e^{+}e^{-}\rightarrow hhZ$ Intermediate Calculations}

We denote the $\left(e^{+},e^{-}\right)$ momenta
by $\left(-l_{1},l_{2}\right)$ and the $\left(h,h,z\right)$ momenta
by $\left(p_{3},p_{4},p_{5}\right)$. Then,
all the terms in the $hhZ$ SM amplitude squared are given by:
\begin{widetext}
\begin{eqnarray}
\frac{1}{4}\sum\left|\mathcal{M}_{SM}^{\left(1\right)}\right|^{2} & = & \frac{4e^{2}m_{Z}^{4}\left(a_{e}^{2}+v_{e}^{2}\right)}{\left(q^{2}-m_{Z}^{2}\right)^{2}v^{4}}t_{1}\,,\\
\frac{1}{4}\sum\left|\mathcal{M}_{SM}^{\left(2\right)}\right|^{2} & = & \frac{16e^{2}m_{Z}^{8}\left(a_{e}^{2}+v_{e}^{2}\right)}{v^{4}\left(q^{2}-m_{Z}^{2}\right)^{2}}
\left(d_{1}^{2}t_{1}+\frac{1}{m_{Z}^{4}}d_{2}+\frac{2}{m_{Z}^{2}}d_{3}\right)\\
\frac{1}{4}\sum\left|\mathcal{M}_{SM}^{\left(3\right)}\right|^{2} & = & \frac{36e^{2}m_{h}^{4}m_{Z}^{4}\left(a_{e}^{2}+v_{e}^{2}\right)}{v^{4}\left(q^{2}-m_{Z}^{2}\right)^{2}}
\frac{t_{1}}{\left(2p_{3}\cdot p_{4}+m_{h}^{2}\right)^{2}} ~,\\
\frac{1}{4}\sum\left(\mathcal{M}_{SM}^{\left(1\right)\ast}\mathcal{M}_{SM}^{\left(3\right)}+
\mathcal{M}_{SM}^{\left(1\right)}\mathcal{M}_{SM}^{\left(3\right)\ast}\right) & = & \frac{24e^{2}m_{h}^{2}m_{Z}^{4}\left(a_{e}^{2}+v_{e}^{2}\right)}{v^{4}\left(q^{2}-m_{Z}^{2}\right)^{2}}
\frac{t_{1}}{2p_{3}\cdot p_{4}+m_{h}^{2}}\\
\frac{1}{4}\sum\left(\mathcal{M}_{SM}^{\left(1\right)\ast}\mathcal{M}_{SM}^{\left(2\right)}+
\mathcal{M}_{SM}^{\left(1\right)}\mathcal{M}_{SM}^{\left(2\right)\ast}\right) & = & \frac{16e^{2}m_{Z}^{6}\left(a_{e}^{2}+v_{e}^{2}\right)}{v^{4}\left(q^{2}-m_{Z}^{2}\right)^{2}}\left(d_{1}t_{1}+\frac{1}{m_{Z}^{2}}d_{4}\right)\label{eq:-147}\\
\frac{1}{4}\sum\left(\mathcal{M}_{SM}^{\left(2\right)\ast}\mathcal{M}_{SM}^{\left(3\right)}+
\mathcal{M}_{SM}^{\left(2\right)}\mathcal{M}_{SM}^{\left(3\right)\ast}\right) & = & \frac{48e^{2}m_{h}^{2}m_{Z}^{6}\left(a_{e}^{2}+v_{e}^{2}\right)}{v^{4}\left(q^{2}-m_{Z}^{2}\right)^{2}}\frac{d_{1}t_{1}+\frac{1}{m_{Z}^{2}}d_{4}}{\left(2p_{3}\cdot p_{4}+m_{h}^{2}\right)}~.
\end{eqnarray}
\end{widetext}
where we introduced the following kinematic variables:
\begin{widetext}
\begin{eqnarray}
d_{1} & = & \frac{1}{2p_{4}\cdot p_{5}+m_{h}^{2}}+\frac{1}{2p_{3}\cdot p_{5}+m_{h}^{2}}\label{eq:-124}\\
d_{2} & = & \frac{t_{2}}{\left(2p_{4}\cdot p_{5}+m_{h}^{2}\right)^{2}}+\frac{t_{2}^{\left(3\right)}}{\left(2p_{3}\cdot p_{5}+m_{h}^{2}\right)^{2}}+\frac{2t_{3}}{\left(2p_{4}\cdot p_{5}+m_{h}^{2}\right)\left(2p_{3}\cdot p_{5}+m_{h}^{2}\right)}\label{eq:-125}\\
d_{3} & = & \frac{t_{4}}{\left(2p_{4}\cdot p_{5}+m_{h}^{2}\right)^{2}}+\frac{t_{4}^{\left(3\right)}}{\left(2p_{3}\cdot p_{5}+m_{h}^{2}\right)^{2}}+\frac{t_{4}+t_{4}^{\left(3\right)}}{\left(2p_{3}\cdot p_{5}+m_{h}^{2}\right)\left(2p_{4}\cdot p_{5}+m_{h}^{2}\right)}\label{eq:-126}\\
d_{4} & = & \frac{t_{4}}{2p_{4}\cdot p_{5}+m_{h}^{2}}+\frac{t_{4}^{\left(3\right)}}{2p_{3}\cdot p_{5}+m_{h}^{2}}\label{eq:-127}
\end{eqnarray}
\end{widetext}
and
\begin{widetext}
\begin{eqnarray}
t_{1} & = & l_{2}\cdot\left(-l_{1}\right)+\frac{2\left(l_{2}\cdot p_{5}\right)\left(\left(-l_{1}\right)\cdot p_{5}\right)}{m_{Z}^{2}}\label{eq:-128}\\
t_{2} & = & \left(2\left(-l_{1}\right)\cdot\left(p_{4}+p_{5}\right)l_{2}\cdot\left(p_{4}+p_{5}\right)-l_{2}\cdot\left(-l_{1}\right)\left(p_{4}+p_{5}\right)^{2}\right)\left(-m_{h}^{2}+\frac{\left(p_{4}\cdot p_{5}\right)^{2}}{m_{Z}^{2}}\right)\label{eq:-129}\\
t_{3} & = & \left[\left(-l_{1}\right)\cdot\left(p_{3}+p_{5}\right)l_{2}\cdot\left(p_{4}+p_{5}\right)+\left(-l_{1}\right)\cdot\left(p_{4}+p_{5}\right)l_{2}\cdot\left(p_{3}+p_{5}\right)-l_{2}\cdot\left(-l_{1}\right)\left(p_{3}+p_{5}\right)\cdot\left(p_{4}+p_{5}\right)\right]\times\nonumber \\
 & \times & \left(-\left(p_{3}\cdot p_{4}\right)+\frac{\left(p_{3}\cdot p_{5}\right)\left(p_{4}\cdot p_{5}\right)}{m_{Z}^{2}}\right)\label{eq:-130}\\
t_{4} & = & 2l_{2}\cdot\left(p_{4}+p_{5}\right)\left(-l_{1}\right)\cdot\left(p_{4}+p_{5}\right)-l_{2}\cdot\left(-l_{1}\right)\left(p_{4}+p_{5}\right)^{2}\nonumber \\
 & - & \left[\left(l_{2}\cdot p_{5}\right)\left(-l_{1}\right)\cdot\left(p_{4}+p_{5}\right)+\left(-l_{1}\right)\cdot p_{5}l_{2}\cdot\left(p_{4}+p_{5}\right)-l_{2}\cdot\left(-l_{1}\right)\left(p_{4}\cdot p_{5}+m_{Z}^{2}\right)\right]\frac{\left(p_{4}\cdot p_{5}+m_{Z}^{2}\right)}{m_{Z}^{2}}\label{eq:-131}\\
t_{2}^{\left(3\right)} & = & t_{2}\left(p_{4}\rightarrow p_{3}\right)\label{eq:-132}\\
t_{4}^{\left(3\right)} & = & t_{4}\left(p_{4}\rightarrow p_{3}\right)\label{eq:-133}
\end{eqnarray}
\end{widetext}

\pagebreak

\section{Single Higgs production  - diagrams}

In this Appendix we depict the SM + NP diagrams for $e^+ e^- \to hZ$ followed by $Z \to x$ for all $h+x$ channels,
(that are calculated by MG5 in section \ref{sec4}) in the presence of $\mathcal{O}_{HL}$. Namely, all
diagrams for the processes $e^{+}e^{-}\rightarrow h\nu_{e}\overline{\nu},hl^{+}l^{-},hb\overline{b}$

The full set of diagrams for $e^{+}e^{-}\rightarrow h\nu_{e}\overline{\nu}_{e}$ is:

\begin{center}
\includegraphics[scale=0.37]{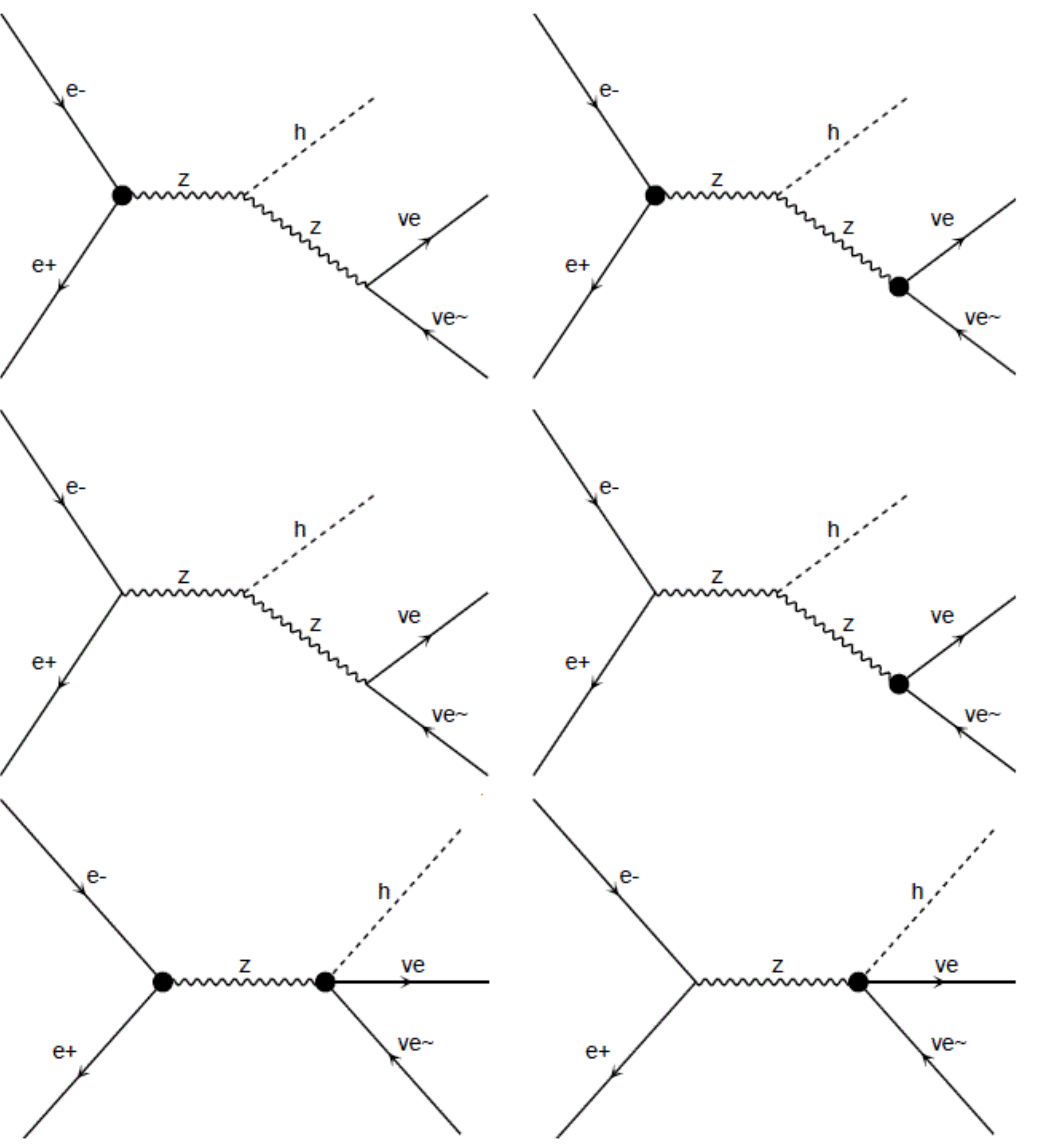}
\includegraphics[scale=0.37]{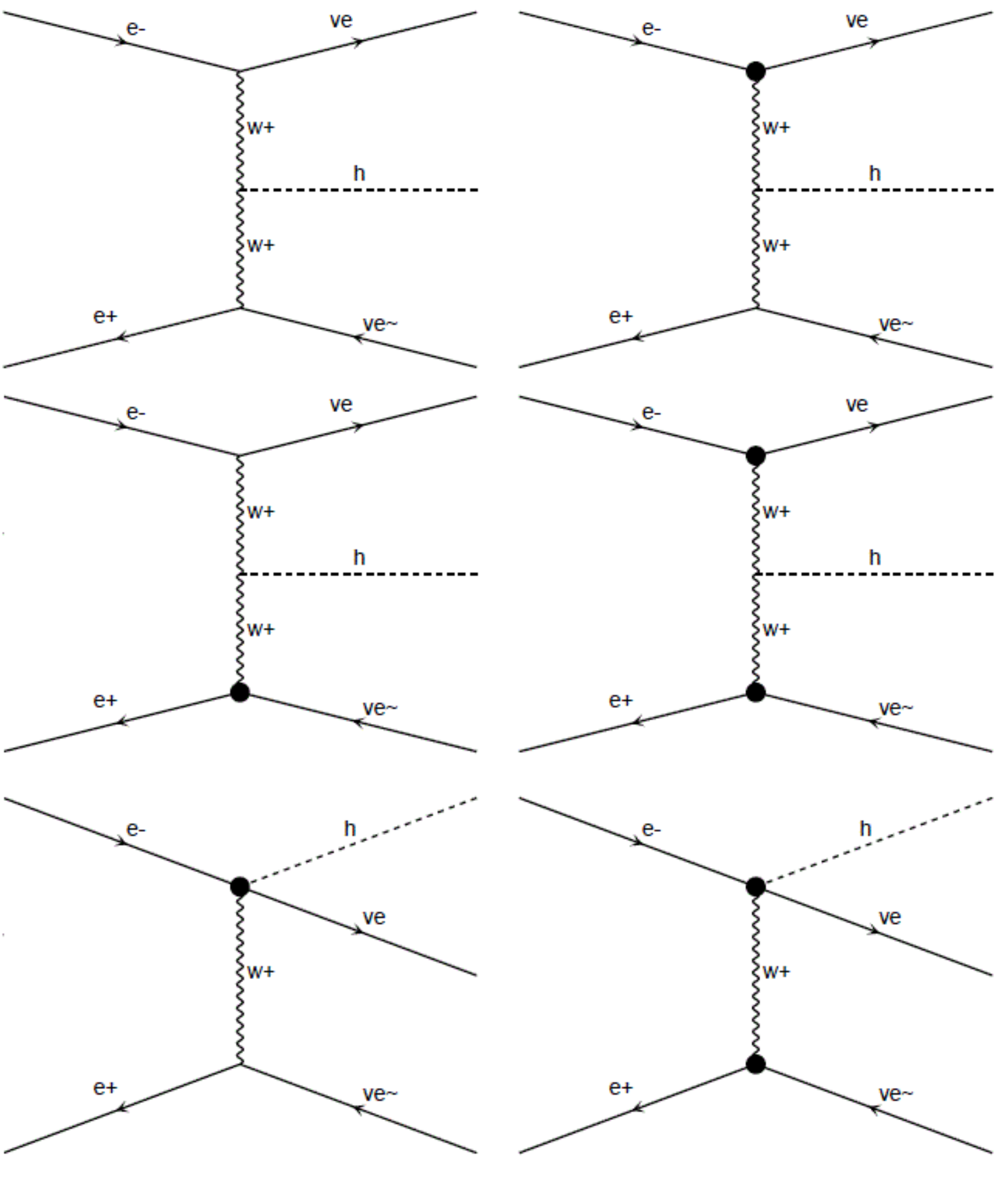}
\includegraphics[scale=0.37]{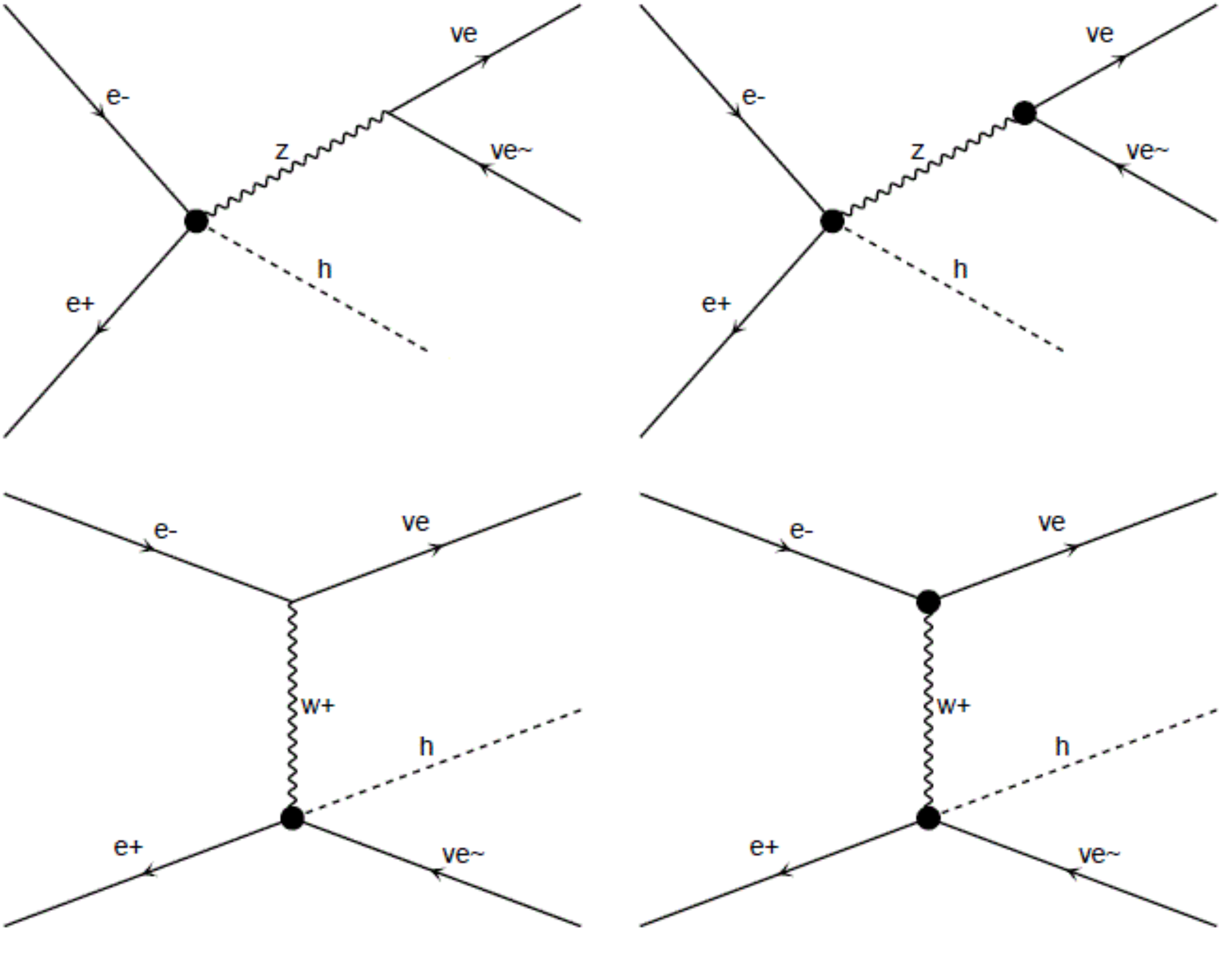}
\end{center}
\pagebreak
The full set of diagrams for $e^{+}e^{-}\rightarrow hl^{+}l^{-}$ is:
\begin{center}
\includegraphics[scale=0.37]{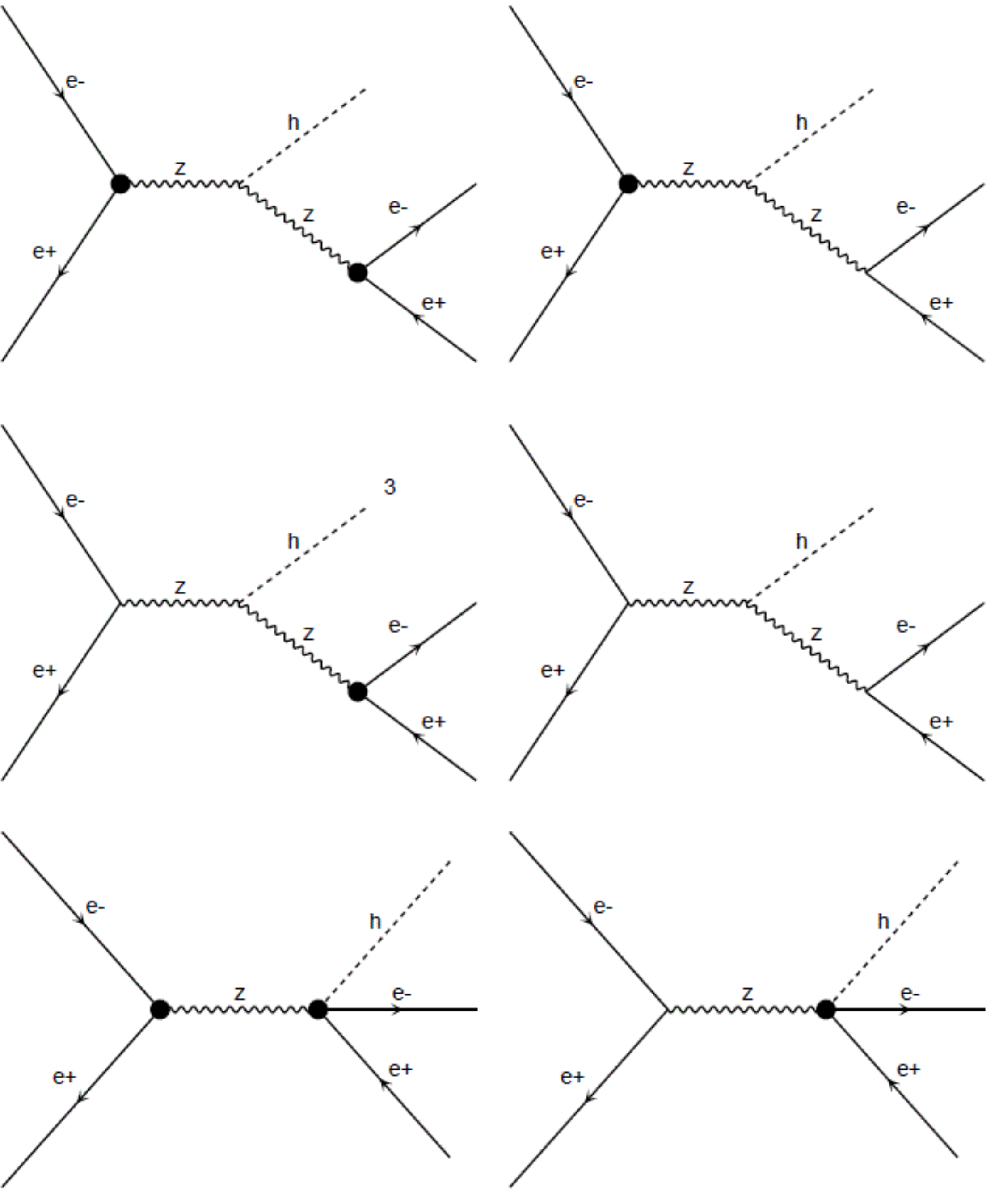}
\includegraphics[scale=0.37]{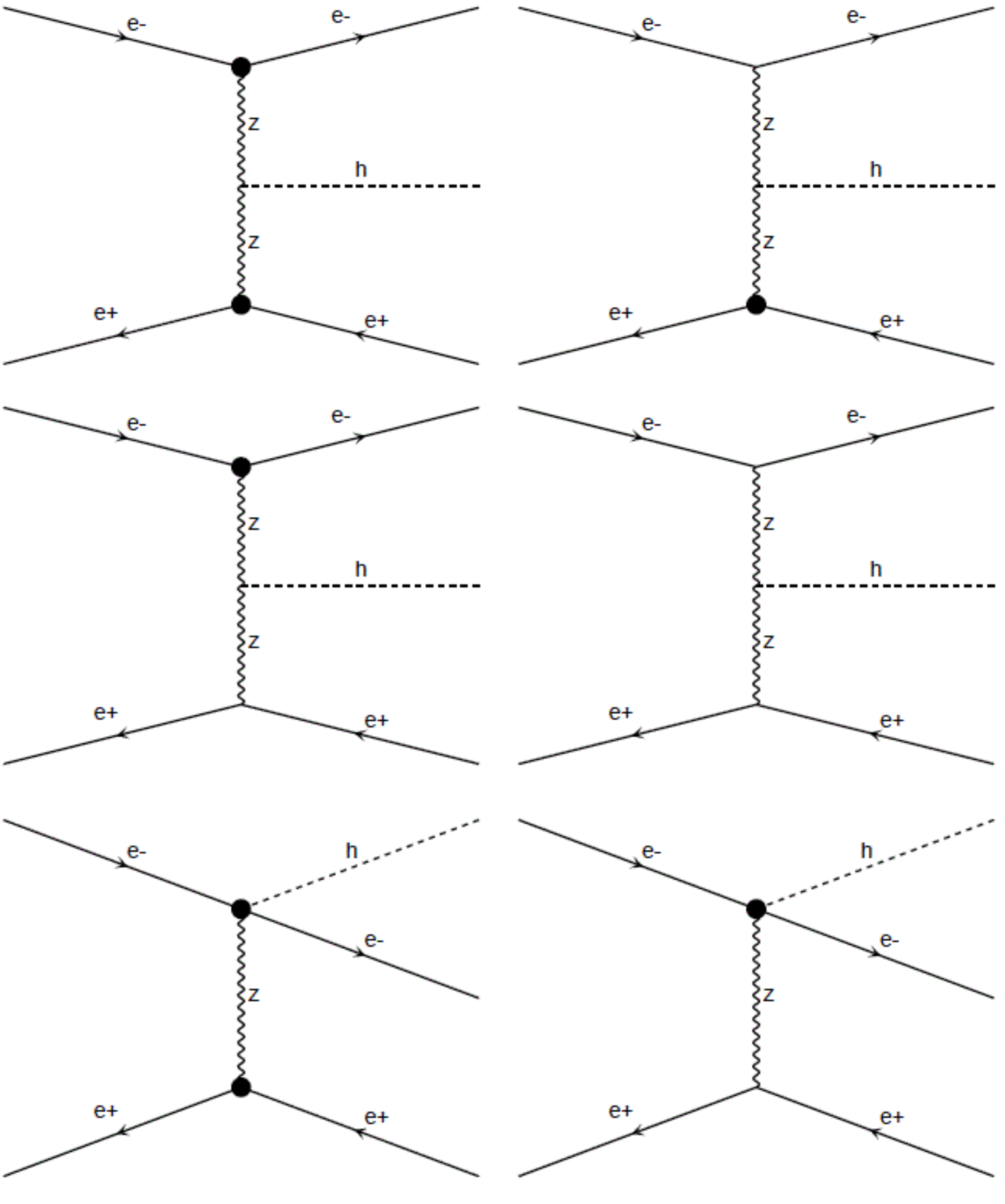}
\includegraphics[scale=0.37]{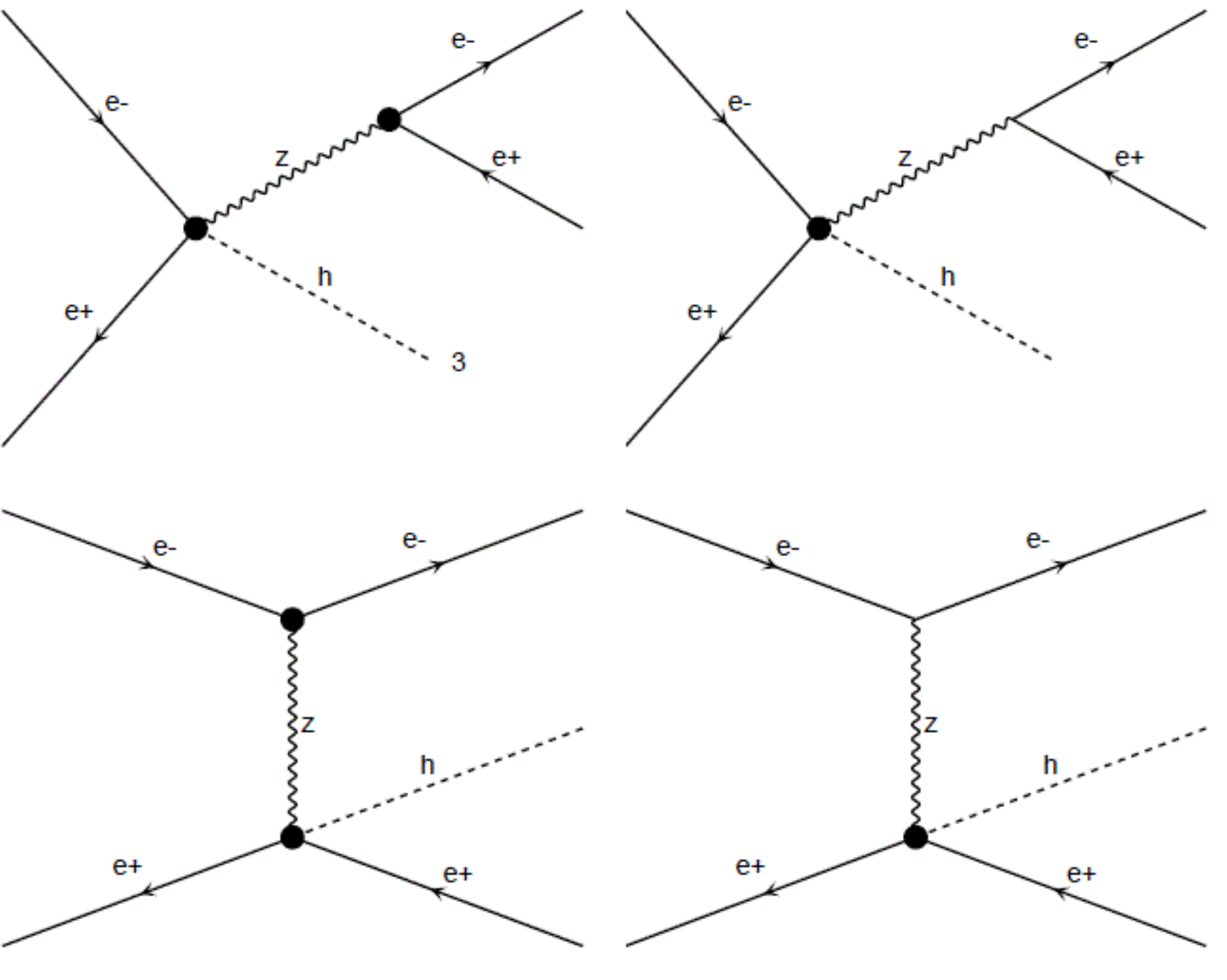}
\includegraphics[scale=0.37]{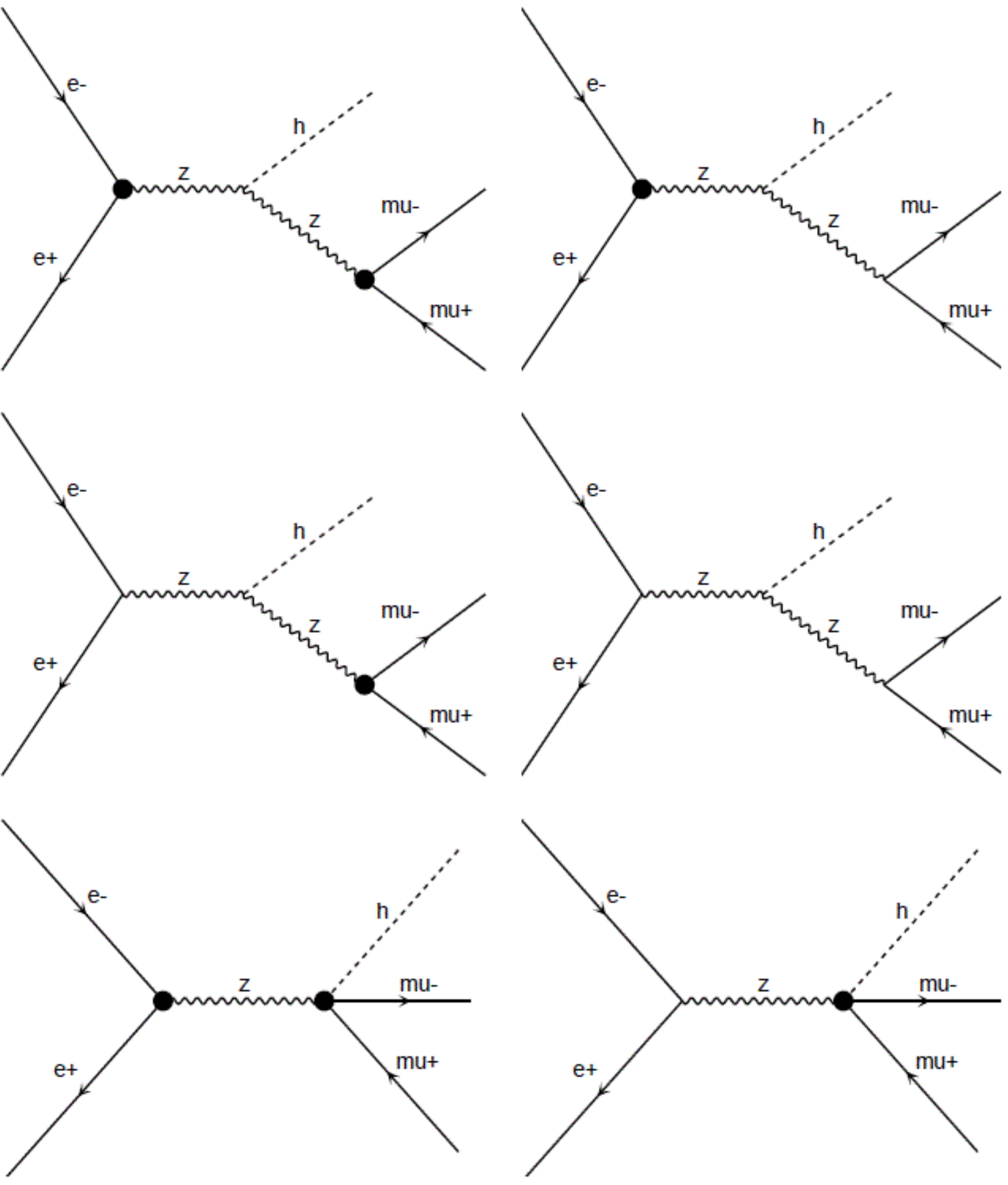}
\includegraphics[scale=0.37]{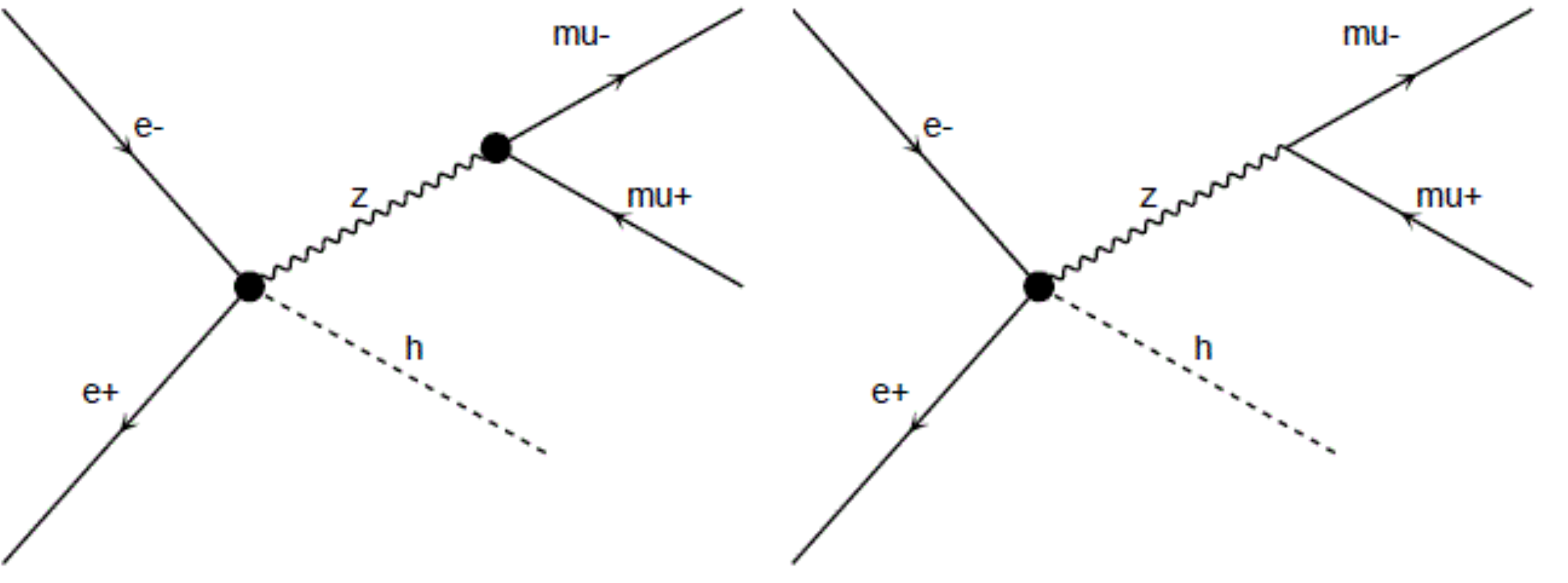}
\end{center}

The full set of diagrams for $e^{+}e^{-}\rightarrow hb\overline{b}$ is:
\begin{center}
\includegraphics[scale=0.37]{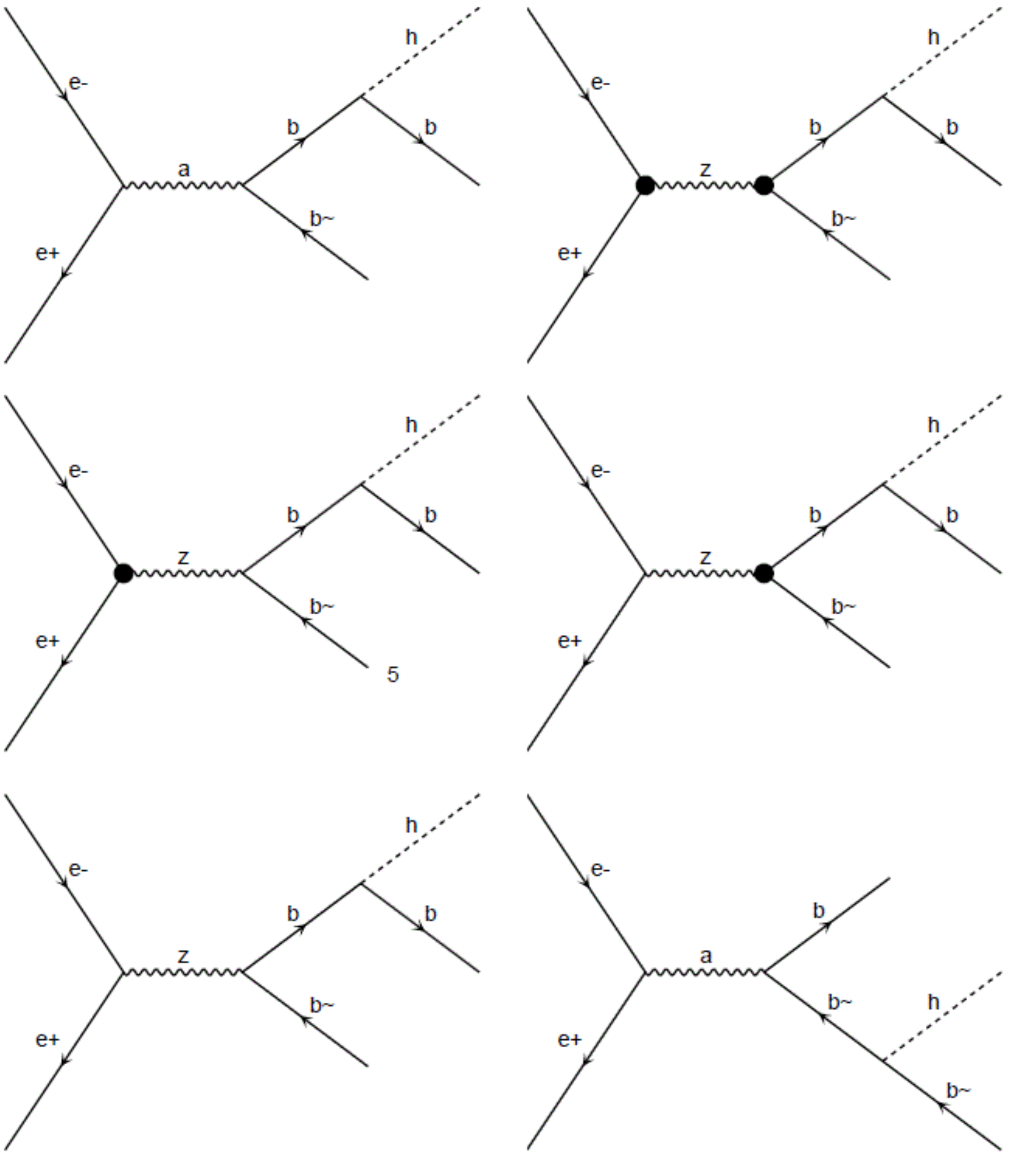} 
\includegraphics[scale=0.37]{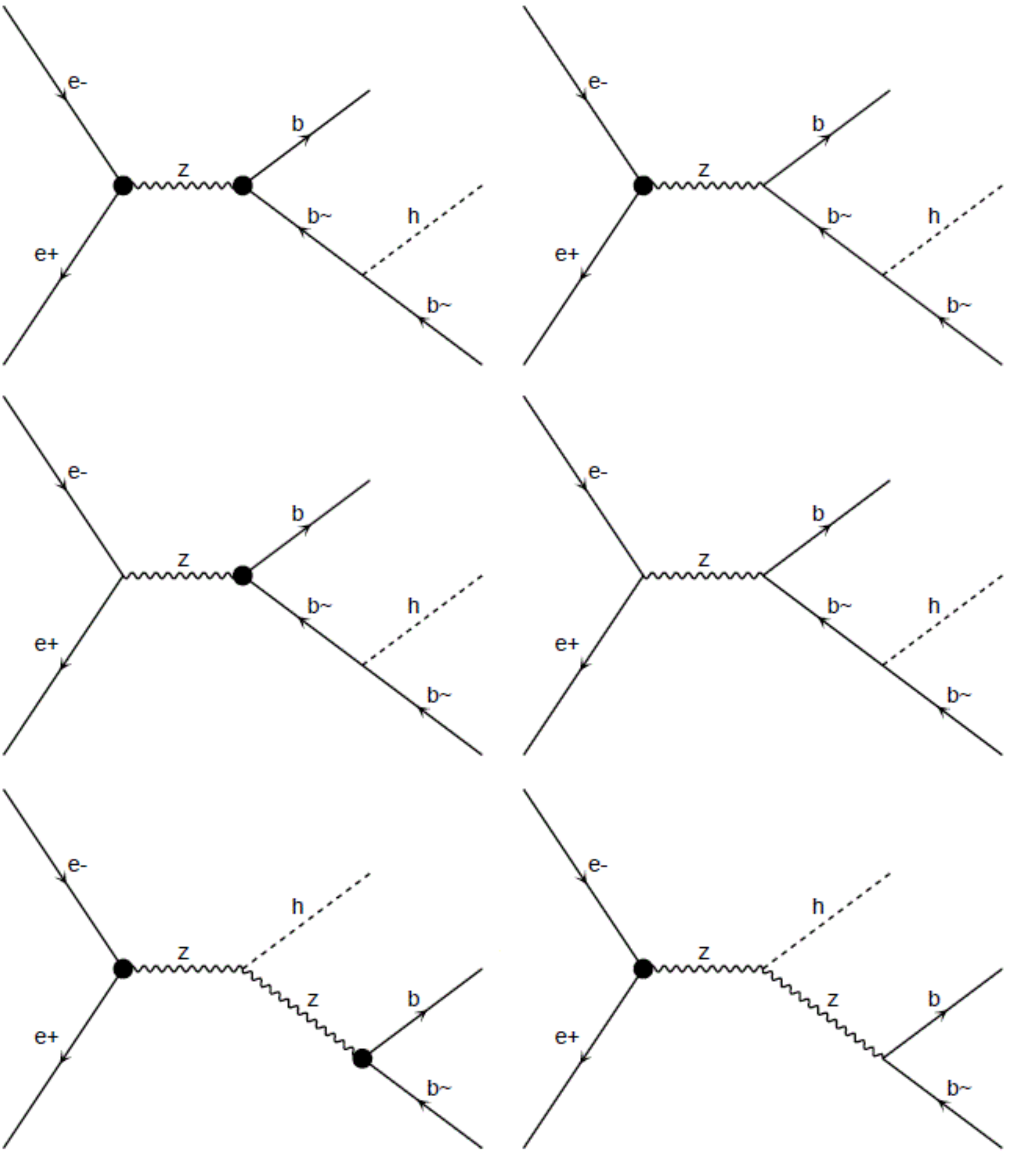} 
\includegraphics[scale=0.37]{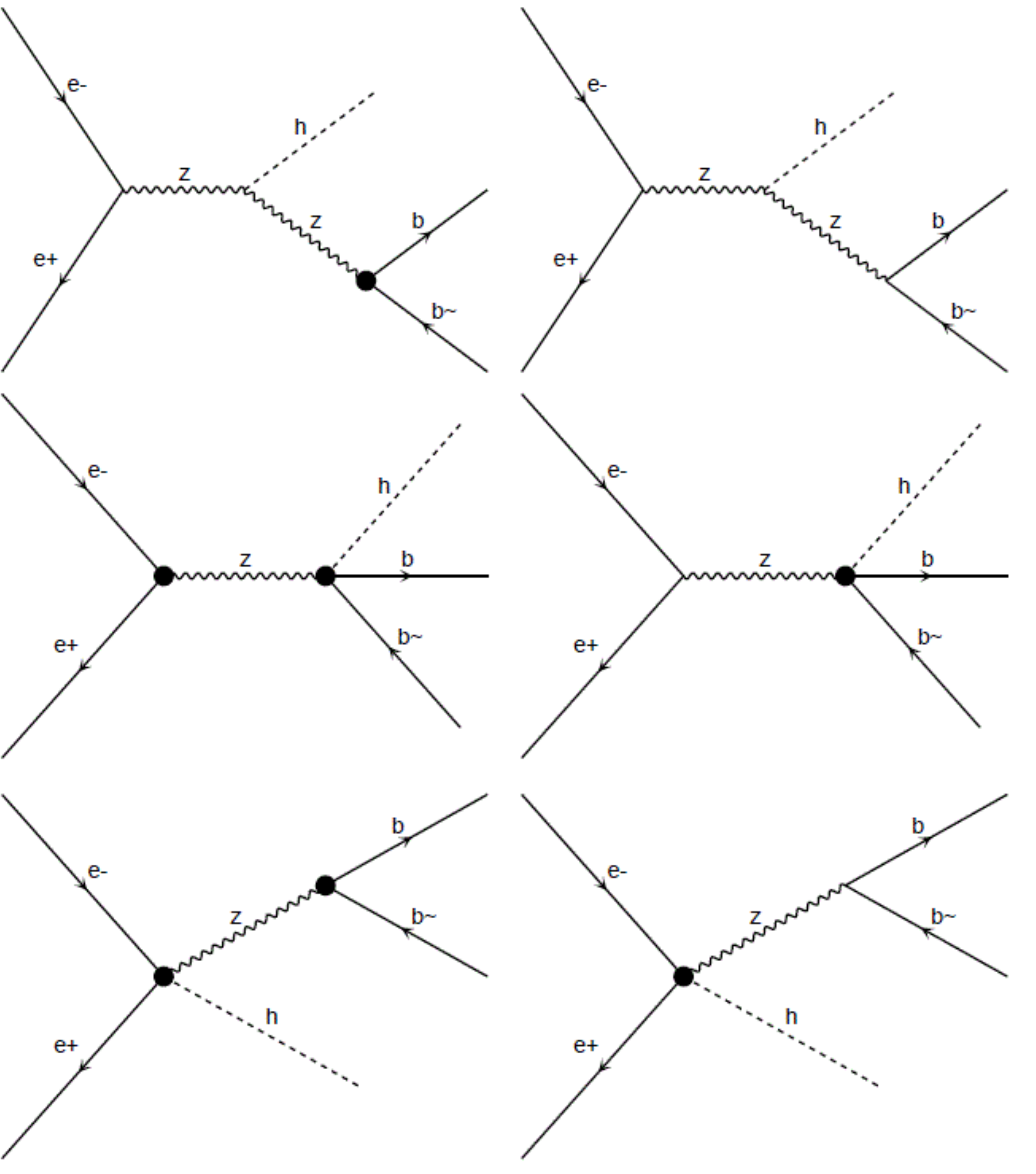}
\end{center}

\pagebreak
\section{Higgs pair production - sample diagrams}

In this Appendix we depict a sample of the SM + NP diagrams for $e^+ e^- \to hhZ$ followed by
$Z \to x$ for all $hh+x$ channels,
(that are calculated by MG5 in section \ref{sec4}) in the presence of $\mathcal{O}_{HL}$. Namely, all
diagrams for the processes $e^{+}e^{-}\rightarrow hh\nu_{e}\overline{\nu},hhl^{+}l^{-},hhb\overline{b}$.

A sample of the full set of diagrams for $e^{+}e^{-}\rightarrow hh\nu_{e}\overline{\nu}_{e}$ is:
\begin{center}
\includegraphics[scale=0.37]{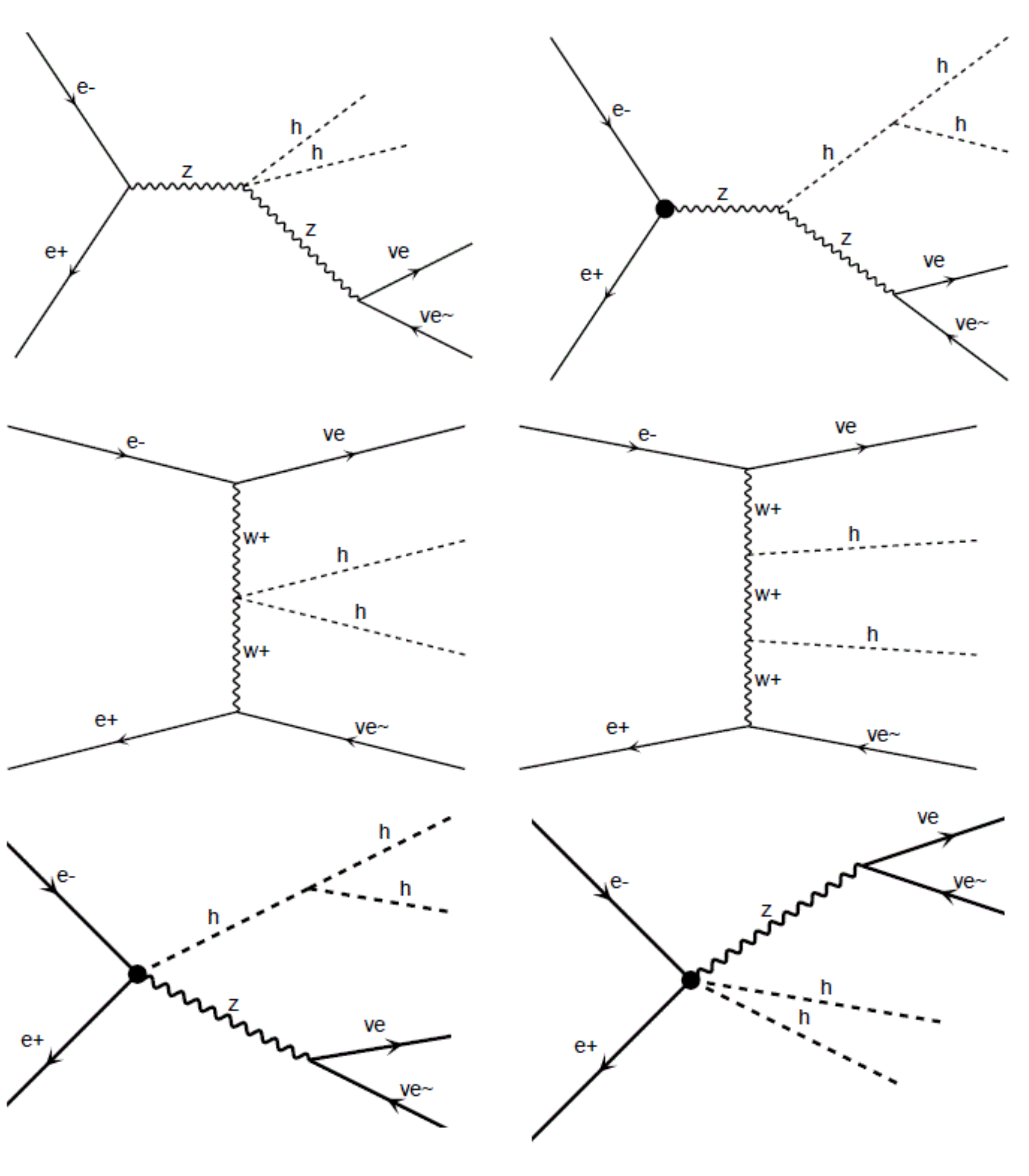}
\end{center}

A sample of the full set of diagrams for $e^{+}e^{-}\rightarrow hhl^{+}l^{-}$ is:
\begin{center}
\includegraphics[scale=0.37]{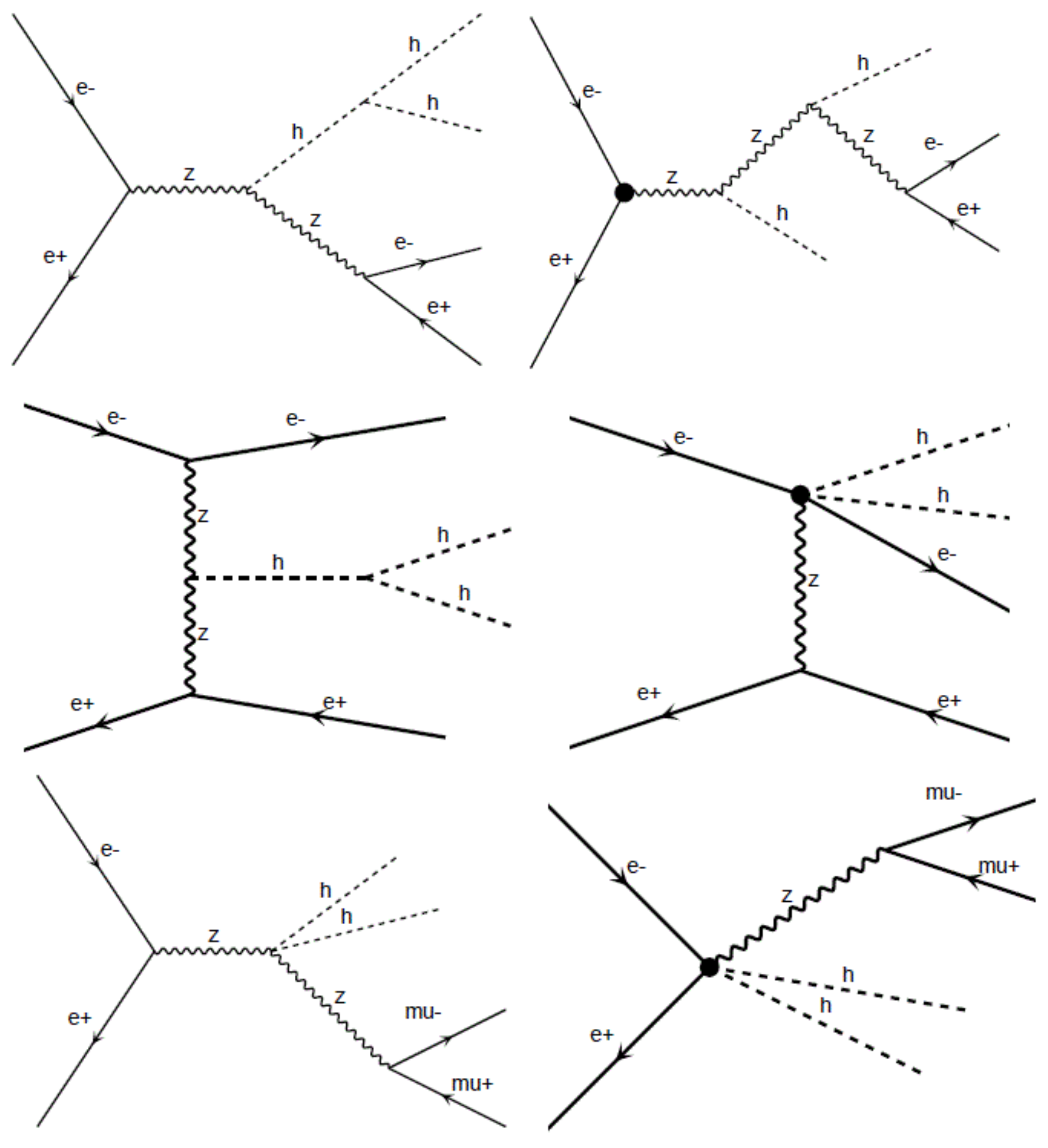}
\end{center}

A sample of the full set of diagrams for $e^{+}e^{-}\rightarrow hhb\overline{b}$ is:
\begin{center}
\includegraphics[scale=0.37]{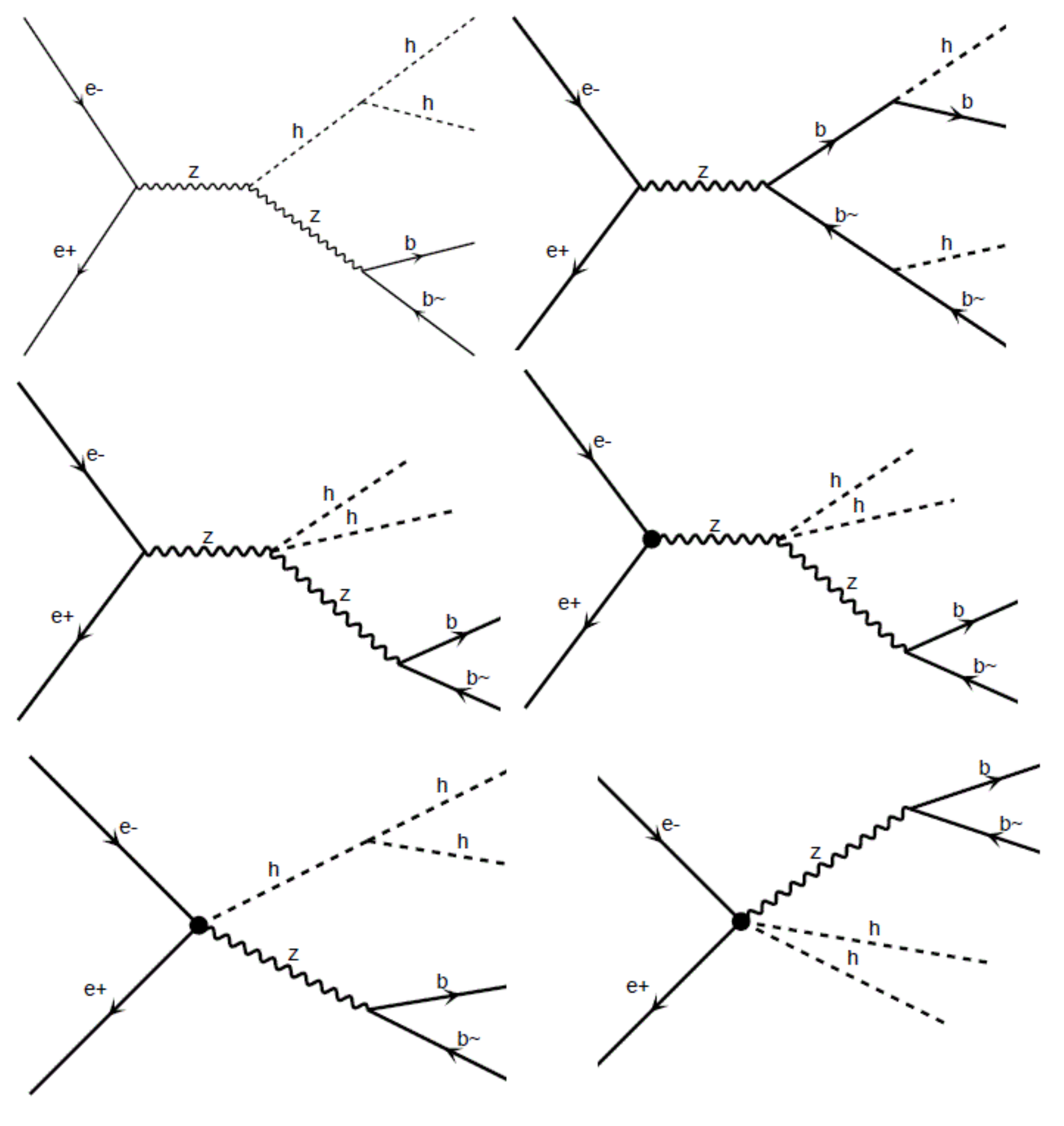}
\end{center}

\end{document}